\documentclass[floatfix,reprint, amsmath,amssymb,aps,pra,superscriptaddress,nofootinbib]{revtex4-2}

\usepackage{graphicx}
\usepackage{dcolumn}
\usepackage{bm}
\usepackage{graphicx}
\usepackage[utf8]{inputenc}
\usepackage[export]{adjustbox}
\usepackage{wrapfig}
\usepackage{amsmath}
\DeclareMathOperator{\Tr}{Tr}

\usepackage{amsfonts} 
\usepackage{hyperref}
\usepackage{braket}
\usepackage{xcolor}

\usepackage[normalem]{ulem}

\usepackage{comment}
\usepackage{amsmath, amssymb, amsthm}
\usepackage{physics}
\usepackage{makecell}
\usepackage{float}
\usepackage{ragged2e}
\usepackage{blindtext}
\usepackage{titlesec}
\usepackage{hyperref}
\usepackage{paralist}
\usepackage{braket}
\usepackage{booktabs}
\usepackage{multirow}

\newcommand{\mH}{\mathcal{H}}

\newcommand{\mT}{\mathcal{T}}



\begin{document}

\preprint{APS/123-QED}

\title{Pre-Distillation of Magic States via Composite Schemes}

\author{Muhammad Erew}%
\affiliation{
 Raymond and Beverly Sackler School of Physics and Astronomy, Tel-Aviv University, Tel-Aviv 6997801, Israel.
 }

\author{Moshe Goldstein}%
\affiliation{
 Raymond and Beverly Sackler School of Physics and Astronomy, Tel-Aviv University, Tel-Aviv 6997801, Israel.
 }
 
\author{Yaron Oz}%
\affiliation{
 Raymond and Beverly Sackler School of Physics and Astronomy, Tel-Aviv University, Tel-Aviv 6997801, Israel.
 }

\author{Haim Suchowski}%
\affiliation{
 Raymond and Beverly Sackler School of Physics and Astronomy, Tel-Aviv University, Tel-Aviv 6997801, Israel.
 }

\date{\today}

\begin{abstract}
Magic state distillation (MSD) is a cornerstone of fault-tolerant quantum computing, enabling non-Clifford gates via state injection into stabilizer circuits.
However, the substantial overhead of current MSD protocols remains a major obstacle to scalable implementations.
We propose a general framework for \emph{pre-distillation}, based on composite pulse sequences that suppress systematic errors in the generation of magic states.
While most composite designs target simple gates such as $X$, $Z$, or Hadamard, our schemes directly implement the non-Clifford $\mT$ gate with enhanced robustness to the targeted systematic errors.
We develop composite sequences tailored to the dominant control imperfections in superconducting, trapped-ion, neutral-atom, and integrated photonic platforms.
To quantify improvement in the implementation, we introduce an operationally motivated fidelity measure specifically tailored to the $\mT$ gate: the T-magic error, which captures the gate’s effectiveness in preparing high-fidelity magic states.
We further show that the error in the channel arising from the injection of faulty magic states scales linearly with the leading-order error of the states.
Within the systematic-error-dominated regimes considered, this approach lowers the number of required distillation levels by up to three across the platforms we study, translating to substantial qubit-overhead savings and offering a practical route toward more resource-efficient universal quantum computation.

\end{abstract}

\maketitle

\section{Introduction}

Quantum computing promises computational advantages over classical methods, with celebrated examples such as Shor’s factoring and Grover’s search algorithms~\cite{Shor1997,Grover1996}. Yet building scalable machines remains challenging because quantum information is fragile, and error correction is restricted by fundamental no-go theorems. In particular, the Eastin-Knill theorem shows that no quantum error-correcting code admits a fully transversal universal gate set~\cite{EastinKnill,GottesmanThesis}, forcing us to seek alternative paths to universality.

A central strategy is to augment the efficiently and fault-tolerantly implementable Clifford operations, realized, for instance, through stabilizer codes, with special non-stabilizer ancilla states known as \emph{magic states}. Injecting these states into Clifford circuits enables the implementation of otherwise inaccessible non-Clifford gates, thereby promoting the Clifford group to universality~\cite{NielsenChuang,GottesmanThesis}. Since such states are typically noisy, they must be purified. This is achieved through \emph{magic state distillation} (MSD)~\cite{BravyiKitaev}, a systematic protocol that consumes multiple noisy copies of a state and, via stabilizer code measurements, produces fewer but higher-fidelity copies.

Yet, the overhead of MSD is severe: each distillation level increases the number of physical qubits exponentially~\cite{BravyiKitaev,BravyiHaah,CampbellOverhead}. For example, preparing a single logical $\ket{T}$ state with error below $10^{-15}$ may require five to ten levels of recursive distillation, consuming thousands to millions of physical qubits depending on the initial error.



To mitigate this cost, we propose a platform-aware strategy of \emph{pre-distillation} based on robust composite gate constructions that reduce the error in magic states prior to entering a distillation protocol. While most composite schemes, which are designed for elementary Clifford gates like $X$, $Z$, or Hadamard \cite{Levitt1979,Shaka1985,Shaka1987,Levitt1986,Timoney2008,Kyoseva2019,9774914,KaplanErew}, our sequences are tailored to directly implement the $\mT$ gate- the gate that produces $\ket{T}$ when acting on $\ket{0}$. By targeting the gate responsible for magic state generation itself, we reduce the physical noise at the source, thereby improving their fidelity and reducing the overhead of distillation.

We develop and analyze composite pulse schemes for a range of platforms to improve the fidelity of magic state preparation prior to distillation. For systems with $X$-$Y$ control Hamiltonians, such as superconducting, trapped-ion, and neutral-atom qubits, we construct symmetric three- and five-segment sequences that implement the $\mT$ gate while canceling first- and second-order global Rabi frequency errors, respectively. In systems governed by $X$-$Z$ dynamics, we show how such gates can be synthesized when they are not directly implementable by single-segment, and design composite sequences that achieve robustness against global detuning-like errors. Finally, in integrated photonics, we apply our framework to directional couplers subject to correlated fabrication errors and design four-segment geometries that reduce the magic state error below the distillation threshold. Across all platforms, we quantify the improvement in terms of the number of required distillation iterations and demonstrate reductions of one to three levels, corresponding to orders-of-magnitude savings in physical qubit overhead. To support this analysis, we also introduce a natural, operationally motivated fidelity measure for the $\mT$ gate, \emph{magic T-gate fidelity}, which directly reflects the success of magic state preparation and guides the estimation of distillation depth. In this way, our work bridges two previously separate directions- robust gate synthesis and magic state distillation- by showing how pulse-level error suppression translates directly into fewer distillation rounds and reduced resource overhead.

The purpose of the present work is not to provide a comprehensive model of the full noise landscape of any specific experimental platform.
Rather, our objective is to demonstrate that when systematic coherent control errors dominate the error budget in magic-state preparation, they can be strongly suppressed at the physical layer through appropriate composite-segment design, and that this suppression directly translates into a reduction of magic-state distillation overhead.
We focus on a practically relevant and experimentally common regime in which coherent systematic control errors constitute the leading contribution to the initial magic-state infidelity.
In this setting, reducing the dominant coherent component at the hardware level improves the injected state quality sufficiently to decrease the required distillation recursion depth, thereby lowering resource costs at the architectural level.
We emphasize that we do not perform a systematic comparison with modern optimal-control strategies, and the present results should therefore be interpreted as demonstrating advantages within the specific systematic-error-dominated regimes considered.

Importantly, the composite-segment framework developed here is not limited to fully correlated error models.
As discussed in our previous work (Ref.~\cite{KaplanErew}), closely related (though not identical) constructions can be adapted to arbitrary error distributions, including stochastic or mixed noise, provided that their statistical structure is known or can be reasonably estimated.
The analytical examples presented in this manuscript are restricted to fully correlated systematic errors in order to maintain conceptual clarity and tractability, and to emphasize the principal novelty of this work: the extension of composite robustness techniques to the preparation of non-Clifford resource states.

This work is organized as follows. In Section~\ref{sec:Distillation and Pre-Distillation}, we review the theoretical background of MSD and the role of the $\mT$ gate in enabling universal fault-tolerant quantum computation. Section~\ref{sec:mT and mH Gates and Physical Realizations} introduces the physical realization of the $\mT$ and $\mH$ gates across multiple quantum platforms, along with their error models. In Section~\ref{sec:X-Y based systems}, we construct composite pulse sequences that implement the $\mT$ and $\mH$ gates in $X$-$Y$ systems while achieving first- and second-order robustness to global Rabi frequency errors. In Section~\ref{sec:Pre-Distillation in $X$-$Z$ Systems ; example}, we extend this approach to $X$-$Z$ systems, providing composite schemes that reduce semi-detuning-induced errors. Section~\ref{sec:Pre-Distillation in Integrated Photonics} applies these ideas to integrated photonic platforms, presenting directional coupler designs that implement the $\mT$ gate with enhanced resilience to global fabrication deviations. In Section~\ref{sec:Errors in Gate-Channel Implementation}, we show that the channel error induced by faulty magic states increases linearly with their leading-order error, confirming that the error analysis presented earlier is sufficient. Finally, in Section~\ref{sec:Conclusion}, we conclude with a discussion of implications for scalable fault-tolerant architectures and potential extensions of our pre-distillation framework.

\section{Background: Distillation and Pre-Distillation}
\label{sec:Distillation and Pre-Distillation}



MSD lies at the heart of most approaches to universal fault-tolerant quantum computation, providing a practical means of suppressing noise in non-Clifford resource states. While the general framework of MSD has been developed in many variants, we illustrate the main principles here through the canonical five-qubit protocol for the $T$-state, which serves as a representative example. A detailed technical discussion of these preliminaries, along with circuit-level constructions, is deferred to Appendix~\ref{app:Background and Preliminaries}.



The canonical magic $T$-state is defined by $\ket{T} = \cos \beta \ket{0} + e^{i\pi/4} \sin \beta \ket{1}$, where $\cos^2(2\beta) = \frac{1}{3}$. This state is an eigenvector of the $T$-gate, $T = e^{i\pi/4} S H$, which its two orthogonal eigenstates are denoted as $\ket{T_0} = \ket{T}, \,
        \ket{T_1} = -e^{-i\pi/4} \sin \beta \ket{0} + \cos \beta \ket{1}$.

\begin{figure}[h]

    \begin{minipage}
    {0.9\linewidth}
    \raggedright (a) \hspace{100pt} (b)
    \end{minipage}

  \centering
  \includegraphics[page=1,width=0.45\linewidth]{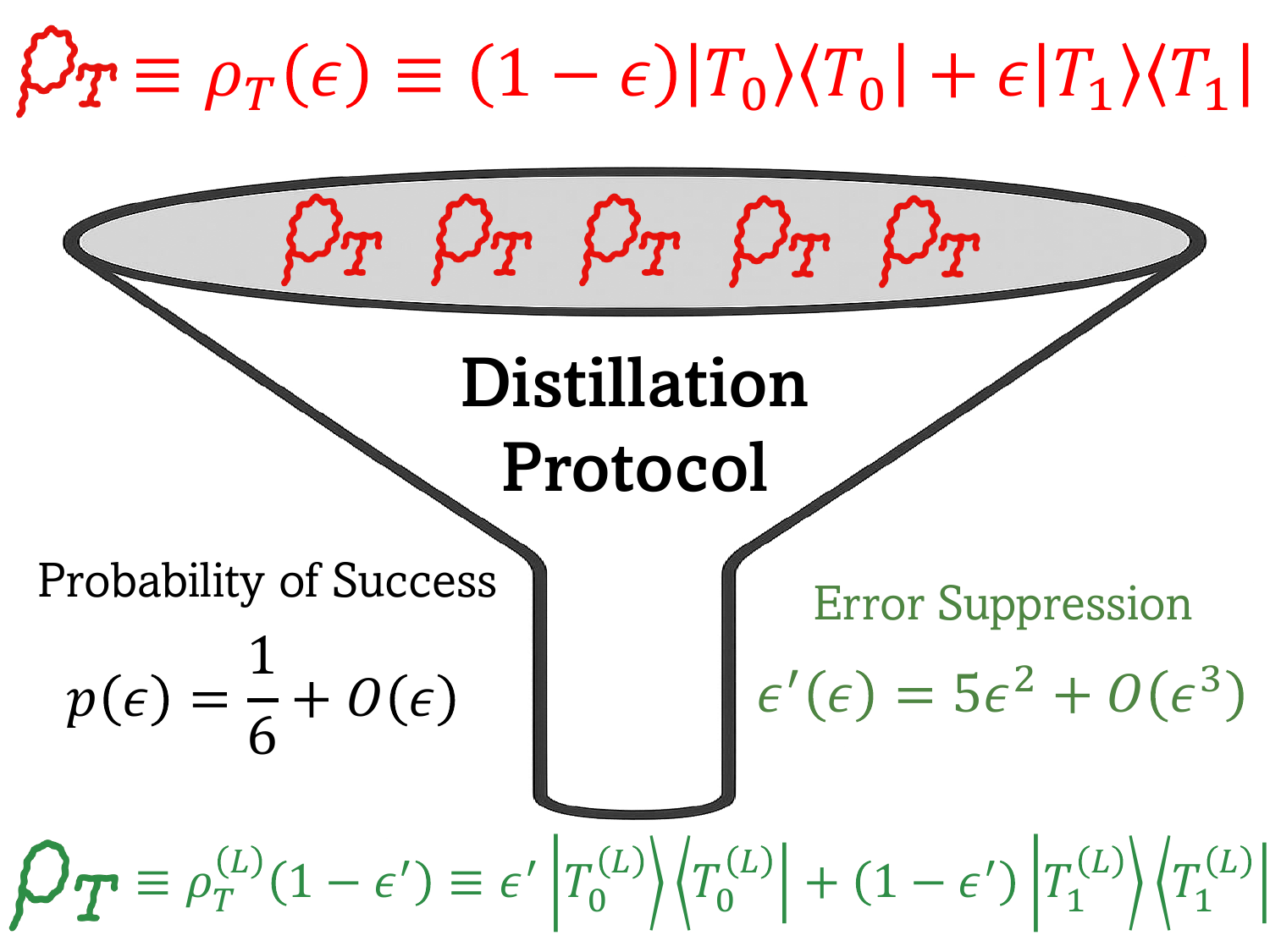}
  \includegraphics[width=0.51\linewidth]{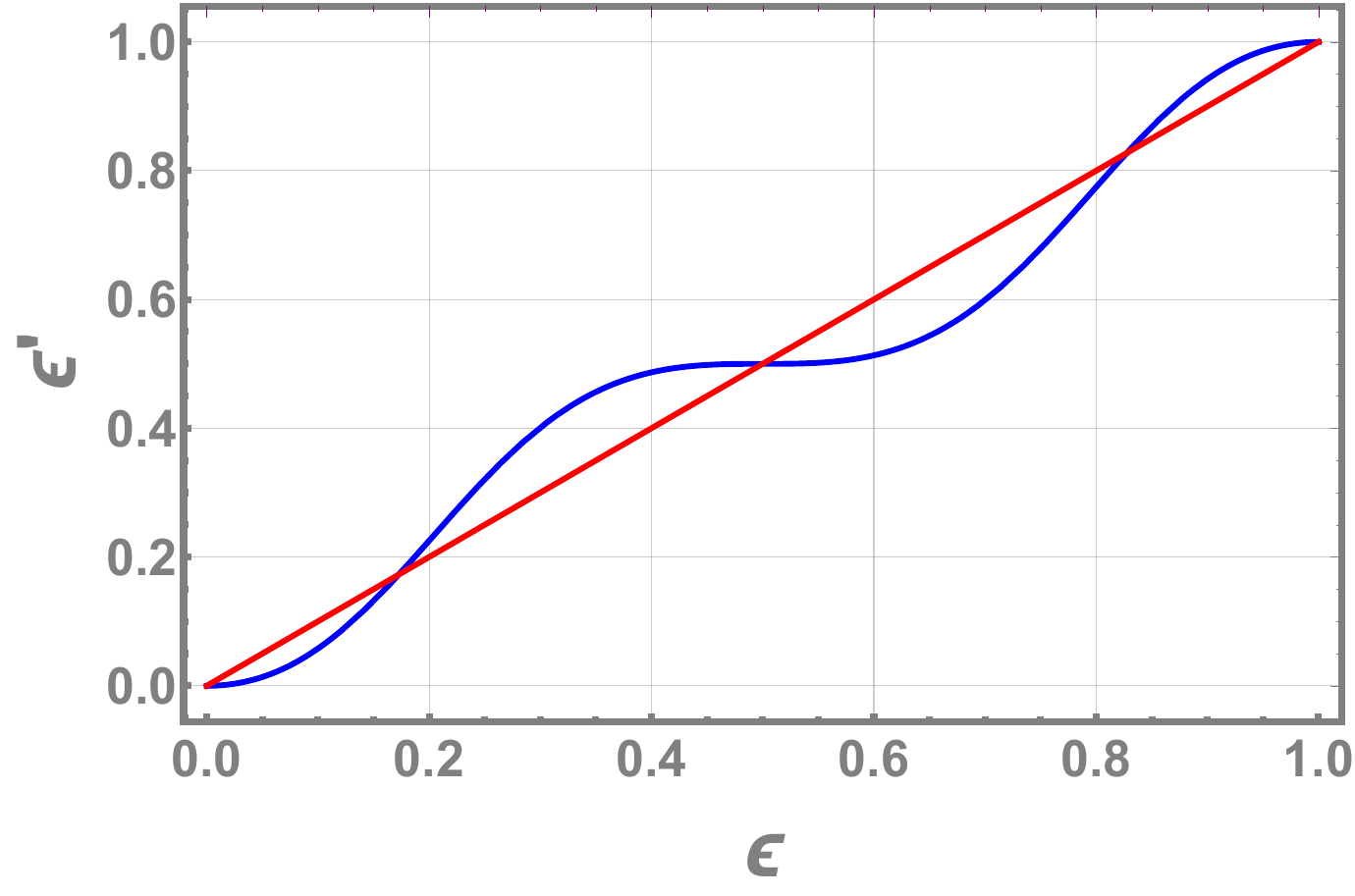}

    \begin{minipage}
    {0.9\linewidth}
    \raggedright (c) \hspace{95pt} (d)
    \end{minipage}

  \includegraphics[width=0.48\linewidth]{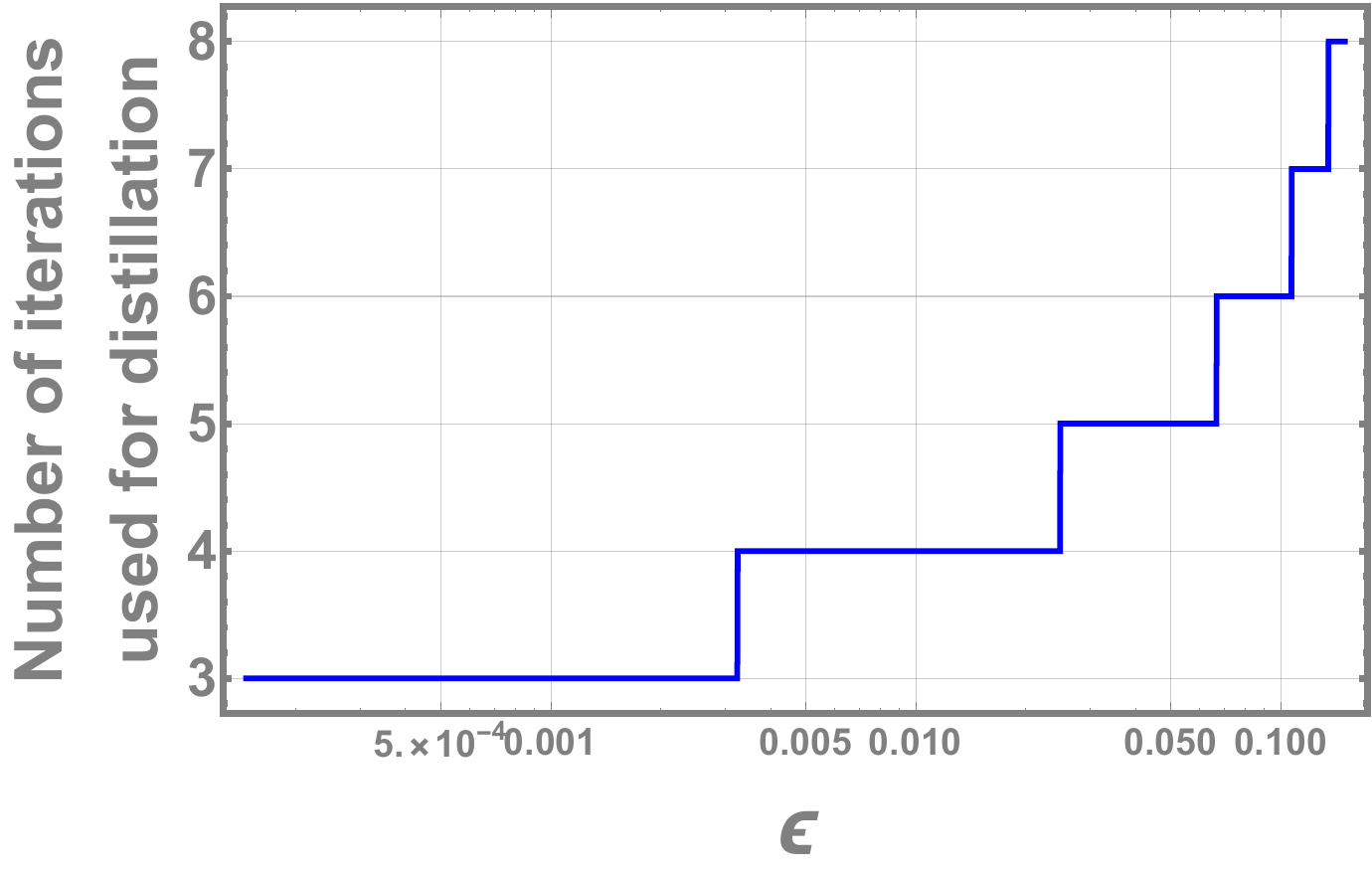}
    \includegraphics[width=0.48\linewidth]{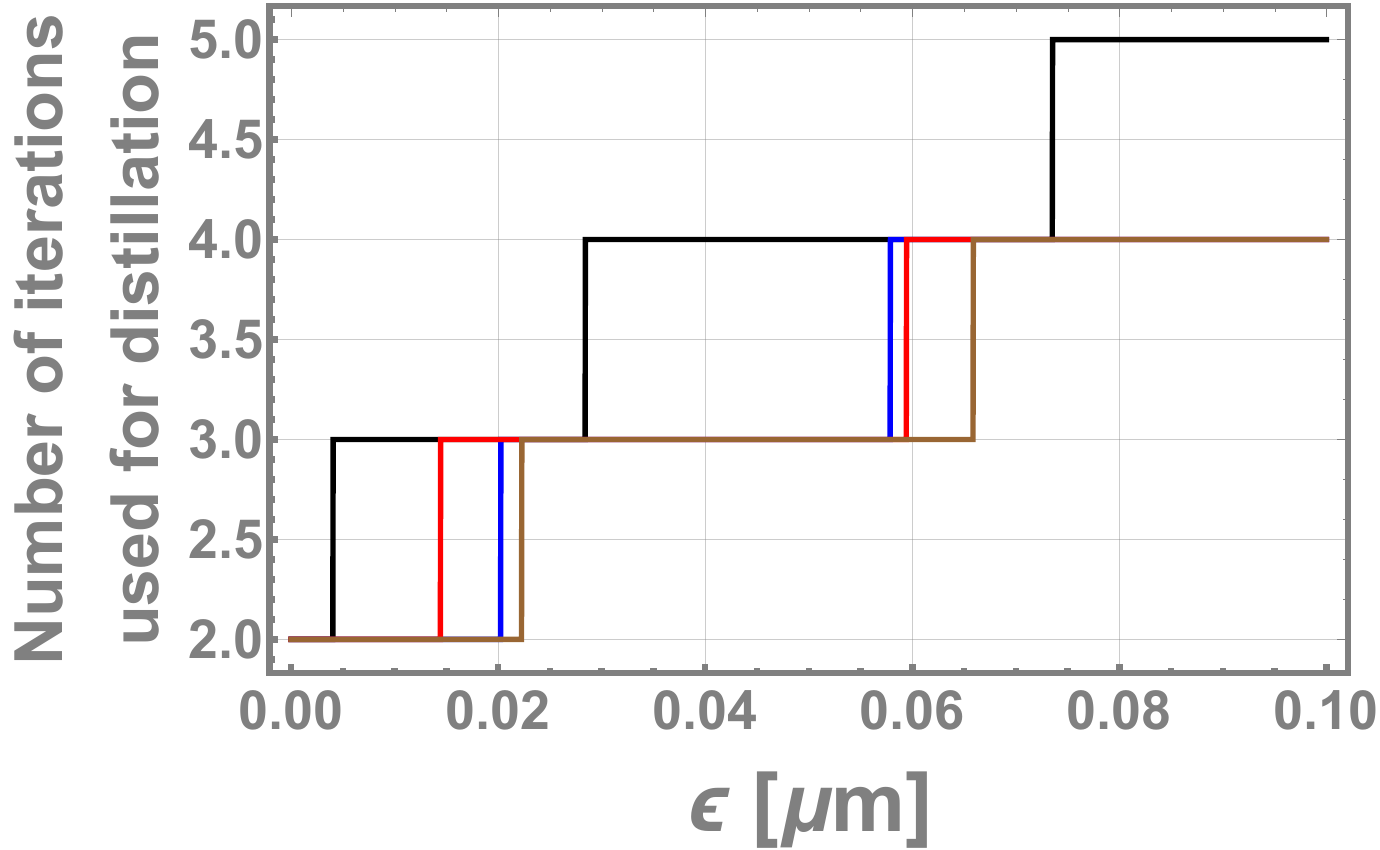}

    \caption{
        Five-qubit $T$-state distillation and its improvement through pre-distillation.
        (a) Conceptual schematic of a single round of the $T$-magic-state distillation protocol. One round consumes five noisy copies of $\rho_T(\epsilon) \equiv (1-\epsilon)\ket{T_0}\bra{T_0} + \epsilon \ket{T_1}\bra{T_1}$ (denoted by $\rho_T$ with heavily wavy red symbols to emphasize noise). With probability $p(\epsilon) = \tfrac{1}{6} + O(\epsilon)$, the protocol outputs the logical state $\rho^{(L)}_T(1-\epsilon')$, whose error is quadratically suppressed, $\epsilon'(\epsilon) = 5\epsilon^2 + O(\epsilon^3)$. We represent this improved state using green symbols with reduced waviness, reflecting its lower noise.
        (b) Distillation curve $\epsilon'(\epsilon)$ (blue): the output error rate as a function of the input error rate, showing quadratic suppression after one round. For reference, the red line shows the identity function $f(\epsilon) = \epsilon$, highlighting the improvement gained over the uncorrected case.
        (c) Required number of recursive distillation rounds to achieve an error below the $10^{-15}$ fault-tolerance threshold, illustrating the exponential scaling of qubit overhead with the initial error $\epsilon$.
        (d) Comparison of $\mathcal{T}$-gate implementations under systematic width errors in directional couplers. The plot shows the number of distillation levels needed to reach $10^{-15}$ error.
        Results are shown for a two-segment design (black) and three composite four-pulse schemes: (A) blue, (B) red, (C) brown.
    }
  \label{fig:Distillation}
\end{figure}

Figure~\ref{fig:Distillation} provides an overview. In Figure~\ref{fig:Distillation}(a), five noisy input states of the form $\rho_T(\epsilon) = (1-\epsilon)\ket{T_0}\bra{T_0} + \epsilon \ket{T_1}\bra{T_1}$ are consumed in a single round of distillation. With probability $p(\epsilon) = \tfrac{1}{6} + O(\epsilon)$, the protocol outputs a logical state $\rho^{(L)}_T(1-\epsilon')$ whose error is quadratically suppressed, $\epsilon'(\epsilon) =\frac{\epsilon^5 + 5\epsilon^2(1 - \epsilon)^3}
 {\epsilon^5 + 5\epsilon^2(1 - \epsilon)^3 + 5\epsilon^3(1 - \epsilon)^2 + (1 - \epsilon)^5} = 5\epsilon^2 + O(\epsilon^3)$.
Thus, whenever the input error lies below a critical threshold, $\epsilon_c \approx 0.173$, the protocol successfully suppresses errors, albeit with only probabilistic success and at the cost of additional resource overhead.
The technical details of the distillation circuit are in Figure~\ref{fig:DistillationAppendix} in Appendix~\ref{app:Background and Preliminaries}.
     
To further suppress the error, the protocol must be recursively applied by nesting and concatenating the code: each physical qubit is elevated to a logical qubit, which in turn is encoded into five new physical qubits. Achieving fault-tolerant thresholds therefore demands multiple rounds of recursion, resulting in an exponential growth of physical-qubit overhead per logical qubit.
Figures~\ref{fig:Distillation}(c-d) characterize the performance of the protocol. Figure~\ref{fig:Distillation}(b) shows the distillation curve $\epsilon'(\epsilon)$, making explicit how the output error depends on the input. Figure~\ref{fig:Distillation}(c) highlights the recursive structure of MSD: by concatenating multiple rounds, one can reduce the error below the fault-tolerance threshold (here taken to be $10^{-15}$). However, the required number of rounds grows rapidly with the initial error, reflecting the exponential qubit overhead inherent to this process.

Therefore, it is more efficient to reduce the error \emph{prior} to encoding, as this can save multiple levels of concatenation and lead to an exponential reduction in the qubit overhead. In this work, we employ composite pulse schemes, traditionally used for implementing gates such as $X$ and Hadamard, to enhance the fidelity of $T$-state preparation.
Figure~\ref{fig:Distillation}(d) illustrates how \emph{pre-distillation techniques} -in this case, composite pulse sequences designed to suppress systematic width errors in directional couplers- can improve overall efficiency. The figure compares several two- and four-pulse composite schemes by the number of distillation levels required to reach an error rate of $10^{-15}$. While some schemes offer modest advantages only at small errors, others consistently outperform across the full range, thereby reducing the depth of concatenation required for fault-tolerant $\mathcal{T}$-gate implementation.

To make the overhead implications of a reduced distillation depth explicit, 
consider the following minimal accounting based on the standard recursive picture.
We work with a $k\!\to\!1$ distillation routine, such as the canonical $5\!\to\!1$ protocol as an illustrative benchmark. After $n$ concatenated levels, the number of raw input magic states required scales as
\begin{equation}
N_{\mathrm{raw}}(n)=k^{\,n} \ .
\label{eq:Nraw}
\end{equation}
If each round succeeds with probability $p(\epsilon)$ when the input error is $\epsilon$, the expected number of raw input states consumed per accepted output is
\begin{equation}
\mathbb{E}\!\left[N_{\mathrm{raw}}(n)\right]\approx
\frac{k^{\,n}}{\prod_{j=0}^{n-1} p(\epsilon_j)}\,,\qquad \epsilon_{j+1}=\epsilon'(\epsilon_j),
\label{eq:ENraw}
\end{equation}
where $\epsilon'(\epsilon)$ is the distillation recursion map.
In the small-error regime relevant for fault-tolerant operation, $p(\epsilon_j)$ varies weakly with $j$ and one may write $p(\epsilon_j)\approx p_0$, yielding
$\mathbb{E}[N_{\mathrm{raw}}(n)]\sim (k/p_0)^n$.
Therefore, reducing the required distillation depth by $\Delta n$ levels implies an  exponential overhead reduction by factors $k^{\Delta n}$ in footprint \footnote{Footprint denotes the peak number of raw magic states, e.g., physical qubits, that must be available in parallel to run an n-level factory.} and approximately $(k/p_0)^{\Delta n}$ in expected consumption.
For the $5\!\to\!1$ example, $N_{\mathrm{raw}}(n)=5^n$, and the success probability is $p(\epsilon)=\tfrac{1}{6}+O(\epsilon)$, so $p_0\simeq 1/6$ and $(k/p_0)=30$ in the small-error regime.
As a concrete illustration, if an initial error requires $n=4$ levels, the parallel raw-state footprint is $5^4=625$; reducing the depth by one (two, three) levels reduces this to $5^3=125$ ($5^2=25$, $5^1=5$), i.e., factors $5\times$ ($25\times$, $125\times$) savings.

\section{\texorpdfstring{$\mT$ and $\mH$ Gates and Physical Realizations}{T and H Gates and Physical Realizations}}
\label{sec:mT and mH Gates and Physical Realizations}

In this section, we examine the explicit unitary operations required to generate the magic states \( \ket{T} \) and \( \ket{H} \) (the eigenstates of the Hadamard gate $H$) from the stabilizer state \( \ket{0} \). We derive the corresponding $\mT$ and $\mH$ gates and analyze how these unitaries can be physically realized across leading quantum hardware platforms. Emphasis is placed on the structure of the underlying control Hamiltonians and the types of systematic errors that arise in each setting.



The terminology surrounding the $T$ gate and associated magic states can be confusing due to inconsistent naming conventions in the literature. Depending on the context, the $T$ gate may refer to either
\(
T = e^{i\pi/4} S H = \frac{e^{i\pi/4}}{\sqrt{2}} \begin{pmatrix} 1 & 1 \\ i & -i \end{pmatrix},
\)
or to the so-called $\pi/4$ gate (or the $\pi/8$ gate historically),
\(
T = \begin{pmatrix} 1 & 0 \\ 0 & e^{-i\pi/4} \end{pmatrix}.
\)
The $\ket{T}$ states are eigenstates of the former and, when injected into a circuit, enable the fault-tolerant implementation of a $\pi/12$ rotation. This differs from the latter $T$ gate (i.e., the $\pi/4$ gate or the $\pi/8$ gate historically), which is used to realize the $Z$-axis rotation $\exp(-i \frac{\pi}{8} Z)$ and is instead implemented via magic state injection of the $\ket{H}$ states- eigenstates of the Hadamard gate $H$.

Here we define new gates, $\mT$ and $\mH$, associated with the $\ket{T}$ and $\ket{H}$ states, respectively. These are the unitary operators that map the stabilizer state $\ket{0}$ to $\ket{T}$ and $\ket{H}$, and they diagonalize the corresponding non-Clifford unitaries. To avoid introducing new ambiguities, we use the notation $\mT$ and $\mH$ rather than overloading existing symbols.

\subsection{\texorpdfstring{$\mT$ and $\mH$ Gates}{T and H Gates}}
\label{sec:mT and mH Gates}

To prepare the \( \ket{T} \) and \( \ket{H} \) magic states, one typically starts with a stabilizer state and applies an appropriate unitary transformation.
In this work, we take the initial state to be \( \ket{0} \), which we assume can be prepared either fault-tolerantly or with negligible error.
Since qubits reside in a two-dimensional Hilbert space, the unitary operator that maps \( \ket{0} \) to a target pure state is uniquely determined up to a global phase. 
The unitaries required to produce \( \ket{T} \) and \( \ket{H} \) from \( \ket{0} \) are given by:
\begin{subequations}
\begin{gather}
\begin{split}
    \mT & = 
\begin{pmatrix}
\cos \beta & -e^{-i\pi/4} \sin \beta \\
 e^{i\pi/4} \sin \beta & \cos \beta
\end{pmatrix} \\ 
& =\begin{pmatrix}
\cos \beta & -i e^{-i \ 3\pi/4} \sin \beta \\
 -i e^{i \ 3\pi/4} \sin \beta & \cos \beta
\end{pmatrix},
\end{split}
 \\
 \begin{split}
 \mH & = 
\begin{pmatrix}
\cos\left(\frac{\pi}{8}\right) & -\sin\left(\frac{\pi}{8}\right) \\
\sin\left(\frac{\pi}{8}\right) & \cos\left(\frac{\pi}{8}\right)
\end{pmatrix} \\ 
& =\begin{pmatrix}
\cos\left(\frac{\pi}{8}\right) & -i e^{-i \pi/2} \sin\left(\frac{\pi}{8}\right) \\
-i e^{i \pi/2} \sin\left(\frac{\pi}{8}\right) & \cos\left(\frac{\pi}{8}\right)
\end{pmatrix},
 \end{split}
\end{gather}
\label{eq:mT snd mH gates}
\end{subequations}
where \( \cos^2(2\beta) = \frac{1}{3} \), respectively.

These gates correspond to specific single-qubit rotations on the Bloch sphere, characterized by nontrivial rotation axes and complex phase factors.
Their physical implementation depends on the underlying quantum hardware platform, which dictates the available control mechanisms and error models.
In the following subsection (\ref{sec:Physical Realizations and Errors}), we examine how various quantum computing platforms realize arbitrary single-qubit unitaries, how these controls can be tailored to implement the specific gates  $\mT$ and $\mH$, and how composite pulse techniques can be employed to mitigate common systematic errors.
In the next sections (\ref{sec:X-Y based systems},\ref{sec:Pre-Distillation in $X$-$Z$ Systems ; example},\ref{sec:Pre-Distillation in Integrated Photonics}), we find composite schemes for them.

\subsection{Physical Realizations and Errors}
\label{sec:Physical Realizations and Errors}

Physical realizations of qubits span several leading experimental platforms, each with distinct control mechanisms, error sources, and robustness strategies. Among the most developed are superconducting circuits, where microwave drives manipulate anharmonic oscillators; trapped ions, where internal states of ions are controlled by laser-driven Raman transitions; neutral atoms in optical tweezers or lattices, where hyperfine ground states are coupled by microwave or Raman fields; and integrated photonic systems, where single photons are guided in waveguide circuits and manipulated through directional couplers. While these platforms differ substantially in hardware implementation, they share common challenges such as systematic amplitude and detuning errors. In many cases, composite pulse or composite device strategies are employed to mitigate these errors in the realization of gates such as $\mT$ and $\mH$. The following overview highlights the key features and error mechanisms of these platforms, forming the basis for the composite schemes developed in subsequent sections. A more detailed technical discussion is provided in Appendix~\ref{app:Realizations}.

\begin{itemize}
    \item \textbf{Superconducting qubits:}~\cite{SuperconductingReview1,SuperconductingReview2} Transmons implement quantum gates via resonant microwave drives. In the rotating frame, the drive produces rotations around equatorial Bloch-sphere axes, with pulse amplitude and phase determining the rotation angle and direction. Gates like $\mT$ and $\mH$ are synthesized through calibrated pulse sequences, often combining $R_x$ and $R_z$ rotations. Dominant errors include amplitude miscalibration, phase errors from control electronics, and leakage into higher excited states. Composite pulse techniques such as BB1 and SK1 mitigate systematic amplitude errors and enhance robustness.
    
    \item \textbf{Trapped ions:}~\cite{TrappedIons,trapped1,trapped2,TrappedIonsReview} Qubits are encoded in long-lived internal states of ions (e.g., $^{171}$Yb$^+$), with laser-driven Raman transitions enabling high-fidelity gates. The laser amplitude, detuning, and phase directly set the unitary, allowing flexible implementation of $\mT$ and $\mH$. Common errors stem from laser intensity noise, detuning from resonance, and AC Stark shifts. Composite control methods such as CORPSE or SCROFULOUS cancel systematic errors by distributing rotations into carefully chosen segments.
    
    \item \textbf{Neutral atoms:}~\cite{NeutralReview1,NeutralReview2} Atoms confined in optical tweezers or lattices encode qubits in hyperfine ground states. Microwave or Raman transitions drive single-qubit rotations, with arbitrary unitaries (including $\mT$ and $\mH$) constructed from combinations of $R_y$ and $R_z$ operations. Errors arise from intensity fluctuations, frequency drifts, and spatial crosstalk between neighboring atoms. Composite schemes such as BB1, CORPSE, and Walsh-modulated pulses are widely applied to suppress coherent amplitude and detuning errors.
    
    \item \textbf{Integrated photonics:}~\cite{IntegratedReview1,IntegratedReview2,KaplanErew,kaplan2024correlation,cohen2025robust} Qubits are encoded in photon paths, with directional couplers implementing coherent mode mixing. The effective Hamiltonian naturally generates $X$ and $Z$ rotations, while gates involving $Y$ (e.g., $\mT$) require multi-segment designs with varied parameters. Photonic devices are limited by static fabrication imperfections in waveguide widths, gaps, and etch depths. Mitigation strategies include post-fabrication tuning, symmetric design (e.g., Mach-Zehnder interferometers), and programmable interferometers. Composite directional coupler designs provide a passive, static route to error resilience, important for scalable and fault-tolerant architectures.
\end{itemize}

Each platform effectively restricts the set of accessible \emph{generators} of $\mathrm{SU}(2)$ in its Hamiltonian.
Consequently, one must adopt an appropriate model for both the Hamiltonian and its associated errors in order to accurately describe the system.
Consider a general single-qubit gate parameterized by
\begin{equation}
    U(\theta,\delta,\phi)= 
    \begin{pmatrix}
        e^{i\delta} \cos(\theta/2) & - i e^{-i\phi} \sin(\theta/2) \\
        -i e^{i\phi} \sin(\theta/2) & e^{-i\delta} \cos(\theta/2)
    \end{pmatrix},
\end{equation}
and examine the effect of dephasing or twirling twirling (see footnote~\ref{footnote-dephasing} in Appendix~\ref{app:Background and Preliminaries}) on the state produced by applying this gate to $\ket{0}$. Define
\begin{equation}
    \rho(\theta,\delta,\phi) = U(\theta,\delta,\phi)\ket{0}\bra{0}U^\dagger(\theta,\delta,\phi).
\end{equation}
The dephased (twirled) state is
\tiny
\begin{equation}
\begin{split}
    D(\rho(\theta,\delta,\phi)) &= \bra{T_0}\rho\ket{T_0} \ket{T_0}\bra{T_0} + \bra{T_1}\rho\ket{T_1} \ket{T_1}\bra{T_1} \\
    &= \frac{1}{6}\Bigl[3 + \sqrt{3}\cos\theta - \sqrt{3}\sin\theta(\cos(\delta{-}\phi) + \sin(\delta{-}\phi))\Bigr] \ket{T_0}\bra{T_0} \\
    &\quad + \frac{1}{6}\Bigl[3 - \sqrt{3}\cos\theta + \sqrt{3}\sin\theta(\cos(\delta{-}\phi) + \sin(\delta{-}\phi))\Bigr] \ket{T_1}\bra{T_1}.
\end{split}
\end{equation}
\normalsize

In the special case of zero detuning ($\delta = 0$) and zero phase ($\phi = 0$), the error rate (the population weight on $\ket{T_1}$) is minimized by optimizing
\begin{equation}
    \frac{3+\sqrt{3}(\sin\theta - \cos\theta)}{6},
\end{equation}
which achieves a minimum value of approximately $9.175 \times 10^{-2}$. According to Figure~\ref{fig:Distillation}(c) (and Eq.~(\ref{eq:thresholds}) in Appendix~\ref{app:Background and Preliminaries}), this error level requires six levels of concatenated distillation using the five-qubit code to reach a fault-tolerant threshold below $10^{-15}$, underscoring the steep resource overhead.

This analysis assumes an idealized implementation. By contrast, if one allows a general phase choice, e.g.~$\phi = -\pi/4$, the desired rotation can be exactly realized in the absence of imperfections. Thus, wherever physically feasible, it is advantageous to exploit the full generality of the gate, even at some implementation cost, as this can significantly reduce the number of distillation rounds by improving the fidelity of the magic state before purification.

With this foundation in place, we next examine how different quantum computing platforms implement arbitrary single-qubit unitaries, how their control parameters can be tuned to realize specific gates such as $\mT$ and $\mH$, and how composite pulse techniques can be employed to mitigate systematic errors.

\section{\texorpdfstring{Composite Pulses Scheme for $X$-$Y$ Systems with Global Rabi Frequency Errors}{Composite Pulses Scheme for X-Y Systems with Global Rabi Frequency Errors}}
\label{sec:X-Y based systems}

This section presents composite pulse schemes for implementing the $\mT$ gate in systems governed by $X$-$Y$ control Hamiltonians, which are common in superconducting qubits, trapped ions, and neutral atoms. For more details, see Appendix~\ref{app:Realizations}. We focus on mitigating global systematic errors in the Rabi frequency- errors that uniformly scale the rotation angle across all segments. By constructing symmetric composite sequences, we analytically derive gate implementations that are robust to first- and second-order errors while exactly realizing the target unitary.

\subsection{Notation}

Our objective is to determine the set of parameters $\{\theta_i, \phi_i\}_{i=1}^N$ that define a composite pulse sequence implementing a target single-qubit gate
$G=U(\theta^*, \phi^*) \equiv \theta^*_{\phi^*} \, ,$
where a general rotation of $\theta_\phi$ is given by
\begin{equation}
\label{eq:general gate}
    \theta_\phi \equiv U(\theta,\phi)= \begin{pmatrix}
        \cos(\theta/2) & - i e^{-i\phi} \sin(\theta/2) \\
        -i e^{i\phi} \sin(\theta/2) & \cos(\theta/2)
    \end{pmatrix} \, .
\end{equation}
Here, $\theta$ denotes the rotation angle and $\phi$ the phase of the driving field, corresponding to a rotation axis in the equatorial plane of the Bloch sphere.
The overall composite gate constructed from $N$ such segments is given by:
\begin{equation}
    U_N \equiv U(\{\theta_i, \phi_i\}_{i=1}^N) = \prod_{i=1}^N U(\theta_i, \phi_i) \, .
\end{equation}

To improve robustness against systematic errors, we optimize the parameters such that the derivatives of the composite gate $U(\{\theta_i, \phi_i\}; \theta^*, \phi^*)$ with respect to a global error parameter vanish.
The performance of the composite gate can be evaluated for example using the Frobenius overlap fidelity,
\begin{equation}
    \mathcal{F}_{\text{frob,2}} = \frac{1}{2} \left| \operatorname{Tr} \left[ U(\{\theta_i, \phi_i\}_i; \theta^*, \phi^*) G^\dagger(\theta^*, \phi^*) \right] \right| \, ,
\end{equation}
which serves as a figure of merit, although it does not fully capture the detailed nature of gate deviations.

\subsection{The Composite Pulses Scheme}

To suppress control errors in such systems, we employ a symmetric sequence of composite pulses (or segmented gates), motivated by numerical evidence supporting its robustness. The unitary transformation is given by:
\begin{equation}
    U_{2n-1} = \theta_{\phi_1} \pi_{\phi_2} \pi_{\phi_3} \cdots \pi_{\phi_{n-1}} \pi_{\phi_n} \pi_{\phi_{n-1}} \cdots \pi_{\phi_3} \pi_{\phi_2} \theta_{\phi_1} \, .
\end{equation}
The use of this symmetric construction enhances robustness against systematic errors in the Rabi frequency or in the overall scaling of the Hamiltonian.
In fact, for three segments, the resulting sequence is equivalent to the well-known SCROFULOUS (Short Composite ROtation For Undoing Length Over- and Under-Shoot) sequence introduced in~\cite{scrofulous}. For a larger number of segments, our construction naturally generalizes this sequence by incorporating additional pulses, thereby enabling the suppression of higher-order error terms.

We denote the $k^{\text{th}}$ derivative of the unitary rotation $U(\theta, \phi)$ with respect to $\theta$ by \( U^{(k,0)}(\theta, \phi) \). These derivatives take the following forms:
\footnotesize
\begin{subequations}
    \begin{align}
        U^{(2k,0)}(\theta, \phi) &= (-1)^k \frac{1}{2^k}
        \begin{pmatrix}
            \cos(\theta/2) & -i e^{-i\phi} \sin(\theta/2) \\
            -i e^{i\phi} \sin(\theta/2) & \cos(\theta/2)
        \end{pmatrix} , \\
        U^{(2k+1,0)}(\theta, \phi) &= (-1)^k \frac{1}{2^{k+1}}
        \begin{pmatrix}
            -\sin(\theta/2) & -i e^{-i\phi} \cos(\theta/2) \\
            -i e^{i\phi} \cos(\theta/2) & -\sin(\theta/2)
        \end{pmatrix} .
    \end{align}
\end{subequations}
\normalsize

Evaluating these expressions at \(\theta = \pi\), we find:
\begin{subequations}
    \begin{align}
        \begin{split}
            U^{(2k,0)}(\pi, \phi) &= (-1)^k \frac{1}{2^k}
        \begin{pmatrix}
            0 & -i e^{-i\phi} \\
            -i e^{i\phi} & 0
        \end{pmatrix}
        \\ & = -(-1)^{k+1} \frac{i}{2^{k+1}} \left( \cos\phi\, X + \sin\phi\, Y \right) ,
        \end{split} \\
        \begin{split}
            U^{(2k+1,0)}(\pi, \phi) &= (-1)^{k+1} \frac{1}{2^{k+1}}
        \begin{pmatrix}
            1 & 0 \\
            0 & 1
        \end{pmatrix}
        \\& = (-1)^{k+1} \frac{1}{2^{k+1}} I .
        \end{split}
    \end{align}
\end{subequations}

Here, \( X \), \( Y \), and \( I \) denote the standard Pauli matrices and the identity operator, respectively. These expressions reveal the periodic structure and symmetry in the higher-order derivatives, which will be useful in analyzing the sensitivity of composite gates to systematic errors.

The error model of the systems we are talking about is $\theta_i \rightarrow \theta_i(1+\epsilon)$ for each pulse or segment.
Therefore, the $k^{\text{th}}$ variation of $U(\theta,\phi)$, denoted by  $\delta^k U(\theta,\phi)$ satisfies
\begin{equation}
    k!\,\delta^k U(\theta,\phi)=\theta^k U^{(k,0)}(\theta,\phi) \, .
\end{equation}

\subsection{\texorpdfstring{Three-Pulse Sequence ($n=2$)}{Three-Pulse Sequence (n=2)}}

For the minimal case $n=2$, the symmetric composite sequence consists of three segments. To achieve robustness to first-order global errors while implementing a desired gate \( U(\theta^*, \phi^*) \), we impose the following conditions:
\begin{subequations}
    \begin{align}
        U^{(0,0)}(\theta,\phi) &= U(\theta^*,\phi^*), \\
        U^{(1,0)}(\theta,\phi) &= 0.
    \end{align}
\end{subequations}
These constraints yield the system:
\begin{subequations}
    \begin{gather}
        -\cos(\phi_1 - \phi_2) \sin(\theta) = \cos \left(\frac{\theta^*}{2}\right), \\
        \begin{split}
            e^{i\phi^*} &\sin\left(\frac{\theta^*}{2}\right) =  \\
            & e^{i \phi_1} \left(
            e^{i(\phi_2 - \phi_1)} \cos^2\left(\frac{\theta}{2}\right)
            - e^{-i(\phi_2 - \phi_1)} \sin^2\left(\frac{\theta}{2}\right)
        \right) , 
        \end{split}
        \\
        \left( 2 \theta \cos(\phi_1 - \phi_2) + \pi \right) \cos(\theta) = 0, \\
        \left( 2 \theta \cos(\phi_1 - \phi_2) + \pi \right) \sin(\theta) = 0.
    \end{gather}
\end{subequations}
This system admits a unique solution given by:
\begin{subequations}
    \begin{gather}
        \frac{\sin\theta}{\theta} = \frac{2}{\pi} \cos\left(\frac{\theta^*}{2}\right), \\
        \cos(\phi_2 - \phi_1) = -\frac{\pi}{2\theta}, \\
        e^{i\phi_{1_\pm}} = \frac{\sin\left(\frac{\theta^*}{2}\right) e^{i\phi^*}}{\cos\theta \cos x \pm i \sin x},
    \end{gather}
    \label{eq:3-pulse solution}
\end{subequations}
\noindent
where \( x = \phi_2 - \phi_1 \). This solution ensures first-order insensitivity to global Rabi frequency errors while exactly implementing the target gate.
As explained before, this is equivalent to the well-known SCROFULOUS sequence introduced in~\cite{scrofulous}.

In Table~\ref{tab:3-pulse parameters ; X-Y}, we present representative solutions for selected values of the target angle $\theta^*$.
Note that for $\theta^* = 1$ and $\phi^* = 0$, the resulting gate is $-iX$, and for $\theta^* = \tfrac{1}{2}$ and $\phi^* = \tfrac{\pi}{4}$, the gate corresponds to $ZH$.

\begin{table}[H]
\centering
\begin{tabular}{| c | c |}
\hline
$\theta^*$ & $(\theta \ ; \ \phi_1-\phi^* \ , \  \phi_2-\phi^* )$ \\
\hline
\hline
$\frac{1}{10}$ & $(0.50614\ ;\ -0.46114 \ ,\ 0.48923)$ \\
\hline
$\frac{1}{5}$  & $(0.52421\ ;\ -0.42464 \ ,\ 0.47824)$ \\
\hline
$\frac{3}{10}$ & $(0.55331\ ;\ -0.39234 \ ,\ 0.46678)$ \\
\hline
$\frac{2}{5}$  & $(0.59226\ ;\ -0.36536 \ ,\ 0.45458)$ \\
\hline
$\frac{1}{2}$  & $(0.63990\ ;\ -0.34419 \ ,\ 0.44129)$ \\
\hline
$\frac{3}{5}$  & $(0.69538\ ;\ -0.32894 \ ,\ 0.42648)$ \\
\hline
$\frac{7}{10}$ & $(0.75827\ ;\ -0.31966 \ ,\ 0.40952)$ \\
\hline
$\frac{4}{5}$  & $(0.82882\ ;\ -0.31661 \ ,\ 0.38952)$ \\
\hline
$\frac{9}{10}$ & $(0.90828\ ;\ -0.32055 \ ,\ 0.36501)$ \\
\hline
$1$ & $(1.00000\ ;\ -0.33333 \ ,\ 0.33333)$ \\
\hline
\end{tabular}
\caption{Composite pulse parameters for three-segment sequences: each row lists the values of $(\theta ;  \phi_1 - \phi^* , \phi_2 - \phi^*)$ corresponding to a target angle $\theta^*$, all expressed in units of $\pi$. The values were obtained by Eqs.~(\ref{eq:3-pulse solution}).}
\label{tab:3-pulse parameters ; X-Y}
\end{table}

A detailed comparison between the three-pulse SCROFULOUS sequence, the three-pulse CORPSE (Compensation for Off-Resonance with a Pulse Sequence) sequence, and the four-pulse BB1 sequence is presented in Refs.~\cite{scrofulous,CompositeComparison}. 
The CORPSE sequence is given by $(\theta_1)_{\phi^*} \cdot (\theta_2)_{\pi+\phi^*} \cdot (\theta_3)_{\phi^*}$, where
\begin{subequations}
    \begin{gather}
        \theta_1 = 2\pi+\frac{\theta^*}{2}-\arcsin\left( \frac{\sin \frac{\theta^*}{2}}{2} \right) ,\\
        \theta_1 = 2\pi-2\arcsin\left( \frac{\sin \frac{\theta^*}{2}}{2} \right) ,\\
        \theta_3 = \frac{\theta^*}{2}-\arcsin\left( \frac{\sin \frac{\theta^*}{2}}{2} \right) .
    \end{gather}
\end{subequations}
The BB1 sequence is given by $\pi_\varphi \cdot (2\pi)_{3\varphi} \cdot \pi_\varphi \cdot \theta^*_{\phi^*}$, where $\phi^*=\pm\arccos\left( - \frac{\theta^*}{4\pi}\right)$.

\subsection{\texorpdfstring{Five-Pulse Sequence ($n=3$)}{Five-Pulse Sequence (n=3)}}

For the case $n=3$, the symmetric composite sequence consists of five segments. To achieve robustness against both first- and second-order global Rabi frequency errors while implementing a target gate \( U(\theta^*, \phi^*) \), we impose the following derivative cancellation conditions:
\begin{subequations}
    \begin{align}
        U^{(0,0)}(\theta,\phi) &= U(\theta^*,\phi^*), \\
        U^{(1,0)}(\theta,\phi) &= 0, \\
        U^{(2,0)}(\theta,\phi) &= 0.
    \end{align}
\end{subequations}
This construction generalizes the well-known three-pulse SCROFULOUS sequence by promoting its error-compensation mechanism to the next higher order.
A detailed comparison between SCROFULOUS, CORPSE, and BB1 in the specific error regimes for which each was designed can be found in Refs.~\cite{scrofulous,CompositeComparison}; here we do not perform systematic benchmarking between these protocols.

We define the phase parameters of the sequence as $\phi_3 = \phi_2 + \alpha$ and $\phi_1 = \phi_2 + \beta$.
This results in the following system of constraints:
\begin{subequations}
    \begin{gather}
        \cos(\alpha+\beta) \sin\theta = \cos \left( \frac{\theta^*}{2} \right), \\
        - e^{i \phi_1} \left(
            \cos\theta\cos(\alpha+\beta)+i \sin(\alpha+\beta)
        \right) = e^{i\phi^*} \sin\left( \frac{\theta^*}{2} \right), \\
        \pi + 2\pi \cos\alpha + 2\theta \cos(\alpha+\beta) = 0, \\
\begin{split}
            4\pi\theta 
        & + 4\pi^2 \cos\beta 
        + 2\pi^2 \cos(\beta-\alpha) 
        + 8\pi\theta \cos\alpha \\
        & + (3\pi^2 + 4\theta^2) \cos(\alpha+\beta) = 0,
\end{split} \\
        4 \sin\beta 
        + 2 \sin(\beta - \alpha) 
        + 3 \sin(\alpha + \beta) = 0.
    \end{gather}
    \label{eq:5 pulses equations}
\end{subequations}

This set comprises five nonlinear equations in four variables: $\alpha$, $\beta$, $\theta$, and $\phi_2$. While the system is not solvable analytically for arbitrary target parameters $\theta^*$ and $\phi^*$, particular solutions can be constructed.
For example, the choice $\alpha = -\arccos(-4/5)$, $\beta = 3\pi/2$, and $\theta = 3\pi/2$ satisfies the equations and yields the implemented gate angle $\theta^* = \pm 2\arccos(3/5)$ with a corresponding relation for $\phi_2$ for every $\phi^*$.

A general parametric solution can be obtained by solving the last three equations first. In particular, Eqs.~(\ref{eq:5 pulses equations}c-e) can be solved in closed form as:
\begin{subequations}
    \begin{align}
        \beta &=  \frac{\pi}{2} \pm \frac{\pi}{2}-\arctan\left( \frac{\sin \alpha}{4 + 5 \cos \alpha} \right), \\
        \theta &= -\frac{\pi}{2} \cdot \frac{1 + 2\cos\alpha}{\cos(\alpha + \beta)}.
    \end{align}
    \label{eq:5 pulses - beta and theta}
\end{subequations}
Substituting these into the first two conditions, we obtain:
\begin{subequations}
    \begin{gather}
        \cos(\alpha + \beta) \sin\theta = \cos \left( \frac{\theta^*}{2} \right), \\
        \phi_2 = \phi^* - \beta + \arctan\left( \frac{\tan(\alpha + \beta)}{\cos \theta} \right) + \frac{\pi}{2} \pm \frac{\pi}{2}.
    \end{gather}
    \label{eq:5 pulses - alpha and phi2}
\end{subequations}

Using Eqs.~(\ref{eq:5 pulses - beta and theta}b) and~(\ref{eq:5 pulses - alpha and phi2}a), we find that $\cos(\alpha + \beta)$ must satisfy the condition:
\begin{equation}
    \cos(\alpha + \beta) = -\frac{\cos \left( \frac{\theta^*}{2} \right)}{\sin \left( \frac{\pi}{2} \cdot \frac{1 + 2\cos\alpha}{\cos(\alpha + \beta)} \right)}.
\end{equation}
Inserting Eq.~(\ref{eq:5 pulses - beta and theta}a) into this, we derive a self-consistency condition for $\alpha$:
\begin{equation}
\begin{split}
    0=\cos & \left( \frac{\theta^*}{2} \right)  + \cos\left( \alpha - \arctan\left( \frac{\sin \alpha}{4 + 5 \cos \alpha} \right) \right) \\
     & 
    \cdot \sin\left( \frac{\pi}{2} \cdot \frac{1 + 2\cos \alpha}{\cos\left( \alpha - \arctan\left( \frac{\sin \alpha}{4 + 5 \cos \alpha} \right) \right)} \right)
\end{split}
\end{equation}

This is a transcendental equation in $\alpha$ that must be solved numerically. Once a solution for $\alpha$ is found, the remaining parameters $\beta$, $\theta$, and $\phi_2$ are determined via Eqs.~(\ref{eq:5 pulses - beta and theta}) and~(\ref{eq:5 pulses - alpha and phi2}).
In Table~\ref{tab:5-pulse parameters ; X-Y}, we present representative solutions for selected values of the target angle $\theta^*$.
Note that for $\theta^* = 1$ and $\phi^* = 0$, the resulting gate is $-iX$, and for $\theta^* = \tfrac{1}{2}$ and $\phi^* = \tfrac{\pi}{4}$, the gate corresponds to $ZH$.

\begin{table}[H]
\centering
\begin{tabular}{| c | c |}
\hline
$\theta^*$ & $(\theta \ ; \ \phi_1-\phi^* \ , \  \phi_2-\phi^* \ , \ \phi_3-\phi^*)$ \\
\hline
\hline
$\frac{1}{10}$ & $(1.50291\ ;\ 0.48163 \ ,\ -0.51210 \ ,\ -0.45592)$ \\
\hline
$\frac{1}{5}$  & $(1.51182\ ;\ 0.46354 \ ,\ -0.52386 \ ,\ -0.41193)$ \\
\hline
$\frac{3}{10}$ & $(1.52730\ ;\ 0.44603 \ ,\ -0.53487 \ ,\ -0.36810)$ \\
\hline
$\frac{2}{5}$  & $(1.55034\ ;\ 0.42955 \ ,\ -0.54464 \ ,\ -0.32444)$ \\
\hline
$\frac{1}{2}$  & $(1.58238\ ;\ 0.41472 \ ,\ -0.55249 \ ,\ -0.28086)$ \\
\hline
$\frac{3}{5}$  & $(1.62536\ ;\ 0.40238 \ ,\ -0.55752 \ ,\ -0.23704)$ \\
\hline
$\frac{7}{10}$ & $(1.68194\ ;\ 0.39372 \ ,\ -0.55852 \ ,\ -0.19226)$ \\
\hline
$\frac{4}{5}$  & $(1.75611\ ;\ 0.39037 \ ,\ -0.55380 \ ,\ -0.14500)$ \\
\hline
$\frac{9}{10}$ & $(1.85530\ ;\ 0.39495 \ ,\ -0.54056 \ ,\ -0.09202)$ \\
\hline
$1$ & $(2.00000\ ;\ 0.41312 \ ,\ -0.51245 \ ,\ -0.02491)$ \\
\hline
\end{tabular}
\caption{Composite pulse parameters for five-segment sequences: each row lists the values of $(\theta ;  \phi_1 - \phi^* , \phi_2 - \phi^* , \phi_3 - \phi^*)$ corresponding to a target angle $\theta^*$, all expressed in units of $\pi$, shown to five decimal digits. The values were obtained by choosing the positive sign in Eqs.~(\ref{eq:5 pulses - beta and theta}a) and the negative sign in Eq.~(\ref{eq:5 pulses - alpha and phi2}b), ensuring physically valid solutions.}
\label{tab:5-pulse parameters ; X-Y}
\end{table}

\subsection{\texorpdfstring{Pre-Distillation: The Improvement\\in the $\mT$ Gate}{Pre-Distillation: The Improvement in the T Gate}}

In quantum computing research, several fidelity and distance measures are used to evaluate how closely an implemented gate approximates its ideal version, each suited for different purposes. The average gate fidelity is the most commonly reported measure, especially in experiments, as it captures how well a gate performs on average across all possible input states and can be efficiently estimated using randomized benchmarking. For more detailed characterization, process fidelity compares the full quantum processes and is often used in quantum process tomography, although it's more resource-intensive to compute. Trace fidelity and trace distance are state-based measures used to assess how distinguishable the output states are, useful in situations like error diagnostics or state preparation. For fault-tolerant quantum computing, the diamond norm distance is the gold standard- it measures the worst-case deviation of the actual gate from the ideal one, accounting for all possible inputs and entanglements with ancillas. While very informative, it's also the most computationally demanding.
In contrast, the Frobenius distance is a simple matrix norm used occasionally in simulations for quick comparisons but lacks the physical rigor of other measures. Each metric balances between ease of computation and how well it captures meaningful quantum behavior, with average fidelity favored in benchmarking, diamond norm in theoretical error thresholds, and trace-based metrics in state analysis.

To assess the improvement in implementing the \( \mT \) gate, we do not rely on standard fidelity or distance measures. Instead, we adopt a natural and operationally motivated metric, denoted by \( \varepsilon(G(\epsilon)) \), which quantifies the deviation of the output state from the ideal \( \ket{T_0} \) when a noisy gate \( G(\epsilon) \), affected by a global control error \( \epsilon \), is applied to the input state \( \ket{0} \). Specifically, we define
\begin{equation}
\label{eq:natural error of the gate}
    \varepsilon(G(\epsilon)) = \bra{T_1} \rho(G(\epsilon)) \ket{T_1} ,
\end{equation}
where \( \rho(G(\epsilon)) \) is the pure state resulting from the gate action:
\footnotesize
\begin{equation}
    \rho(G(\epsilon)) = \left( G_{1,1}(\epsilon) \ket{0} + G_{2,1}(\epsilon) \ket{1} \right) \left( G^*_{1,1}(\epsilon) \bra{0} + G^*_{2,1}(\epsilon) \bra{1} \right) .
\end{equation}
\normalsize
We Call $\varepsilon(G(\epsilon))$ \emph{the T-magic error} of the gate. This measure is particularly relevant for applications in magic state injection into stabilizer codes, where the output of \( G(\epsilon)\ket{0} \) is intended to serve as a noisy magic state. The metric \( \varepsilon(G(\epsilon)) \) reflects the overlap with the orthogonal state \( \ket{T_1} \), thereby capturing the component of the error that undermines distillability via, for example, the five-qubit code. Our choice of this metric is guided by the operational considerations discussed in Section~\ref{sec:mT and mH Gates}.

We introduce a natural and operationally motivated definition of the \emph{magic \(T\)-gate fidelity}:
\begin{equation}
    \mathcal{F}_{\text{magic } T\text{-gate}} = 1 - \sqrt{\varepsilon(G(\epsilon))}.
\end{equation}
This measure quantifies the success of preparing the ideal \( \ket{T} \) state using the noisy gate \( G(\epsilon) \), thereby providing a direct estimate of the number of distillation levels or concatenation rounds required for fault-tolerant encoding.
Indeed, this is the central figure of merit we will plot throughout our analysis, whereas the T-magic fidelity itself is plotted only in Appendix~\ref{app:Plots for the Frobenius Fidelity and the T-Magic Fidelity}, to illustrate its nontrivial behavior and its lack of monotonicity with respect to other standard gate fidelity metrics.
It is important to note that \( \mathcal{F}_{\text{magic } T\text{-gate}} \) does not capture the full fidelity of the gate \( G(\epsilon) \) when applied to arbitrary input states or within general qudit or multi-qubit contexts. Rather, its definition is tailored to assess the quality of the output state obtained when the noisy gate acts on a standard stabilizer input, specifically in the context of MSD.
For completeness, we also recall two commonly used fidelity measures for \textbf{gates}- the Frobenius distance fidelity and the Frobenius overlap fidelity:
\begin{subequations}
    \begin{gather}
        \mathcal{F}_{\text{frob,1}}(G(\epsilon), \mT) = 1 - \sqrt{\frac{1}{4} \text{Tr} \left[ \left(G(\epsilon) - \mT\right)\left(G^\dagger(\epsilon) - \mT^\dagger\right)\right]} \, , \\
        \mathcal{F}_{\text{frob,2}}(G(\epsilon), \mT) = \frac{1}{2} \left| \text{Tr} \left( G(\epsilon) \mT^\dagger\right) \right| \, .
    \end{gather}
\end{subequations}

At this stage of the analysis, we emphasize that the natural error metric of the $\mT$-gate introduced in Eq.~\eqref{eq:natural error of the gate} coincides with \emph{Jozsa’s fidelity} for quantum states~\cite{Jozsa01121994}.
\footnote{
Jozsa’s fidelity between two quantum states $\rho$ and $\sigma$, represented as density matrices, is defined as
\[
F(\rho, \sigma) = \left( \Tr \sqrt{\sqrt{\rho}\ \sigma \, \sqrt{\rho}}\right)^2 .
\]
}
Specifically, our metric is the fidelity between the two pure states
$G(\epsilon)\ket{0}$ and $\ket{T_1}$, namely
\[
F\left(G(\epsilon)\ket{0}\bra{0}G^\dagger(\epsilon),\ket{T_1}\bra{T_1}\right).
\]
For arbitrary density operators $\rho$ and $\sigma$, fidelity and trace distance satisfy the well-known inequalities~\cite{FidelityComparison}
\[1 - \sqrt{F(\rho,\sigma)} \leq D(\rho,\sigma) \leq \sqrt{1 - F(\rho,\sigma)},
\]
where the trace distance is defined by
\[ D(\rho,\sigma) \equiv \frac{1}{2} \left\| \rho - \sigma \right\|_1,
\qquad \left\|A\right\|_1 \equiv \Tr \sqrt{A^\dagger A}.
\]
Moreover, when at least one of the states is pure, the lower bound strengthens to
\[
1 - F(\rho,\sigma) \leq D(\rho,\sigma) \leq \sqrt{1 - F(\rho,\sigma)}.
\]
Thus, in the present setting, where the ideal target state is pure, the fidelity-based error metric directly controls the operationally meaningful trace distance between the implemented and ideal outputs.

Moreover, we emphasize that the figure of merit employed in this work is \emph{not} a conventional gate fidelity.
Instead, we evaluate a \emph{state fidelity} associated with a specific operational task: the generation of the target state $\ket{T_0}$ from the fixed input state $\ket{0}$.
Our objective is therefore not to quantify the global proximity of the implemented unitary to the ideal gate over the entire Hilbert space, but rather to assess how accurately the physical operation performs the concrete transformation $\ket{0} \longmapsto \ket{T_0}$.
Accordingly, the relevant performance metric is the fidelity between the actual output state and the desired target state $\ket{T_0}$.

The evaluation procedure consists of two conceptually distinct steps.
First, we restrict attention to the state produced by applying the (possibly imperfect) gate to the specific input $\ket{0}$.
This reflects the operational fact that, within the present protocol, the sole required functionality of the gate is the preparation of $\ket{T_0}$.
Deviations in other regions of the Bloch sphere are therefore irrelevant for the task at hand.
Second, we exploit the structure of the error after dephasing in the $\{ \ket{T_0}, \ket{T_1} \}$ basis.
After dephasing, the output state takes a classical mixture of $\ket{T_0}\bra{T_0}$ and $\ket{T_1}\bra{T_1}$, where the off-diagonal coherence terms vanish. 
Consequently, the only remaining error contribution is the population leakage into the orthogonal component $\ket{T_1}$.
We therefore evaluate the weight of the orthogonal complement after dephasing and define the effective success probability as one minus this error weight. 
This procedure isolates precisely the physically relevant deviation, namely, the population transferred to $\ket{T_1}$, while eliminating coherent off, diagonal contributions that do not affect the final probabilistic interpretation following dephasing.
In light of these considerations, the adopted metric directly captures the operational success probability of preparing $\ket{T_0}$ from $\ket{0}$, rather than the global distance between the implemented and ideal unitaries.

The parameters defining the target $\mT$ gate are given by (see Eqs.~(\ref{eq:mT snd mH gates}) and~(\ref{eq:general gate})):
\begin{equation}
    \theta^*=\arccos\left( \frac{1}{\sqrt{3}} \right) \quad , \quad \phi^*=\frac{3\pi}{4} \, .
\end{equation}
The parameters \(\theta, \phi_1, \phi_2\) for the three-pulse composite implementation of the $\mT$ gate are (see Eqs.~(\ref{eq:3-pulse solution})):
\begin{equation}
    \theta\approx 1.74270 \quad , \quad \phi_1 \approx 1.12743  \quad , \quad \phi_2 \approx 3.82112 \, .
\end{equation}
For the five-pulse composite implementation, the corresponding parameters \(\theta, \phi_1, \phi_2, \phi_3\) are:
\begin{equation}
    \theta \approx 4.80062 \ , \ \phi_1 \approx 3.75525 \ , \ \phi_2 \approx 0.67449 \ , \ \phi_3 \approx 1.20539 \, .
\end{equation}
A numerically obtained solution for a seven-pulse composite sequence that achieves third-order robustness to control errors in the gate implementation is given by:
\begin{equation}
\begin{split}
    \theta \approx 1.78928 \ , \ \phi_1 & \approx 3.4837 \ , \ \phi_2 \approx 4.23899 \ ,
    \\ \ \phi_3 \approx 1.15951 \, &, \ \phi_4 \approx 0.894556 \, .
\end{split}
\end{equation}

\begin{figure}[h!]
    \centering
    \begin{minipage}[b]{0.48\textwidth}
        \centering

            \centering
            \includegraphics[width=\textwidth]{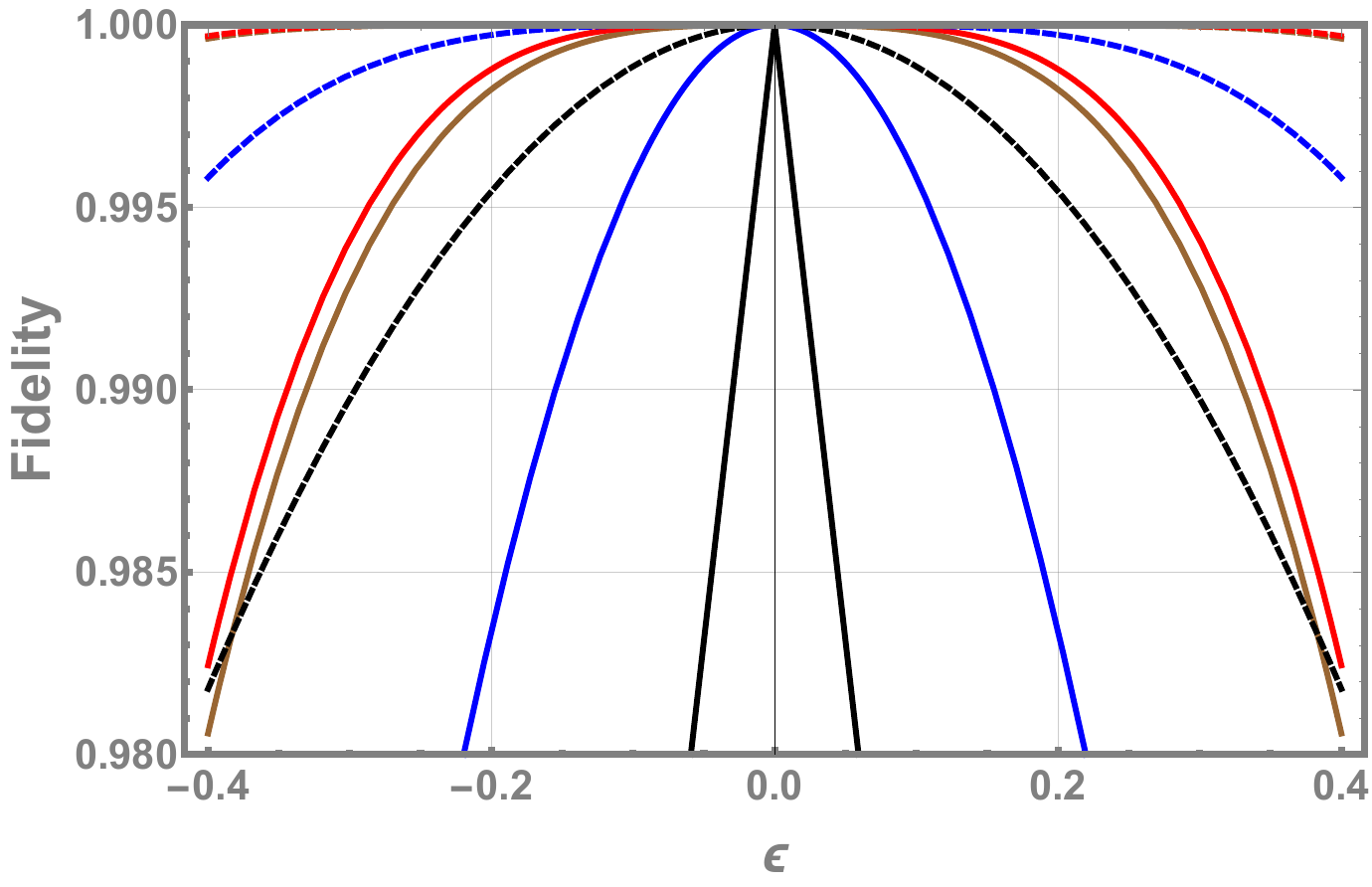}

    \end{minipage}

    \begin{minipage}
    {0.9\linewidth}
    \raggedright (a) Frobenius distance fidelity $\mathcal{F}_{\text{frob,1}}$ for different composite pulse schemes.
    \end{minipage}
    
    \hfill
    \begin{minipage}[b]{0.48\textwidth}
        \centering

            \centering
            \includegraphics[width=\textwidth]{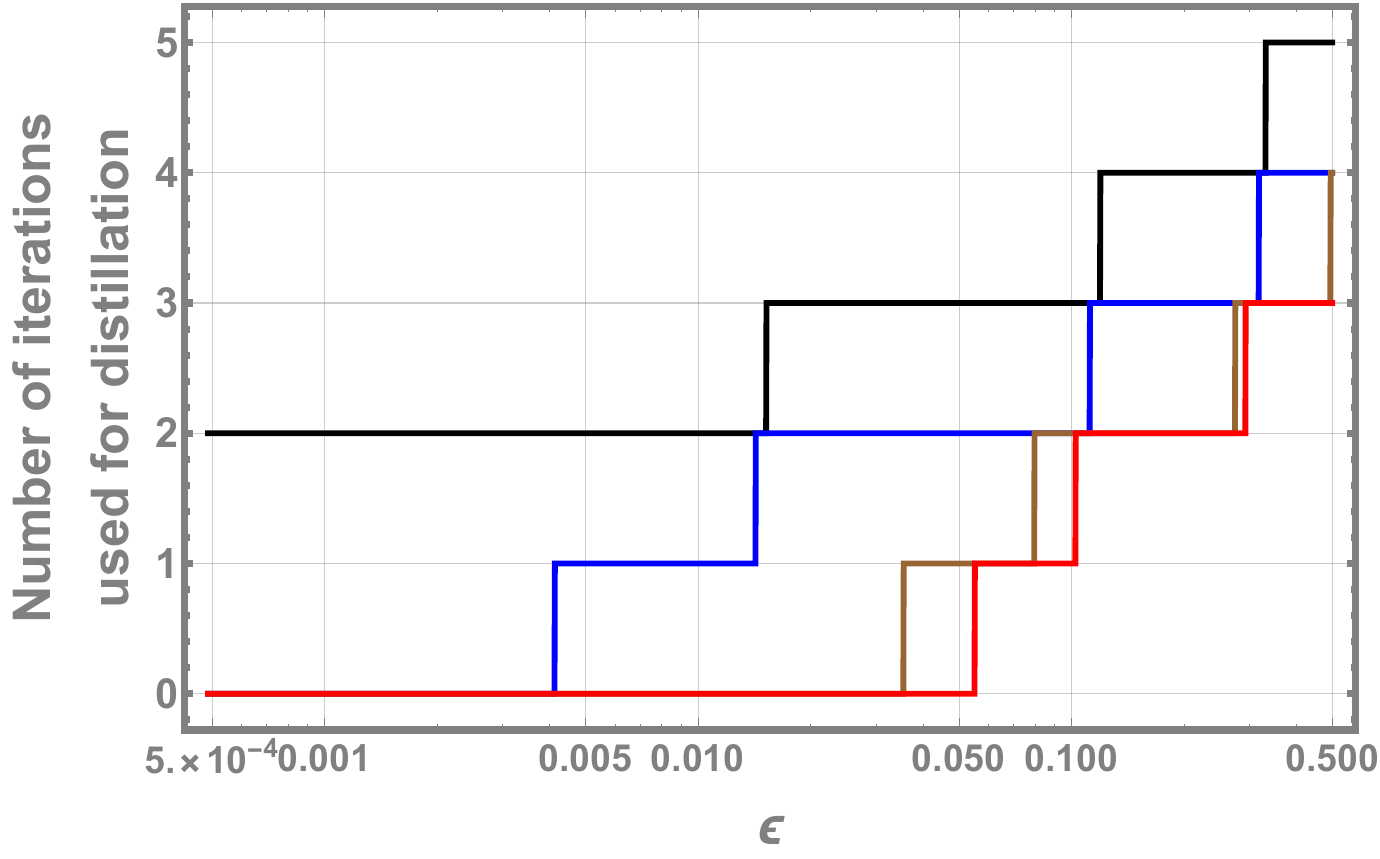}

    \end{minipage}

    \begin{minipage}
    {0.9\linewidth}
    \raggedright (b) Number of distillation levels required to reach an error of $10^{-15}$.
    \end{minipage}
    
    \caption{
    Improvements enabled by composite pulse sequences for implementing the $\mT$ gate in $X$-$Y$ Systems with Global Rabi Frequency Errors.
    The black curves correspond to the single-pulse implementation, while the blue, red, and brown curves correspond to the three-, five-, and seven-pulse composite gates.
    (a) Comparison of Frobenius distance fidelity $\mathcal{F}_{\text{frob,1}}$ (solid lines) and Frobenius overlap fidelity $\mathcal{F}_{\text{frob,2}}$ (dashed lines) for single-, three-, five-, and seven-pulse implementations under global systematic errors. 
    (b) Reduction in the number of MSD iterations required to achieve a target gate error of \(10^{-15}\), where composite sequences reduce the required levels by one, two, or even three in some error regimes.
    }
    \label{fig:mT fidelity and distillation improvement}
\end{figure}

Figure~\ref{fig:mT fidelity and distillation improvement} illustrates the effectiveness of composite pulse sequences in improving the fidelity of the $\mT$ gate and in reducing the resource overhead required for MSD.
The black curves correspond to the single-pulse implementation, while the blue, red, and brown curves correspond to the three-, five-, and seven-pulse composite gates.
Panel~(a) compares two fidelity measures, Frobenius distance fidelity (solid lines) and trace distance fidelity (dashed lines), for single-, three-, five-, and seven-pulse implementations under global systematic errors. The multi-pulse sequences clearly enhance gate robustness, reducing sensitivity to over- and under-rotations.  
Panel~(b) highlights the practical significance of this improvement by showing the number of distillation iterations (of applying $\epsilon'(\epsilon)$ given in Section~\ref{sec:Distillation and Pre-Distillation} iteratively) required to reach a target gate error of \(10^{-15}\). In several physical error regimes, the use of composite sequences reduces the number of required distillation levels by one, two, or even three, substantially lowering the overhead for fault-tolerant quantum computation.

\subsection{Pre-distillation of $H$-states}

We now apply the pre-distillation scheme presented in Tables~\ref{tab:3-pulse parameters ; X-Y}, \ref{tab:5-pulse parameters ; X-Y} to the case of $H$-states. 
Although $T$-states are considered to possess more magic than $H$-states, exhibiting higher symmetry and lower stabilizer fidelity, the threshold fidelity required for any protocol that distills $H$-states is larger than that of $T$-states~\cite{BravyiKitaev}.

Several magic-state distillation protocols for $H$-type states are known. In particular, smaller constructions exist, most notably the 7-qubit protocol introduced by Reichardt~\cite{Reichardt2005}, which provides linear error suppression. However, when considering protocols that correct arbitrary single-qubit errors while achieving high-order suppression, the standard comparison is the 15-to-1 encoding~\cite{BravyiKitaev}, which yields cubic suppression of the form
\[
\epsilon' = 35\epsilon^3 + O(\epsilon^4),
\]
compared to 
\[
\epsilon' = 5\epsilon^2 + O(\epsilon^3)
\] 
for $T$-states~\cite{BravyiKitaev}.
This cubic suppression makes $H$-states a more efficient resource in certain regimes, highlighting the practical motivation for pursuing $H$-state pre-distillation.

The parameters defining the target $\mH$ gate are given by (see Eqs.~(\ref{eq:mT snd mH gates}) and~(\ref{eq:general gate})):
\begin{equation}
    \theta^*= \frac{\pi}{4} \quad , \quad \phi^*=\frac{\pi}{2} \, .
\end{equation}
The parameters \(\theta, \phi_1, \phi_2\) for the three-pulse composite implementation of the $\mT$ gate are (see Eqs.~(\ref{eq:3-pulse solution})):
\begin{equation}
    \theta\approx 1.68846 \quad , \quad \phi_1 \approx 0.28941  \quad , \quad \phi_2 \approx 3.05547 \, .
\end{equation}
For the five-pulse composite implementation, the corresponding parameters \(\theta, \phi_1, \phi_2, \phi_3\) are:
\begin{equation}
\begin{split}
    \theta \approx 4.77109 \ , \ \phi_1 & \approx 2.99923 \ , \\ \phi_2 \approx -0.09264 \ &, \ \phi_3 \approx 0.34559 \, .
\end{split}
\end{equation}
A numerically obtained solution for a seven-pulse composite sequence that achieves third-order robustness to control errors in the gate implementation is given by:
\begin{equation}
\begin{split}
    \theta \approx 1.72181 \ , \ \phi_1 & \approx 2.76539 \ , \ \phi_2 \approx 3.39854 \ ,
    \\ \ \phi_3 \approx 0.30736 \, &, \ \phi_4 \approx 0.08854 \, .
\end{split}
\end{equation}

\begin{figure}[h!]
    \centering
    \begin{minipage}[b]{0.48\textwidth}
        \centering

            \centering
            \includegraphics[width=\textwidth]{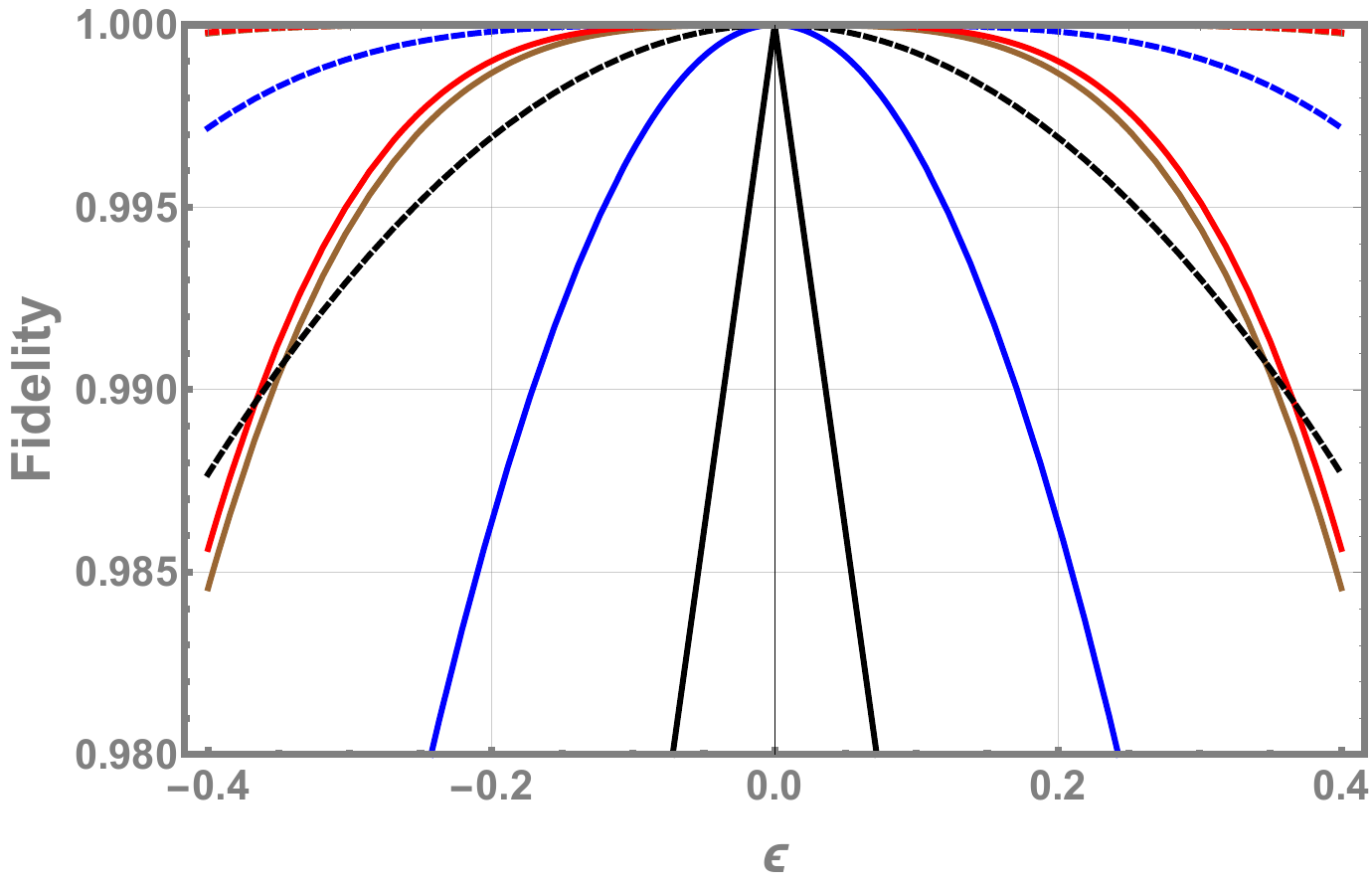}

    \end{minipage}

    \begin{minipage}
    {0.9\linewidth}
    \raggedright (a) Frobenius distance fidelity $\mathcal{F}_{\text{frob,1}}$ for different composite pulse schemes.
    \end{minipage}
    
    \hfill
    \begin{minipage}[b]{0.48\textwidth}
        \centering

            \centering
            \includegraphics[width=\textwidth]{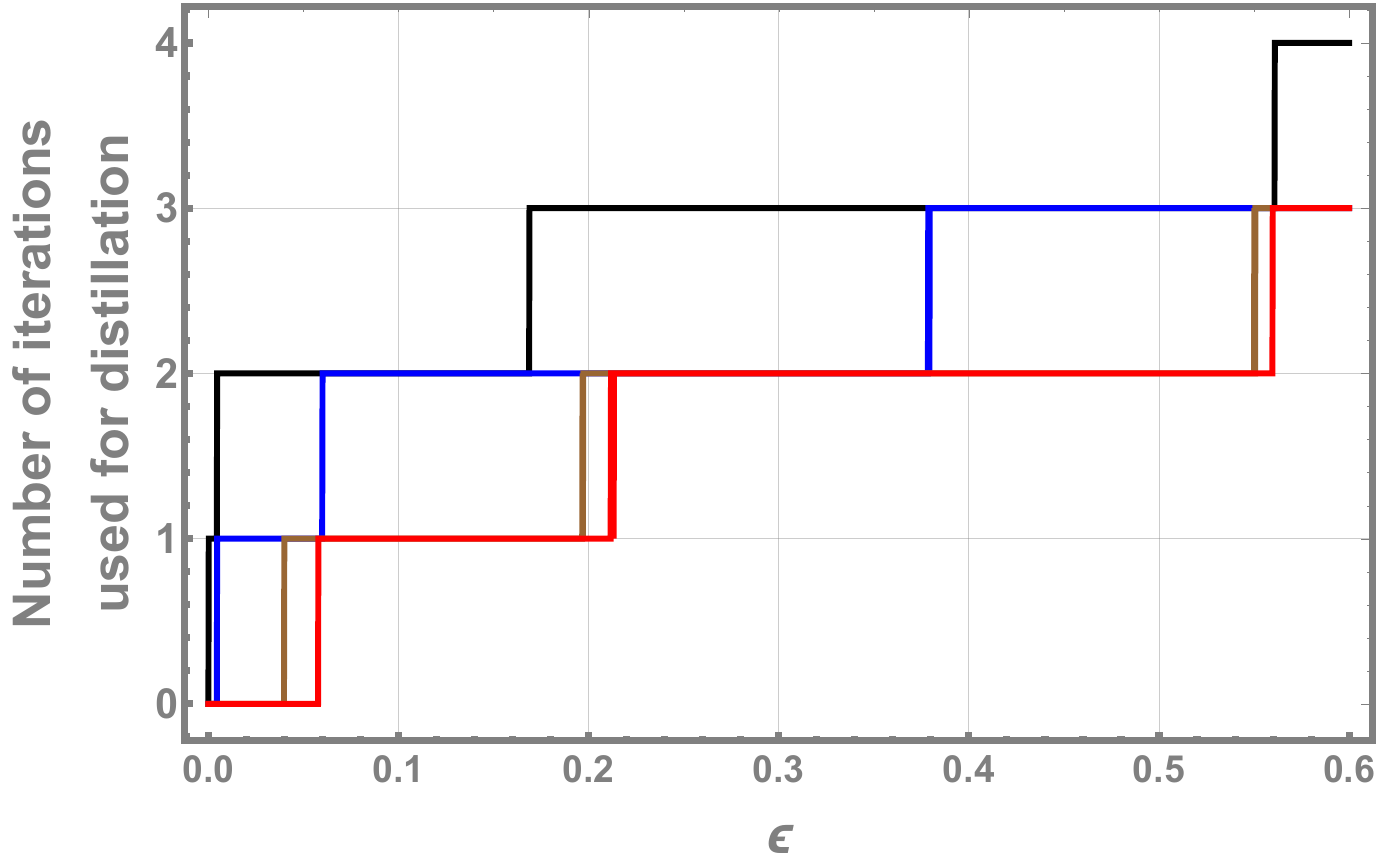}

    \end{minipage}

    \begin{minipage}
    {0.9\linewidth}
    \raggedright (b) Number of distillation levels required to reach an error of $10^{-15}$.
    \end{minipage}
    
    \caption{
    Improvements enabled by composite pulse sequences for implementing the $\mH$ gate in $X$-$Y$ Systems with global Rabi frequency errors.
    The black curves correspond to the single-pulse implementation, while the blue, red, and brown curves correspond to the three-, five-, and seven-pulse composite gates.
    (a) Comparison of Frobenius distance fidelity $\mathcal{F}_{\text{frob,1}}$ (solid lines) and Frobenius overlap fidelity $\mathcal{F}_{\text{frob,2}}$ (dashed lines) for single-, three-, five-, and seven-pulse implementations under global systematic errors. 
    (b) Reduction in the number of MSD iterations required to achieve a target gate error of \(10^{-15}\). In certain error regimes, composite sequences lower the required distillation levels by one or two.
    }
    \label{fig:mH fidelity and distillation improvement}
\end{figure}

The error suppression achieved by the 15-1 code is given in~\cite{BravyiKitaev} as:
\begin{equation}
    \epsilon'(\epsilon) = 
\frac{1 - 15(1 - 2\epsilon)^{7} + 15(1 - 2\epsilon)^{8} - (1 - 2\epsilon)^{15}}
     {2 \bigl(1 + 15(1 - 2\epsilon)^{8}\bigr)}.
\end{equation}
Figure~\ref{fig:mH fidelity and distillation improvement} illustrates the effectiveness of composite pulse sequences in improving the fidelity of the $\mH$ gate and in reducing the resource overhead required for MSD.

\section{\texorpdfstring{Example: Pre-Distillation in $X$-$Z$ Systems\\with Global $Z$ Errors}{Example: Pre-Distillation in X-Z Systems with Global Z Errors}}
\label{sec:Pre-Distillation in $X$-$Z$ Systems ; example}

As discussed in Section~\ref{sec:Physical Realizations and Errors} (and in Appendix~\ref{app:Realizations}), errors in integrated photonic platforms typically arise in both the Rabi frequency and the detuning parameters.
For more details, see Appendix~\ref{app:Realizations}.
To demonstrate the effectiveness of composite pulse schemes in constructing the $\mT$ gate, we consider a simplified error model previously studied in Ref.~\cite{KaplanErew}. In this model, the system is restricted to real-valued Rabi frequencies $\Omega$, and we focus on a single error parameter: a systematic $Z$ error denoted by $\epsilon_k = \epsilon$ for all segments $k = 1, \dots, N$. In this model, we neglect errors in the Rabi frequency $\Omega$ and assume that the detuning errors are fully correlated across all segments.

This model captures the essential features of error behavior in a variety of physical platforms governed by $\mathrm{SU}(2)$ dynamics, including trapped-ion qubits~\cite{trapped1,trapped2}, nonlinear optical processes such as sum-frequency generation~\cite{boyd2020nonlinear}, and atomic systems~\cite{atomic1,atomic2,atomic3}.

The total unitary evolution under an $N$-segment control sequence is given by
\begin{equation}
    U_N = \prod_{k=1}^N U\left(\theta_k, \phi_k, \epsilon\right),
\end{equation}
where each segment corresponds to the unitary
\begin{equation}
\label{eq:X-Z error}
    U\left(\theta, \phi, \epsilon\right) = \exp\left[i \theta \left( \cos\phi X + \sin\phi Z \right) + i \epsilon Z \right],
\end{equation}
and where $\theta_k$ and $\phi_k$ denote the parameters of the $k^\text{th}$ segment, respectively.
The error term here should not be interpreted as a discrete phase kick applied at the boundary of each segment, which would instead be modeled by a factorized form such as
\[
\exp\!\left[i\epsilon Z\right]\,
\exp\!\left[i\theta\left(\cos\phi X+\sin\phi Z\right)\right].
\]
Rather, the contribution $i\epsilon Z$ in Eq.~\eqref{eq:X-Z error} represents a global systematic detuning error in the underlying Hamiltonian, present continuously during the implementation of every control segment.
We model the detuning error as identical for all segments of the composite sequence.
Since the segment durations are free design parameters that are fixed only after the optimal pairs $\{\theta_k,\phi_k\}$ are determined, the effective error parameter $\epsilon$ appearing in each segment propagator can be treated as a uniform, segment-independent quantity.
More precisely, each control segment possesses three independent design parameters.
The composite-sequence constraints determine only two of them, namely the intended rotation angle $\theta_k$ and axis parameter $\phi_k$.
Consequently, one degree of freedom per segment remains.
This residual freedom can be used to select the segment durations such that the detuning accumulated during each segment produces the same additive contribution $i\epsilon Z$ in every propagator.
In this way, the effective detuning term appearing in each segment unitary can consistently be modeled by a common parameter $\epsilon$, thereby justifying the segment-independent error structure assumed in Eq.~\eqref{eq:X-Z error}.

For the purpose of facilitating numerical optimization in the next section, which also deals with an $X-Z$ system, we redefine the $\mT$ gate and the associated $T$-states as follows.
\begin{subequations}
    \begin{gather}
        \ket{T_0}= \cos\beta\, \ket{0} - e^{-i\pi/4} \sin\beta\, \ket{1}, \\
        \ket{T_1}= e^{i\pi/4} \sin\beta\, \ket{0} + \cos\beta\, \ket{1}, \\
        \mT = 
        \begin{pmatrix}
            \cos\beta & e^{i\pi/4} \sin\beta \\
            - e^{-i\pi/4} \sin\beta & \cos\beta
        \end{pmatrix}.
    \end{gather}
\end{subequations}
These states are Clifford-equivalent to the previously defined $T$-states, and thus this alternative representation is fully consistent for our purposes.

Since the $\mT$ gate involves a non-zero component along the $Y$ axis, it cannot be implemented using a single segment with real-valued control fields. At least two segments are required to synthesize such a unitary. Solving the constraint equation \( U_2 = \mT \) yields the following relations:
\begin{equation}
\begin{split}
        \cot \theta_1 = - & \sin \phi_1 \quad , \quad \cot \theta_2 = \sin \phi_2, \\ \cot \phi_2 =& \frac{(1 + \sqrt{3}) \cot \phi_1 \pm 2}{(1 + \sqrt{3}) \mp 1 \cot \phi_1},
\end{split}
\end{equation}
where $\phi_1$ can be chosen freely as a parameter.

To achieve first-order robustness against detuning errors, we design a three-segment composite pulse sequence by solving the coupled conditions \( U_3 = \mT \) and \( \left.\frac{dU_3}{d\epsilon}\right|_{\epsilon=0} = 0 \).
This leads to a continuous family of solutions that cannot be expressed analytically.
From this family, we select the following three representative solutions given in Table~\ref{tab:composite angles for the example} to demonstrate the feasibility such robust constructions.
\begin{table}[h!]
\centering
\resizebox{\linewidth}{!}{%
\begin{tabular}{c|ccc|ccc}
\toprule
\textbf{Sequence} & $\phi_1$ & $\phi_2$ & $\phi_3$ & $\theta_1$ & $\theta_2$ & $\theta_3$ \\
\midrule
(a) & 3.31558 & $-3.14159$ & $-0.17399$ & 1.26121 & $-2.86242$ & $-1.26121$ \\
(b) & 0.49741 & $-0.06006$ & $-0.41585$ & 3.25204 & $-1.56495$ & $-1.35397$ \\
(c) & 0.55004 & $-0.08606$ & $-3.58545$ & $-3.02982$ & $-1.56520$ & 1.35359 \\
\bottomrule
\end{tabular}
}
\caption{Optimized pulse parameters $(\theta_k, \phi_k)$ for three-segment composite sequences achieving first-order robustness to detuning errors. All values are in radians, shown to five decimal digits.}
\label{tab:composite angles for the example}
\end{table}

\begin{figure}[h!]
    \centering

        \centering
        \includegraphics[width=\linewidth]{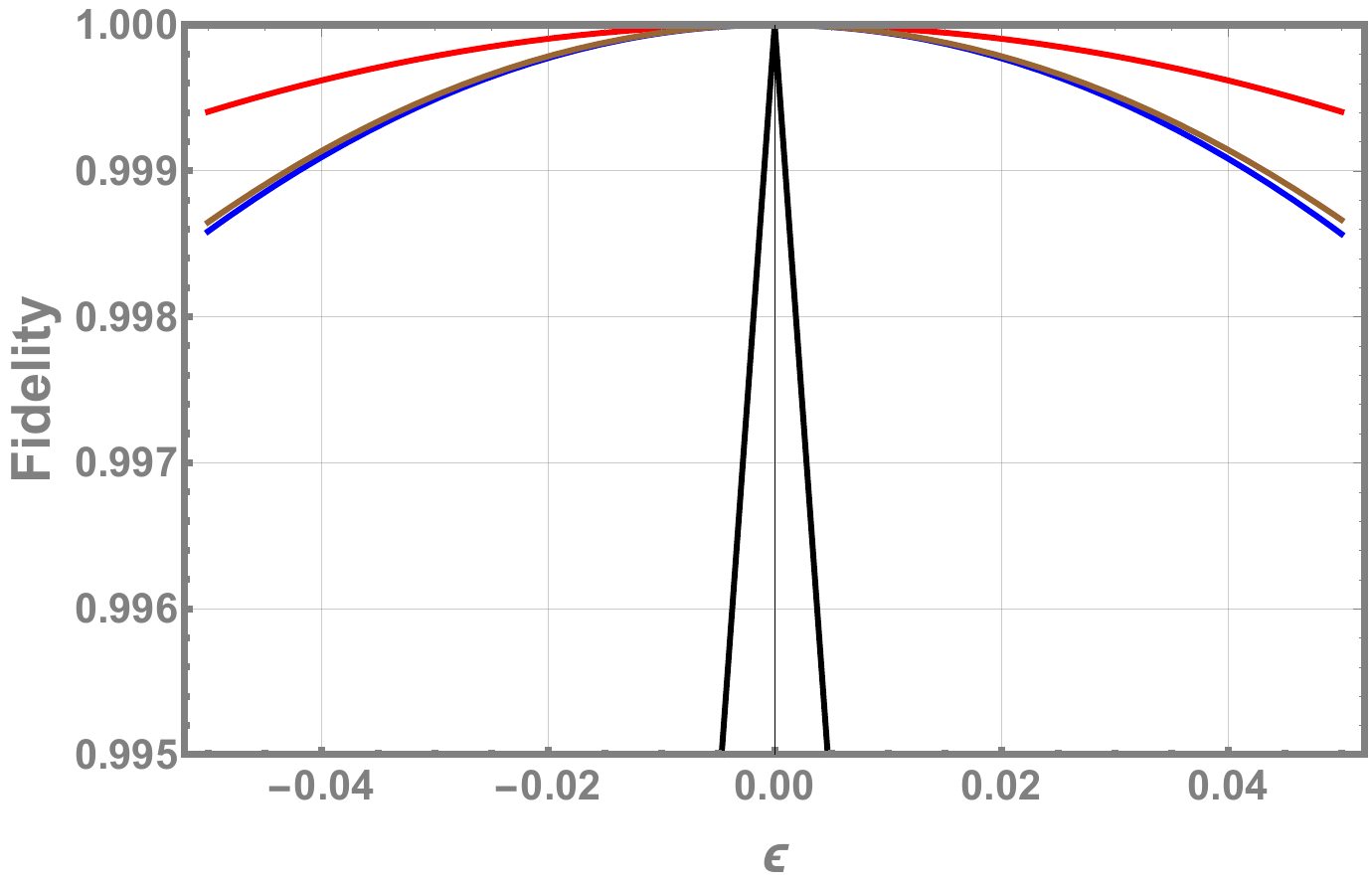}

    \begin{minipage}
    {0.9\linewidth}
    \raggedright (a) Frobenius distance fidelity $\mathcal{F}_{\text{frob,1}}$ for different composite pulse schemes.
    \end{minipage}
    
    \hfill

        \centering
        \includegraphics[width=\linewidth]{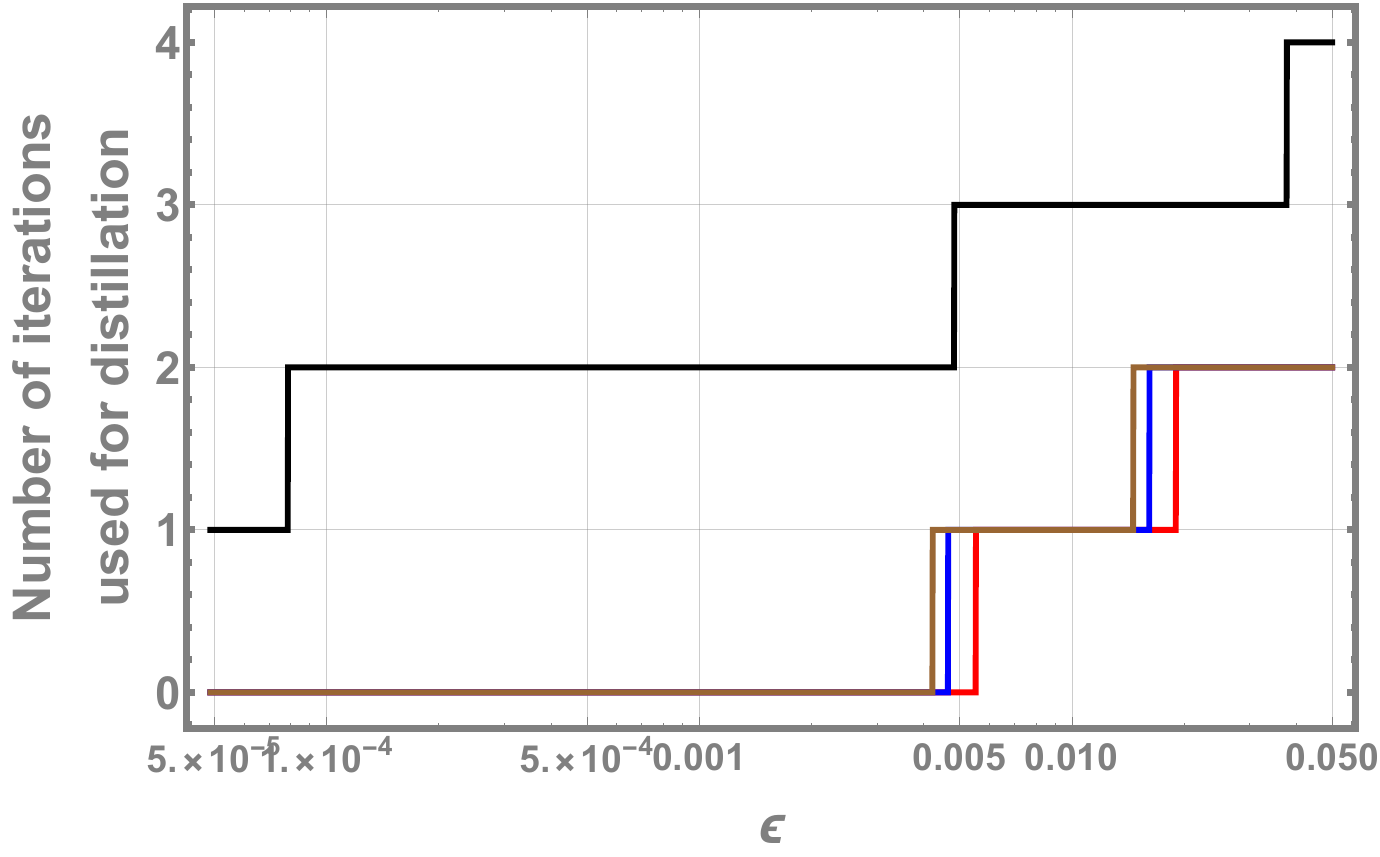}

    \begin{minipage}
    {0.9\linewidth}
    \raggedright (b) Number of distillation levels required to reach an error of $10^{-15}$.
    \end{minipage}
    
    \caption{
        Performance of $\mT$ gate implementations using composite pulse sequences under correlated systematic $Z$ errors in $X$-$Z$ Hamiltonians. The black curve shows the two-segment baseline, while the red, blue, and brown curves correspond to three different three-pulse designs labeled (a), (b), and (c). 
        Panel~(a) presents the Frobenius distance fidelity $\mathcal{F}_{\text{frob,1}}$ across these schemes.
        Panel~(b) shows the corresponding reduction in the number of magic state distillation levels required to achieve a gate T-magic error below \(10^{-15}\), based on Eq.~(\ref{eq:natural error of the gate}).
    }
    \label{fig:The improvement in mT gate for the example}
\end{figure}

Figure~\ref{fig:The improvement in mT gate for the example} illustrates the effectiveness of these composite pulse schemes in enhancing the fidelity of the $\mT$ gate under correlated systematic $Z$ errors, as arise from global imperfections in coupler widths in $X$-$Z$ control systems. The black curve corresponds to the baseline two-segment implementation, while the red, blue, and brown curves represent three distinct three-segment composite sequences labeled (a), (b), and (c), respectively.

Panel~\ref{fig:The improvement in mT gate for the example}(a) displays the Frobenius distance fidelity $\mathcal{F}_{\text{frob,1}}$ as a function of the global error parameter, highlighting the superior robustness of the composite designs. Panel~\ref{fig:The improvement in mT gate for the example}(b) quantifies the resulting resource savings by plotting the number of distillation levels required to reach a target gate error of \(10^{-15}\), based on the T-magic error of the gate- the operational error metric defined in Eq.~(\ref{eq:natural error of the gate}). Notably, in wide regions of the error domain, the composite pulse schemes reduce the required number of distillation rounds by up to two levels- achieving this improvement solely through a more robust gate synthesis using three pulses instead of two.

\section{Pre-Distillation in Integrated Photonics}
\label{sec:Pre-Distillation in Integrated Photonics}

As discussed in Section~\ref{sec:Physical Realizations and Errors} (and in Appendix~\ref{app:Realizations}), we model the dominant source of error in integrated photonic implementations as a fully correlated global perturbation in the waveguide widths across all segments. Specifically, each segment experiences a common deviation $\delta w$ added to both waveguide widths.
For more details, see Appendix~\ref{app:Realizations}.

\subsection{Optimization Procedure and Assumptions}

The integrated-photonics composite schemes presented in this section build upon the directional-coupler framework developed in Ref.~\cite{KaplanErew}. In particular, the numerical extraction of the effective Hamiltonian parameters and their dependence on the physical waveguide geometry follow the methodology established there. The new ingredient introduced in the present work is the application of this framework to the robust implementation of the non-Clifford $\mT$ gate and to the analysis of its impact on magic-state distillation overhead.

The effective coupling and detuning parameters, denoted by $\Omega(w_1,w_2)$ and $\Delta(w_1,w_2)$, were obtained from using coupled-mode theory for analytical guidance, together with full-wave electromagnetic simulations performed in the \textit{Lumerical} software package for waveguide widths in the range
\[
350\,\mathrm{nm}\leq w_{1,2}\leq450\,\mathrm{nm},
\]
while keeping the remaining physical parameters fixed, including the waveguide gap, wavelength, temperature, and etching depth. Smooth polynomial interpolation functions were then constructed from the discrete simulation data, allowing efficient evaluation of the effective Hamiltonian parameters throughout the optimization procedure without repeatedly invoking full-wave simulations.

The optimization variables in the present work consist of the segment widths and segment lengths defining the composite directional-coupler structure. The widths were restricted to the interpolation range above, within which the polynomial fits for $\Omega(w_1,w_2)$ and $\Delta(w_1,w_2)$ remain valid and accurately reproduce the underlying numerical simulations.

The optimization objective was to realize the target $\mT$ gate while minimizing the corresponding T-magic error defined in Eq.~(\ref{eq:natural error of the gate}). Robustness against correlated fabrication imperfections was imposed through derivative-cancellation conditions with respect to the global fabrication-error parameter, thereby suppressing the leading-order sensitivity of the implemented unitary to systematic width deviations.

Several physical effects were neglected in the present analysis. In particular, we do not include propagation loss, incoherent decoherence processes, mode-dependent absorption, polarization-mixing effects, thermal fluctuations, or temporally varying stochastic noise. The analysis is therefore restricted to the systematic-error-dominated regime in which correlated fabrication imperfections constitute the leading contribution to the implementation error.

\subsection{Results}

The total unitary evolution of an $N$-segment control sequence is given by
\begin{equation}
    U_N = \prod_{k=1}^N U(z_k, w_{1,k}, w_{2,k}; \delta w),
\end{equation}
where $z_k$ is the length of the $k^\text{th}$ segment, $w_{1,k}$ and $w_{2,k}$ are the nominal widths of the two waveguides, and $\delta w$ is the global width error. The unitary for each segment takes the form
\footnotesize
\begin{equation}
\begin{split}
    U(z_k, w_{1,k}, w_{2,k}; \delta w) 
    &= \exp\Bigg[ -i \frac{z_k}{2} \Big( 
        \Omega(w_{1,k} + \delta w, w_{2,k} + \delta w) X \\
    &\quad - \Delta(w_{1,k} + \delta w, w_{2,k} + \delta w) Z 
    \Big) \Bigg],
\end{split}
\end{equation}
\normalsize
where $\Omega_k$ and $\Delta_k$ denote the effective coupling strength and detuning for the $k^\text{th}$ segment, determined by the geometry.

The interpolation functions $\Omega(w_1,w_2)$ and $\Delta(w_1,w_2)$, which map the physical waveguide widths to the effective Hamiltonian parameters, were obtained through polynomial fitting of the numerical simulation data reported in Ref.~\cite{KaplanErew} for waveguide widths in the range $350\,\mathrm{nm}\leq w_{1,2}\leq450\,\mathrm{nm}$. The resulting fitted expressions, valid within this parameter regime, are given below, where the widths are expressed in micrometers and the resulting values are in units of $1/\mu\mathrm{m}$:
\small
\begin{subequations}
\begin{equation}
\begin{split}
\Delta(w_1, w_2) &= 3.94502808 \, w_2 
- 18.0203544 \, w_2^2 \\
& + 27.94843595 \, w_2^3 
- 15.42295066 \, w_2^4 \\
& - 3.94502818 \, w_1 
+ 18.02035521 \, w_1^2 \\
& - 27.94843797 \, w_1^3 
+ 15.42295233 \, w_1^4,
\end{split}
\end{equation}
\begin{equation}
\begin{split}
\Omega(w_1, w_2) &= 0.38044405 
- 1.48138422 (w_1 + w_2) \\
& + 2.51783632 (w_1 + w_2)^2 
 - 1.9993113 (w_1 + w_2)^3 \\
& + 0.60771393 (w_1 + w_2)^4.
\end{split}
\end{equation}
\end{subequations}
\normalsize

To ensure compatibility with fabrication constraints, we restrict all geometric parameters, specifically, the waveguide widths and segment lengths, to integer multiples of 1~nm. For each candidate solution to gate synthesis, we round these values up or down and select the combination that yields the highest fidelity to the target unitary. The fidelity is evaluated by comparing the synthesized gate to the desired one, and the optimal configuration is retained. For a detailed description of the physical system, the underlying error model, and the optimization procedure, we refer the reader to Ref.~\cite{KaplanErew}. The present analysis follows the same framework in order to construct the solutions and evaluate their robustness.

Table~\ref{tab:composite T gate solutions in couplers} summarizes the implementations of the $\mT$ gate using both two- and four-segment pulse sequences. The two-segment design directly implements the target unitary without robustness considerations. In contrast, the four-segment case yields a continuous family of solutions that achieve first-order robustness against variations in the coupler width, $\delta w$. Although this family cannot be succinctly described in closed form, we identify three representative configurations that exhibit high fidelity and practical feasibility.

\begin{table}[h!]
\centering
\resizebox{\linewidth}{!}{%
\begin{tabular}{c||cccccc}
\toprule
\textbf{Design} & \textbf{Segment} & $w_1$ (nm) & $w_2$ (nm) & $z$ ($\mu$m) \\
\midrule
\multirow{2}{*}{Two-segment} 
    & 1 & 449 & 356 & 23.813 \\
&   2 & 356 & 449 & 23.813 \\
\midrule
\multirow{4}{*}{Four-segment: Design (a)} 
    & 1 & 449 & 387 & 23.220 \\
&   2 & 424 & 448 & 17.190 \\
&   3 & 450 & 351 & 21.152 \\
&   4 & 353 & 450 & 36.094 \\
\cmidrule(lr){1-5}
\multirow{4}{*}{Four-segment: Design (b)} 
    & 1 & 448 & 387 & 23.822 \\
&   2 & 395 & 422 & 15.660 \\
&   3 & 449 & 350 & 21.215 \\
&   4 & 355 & 449 & 37.081 \\
\cmidrule(lr){1-5}
\multirow{4}{*}{Four-segment: Design (c)} 
    & 1 & 450 & 389 & 22.675 \\
&   2 & 395 & 407 & 18.369 \\
&   3 & 449 & 350 & 20.343 \\
&   4 & 359 & 450 & 39.073 \\
\bottomrule
\end{tabular}
}
\caption{Segment parameters for two-, three-, and four-segment composite implementations of the $\mT$ gate. Widths $w_1$, $w_2$ are in nm, lengths $z$ in $\mu$m. The three four-segment designs are representative samples from a continuous robust family.}
\label{tab:composite T gate solutions in couplers}
\end{table}

\begin{figure}[h!]
    \centering

        \centering
        \includegraphics[width=\linewidth]{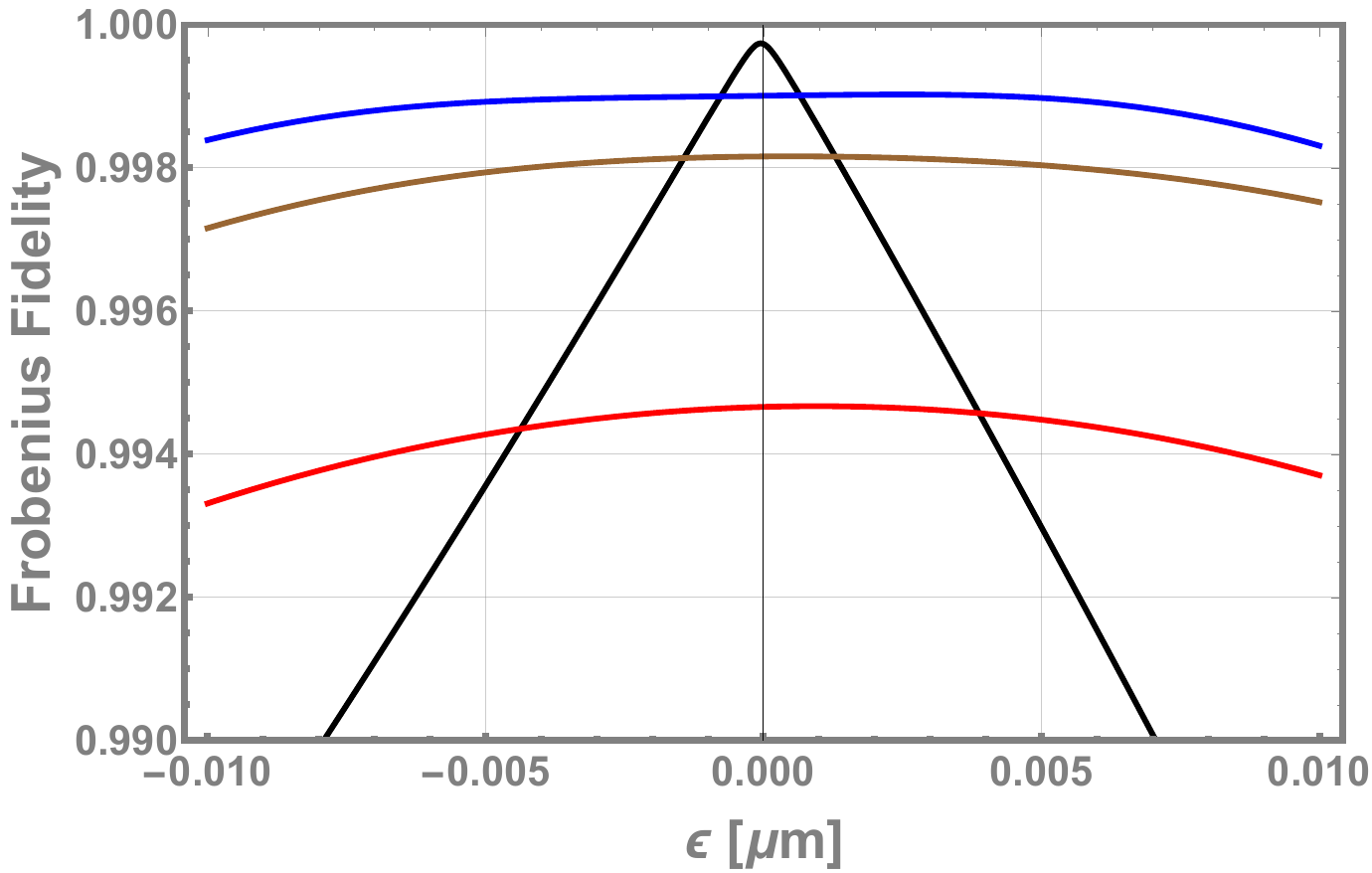}

    \begin{minipage}
    {0.9\linewidth}
    \raggedright (a) Frobenius distance fidelity $\mathcal{F}_{\text{frob,1}}$ for different composite pulse schemes.
    \end{minipage}
    
    \hfill

        \centering
        \includegraphics[width=\linewidth]{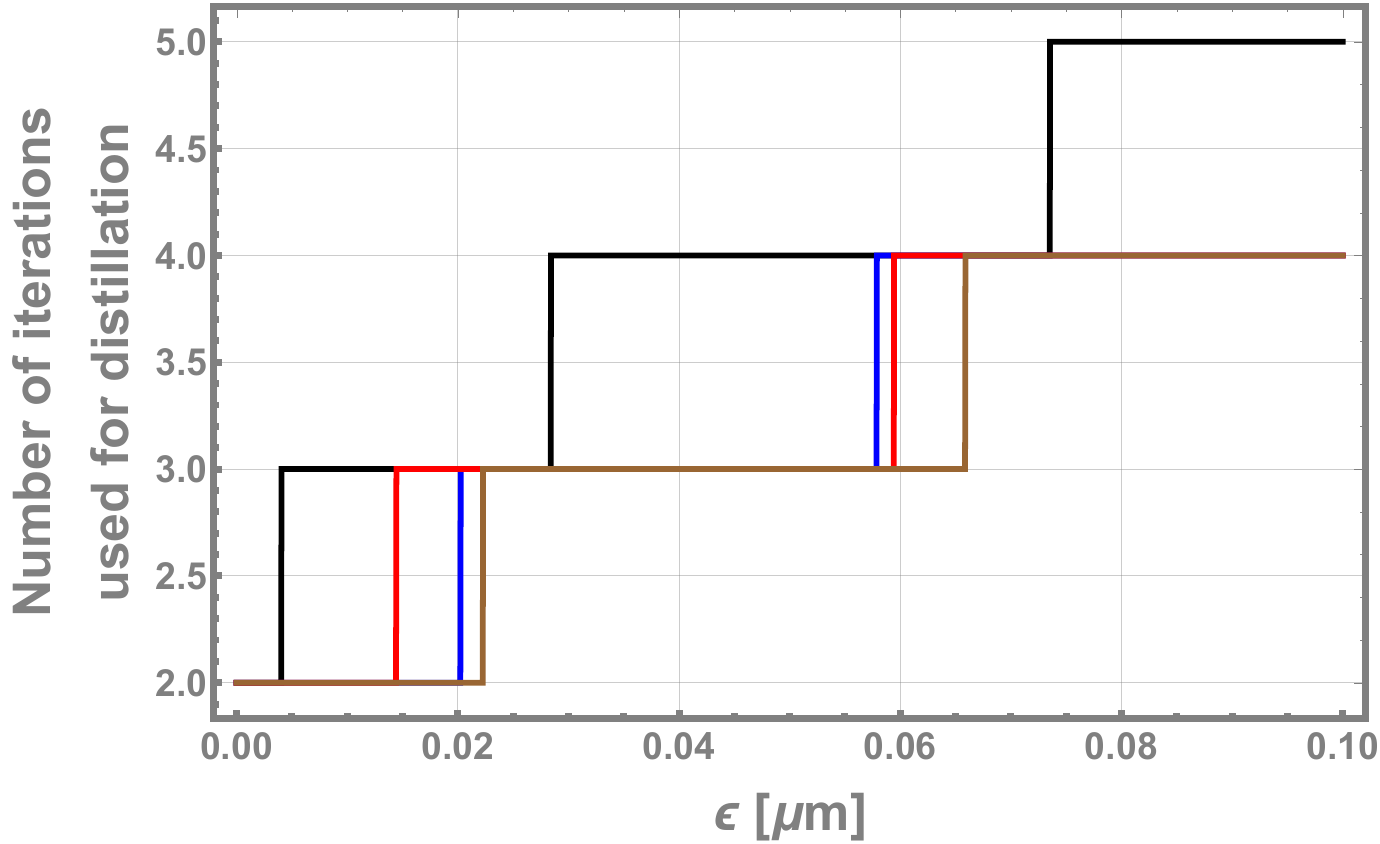}

    \begin{minipage}
    {0.9\linewidth}
    \raggedright (b) Number of distillation levels required to reach an error of $10^{-15}$.
    \end{minipage}
    
    \caption{
        Performance of $\mT$ gate implementations using composite pulse sequences under global systematic width errors in directional couplers. The two-segment design (black) is compared with three four-pulse composite schemes: (a) blue, (b) red, and (c) brown. While scheme (a) shows an initial advantage in fidelity, scheme (c) offers superior performance over the entire error range in terms of the number of concatenation levels.
    }
    \label{fig:The improvement in mT gate COUPLERS}
\end{figure}

Figure~\ref{fig:The improvement in mT gate COUPLERS} illustrates the performance of composite pulse schemes in enhancing the fidelity of the $\mT$ gate under global systematic errors in the coupler widths. The black curve corresponds to the two-segment implementation, while the blue, red, and brown curves represent three distinct four-pulse schemes labeled (a), (b), and (c), respectively.

Panel~\ref{fig:The improvement in mT gate COUPLERS}(a) compares the Frobenius distance fidelity $\mathcal{F}_{\text{frob,1}}$ across these schemes. Due to limitations in the available generators and fabrication constraints, the fidelity achieved by the four-pulse composite schemes can be slightly lower than that of the two-segment design for very small error values. However, across a broader error range, the composite sequences demonstrate superior robustness.

Panel~\ref{fig:The improvement in mT gate COUPLERS}(b) shows the corresponding reduction in the number of distillation levels required to achieve a target gate T-magic error of \(10^{-15}\), as defined in Eq.~(\ref{eq:natural error of the gate}). While the fidelity differences observed at small error rates might suggest an advantage for the two-segment scheme, all implementations ultimately require only two levels of MSD in this regime. Notably, although scheme (a) initially appears advantageous, among the four-segment designs examined here scheme (c) yields the lowest number of distillation levels across the error range considered. We do not claim that scheme (c) is globally optimal; rather, it serves as a representative example of the robustness achievable within our composite-segment framework.

\section{Errors in Gate-Channel Implementation}
\label{sec:Errors in Gate-Channel Implementation}

As presented in Section~\ref{sec:Distillation and Pre-Distillation} and Appendix~\ref{app:Background and Preliminaries}, distilled magic states are injected into Clifford-based, fault-tolerant circuits to realize non-Clifford gates. These gates complete the universal gate set required for fault-tolerant quantum computation. So, in practice, one must also evaluate the error of the implemented gate, since imperfections in the injected magic state directly propagate into the non-Clifford operation and may compromise fault tolerance. Here we quantify how imperfections in state preparation propagate through standard injection circuits and enter distillation recursion maps and gate-channel errors at leading order. We show that the error in the operation scales linearly with the leading-order error of the noisy prepared $T$-states.

A standard technique of the injection employs an ancilla-assisted circuit that applies the desired non-Clifford gate probabilistically via postselection. Consider the ancilla state
\begin{equation}
    \ket{A_\theta} = \frac{1}{\sqrt{2}}\left(\ket{0} + e^{i\theta}\ket{1}\right),
\end{equation}
and assume the initial state is \( \ket{\psi} \otimes \ket{A_\theta} \).  
A stabilizer measurement of \( \sigma_z \otimes \sigma_z \) is performed, followed by a CNOT gate with the first qubit as control.  

We define the single-qubit rotation
\begin{equation}
    \Lambda(e^{i\theta}) \equiv 
    \begin{pmatrix}
        1 & 0 \\
        0 & e^{i\theta}
    \end{pmatrix}.
\end{equation}
The measurement outcomes yield the following transformations:
\begin{itemize}
    \item If the outcome is \( +1 \), the resulting state is
    \(
    \Lambda(e^{i\theta}) \ket{\psi} \otimes \ket{0}.
    \)
    \item If the outcome is \( -1 \), the resulting state is
    \(
    e^{i\theta}\Lambda(e^{-i\theta}) \ket{\psi} \otimes \ket{1}.
    \)
\end{itemize}
Hence, the rotation $\Lambda(e^{i\theta})$ is realized with probability $1/2$.  
At this point, the second qubit can be discarded, i.e., traced out.

In particular, given access to the magic state \( \ket{T} \), one can prepare the two-qubit state \( \ket{T} \otimes \ket{T} \), perform a \( \sigma_z \otimes \sigma_z \) measurement, apply a CNOT, and then a Hadamard gate on the first qubit.  
For a \( +1 \) measurement outcome, the postselected state is
\[
    e^{i\gamma} \ket{A_{-2\gamma},0}, \quad \text{with } \gamma = \tfrac{\pi}{12},
\]
which enables the implementation of the non-Clifford unitary
\(
\Lambda(e^{-2i\gamma}).
\)
Again, the second qubit can be discarded here, i.e., traced out.

Adopting the Bravyi-Kitaev model, we assume the ability to prepare $\ket{0}$ states ideally, to apply Clifford group operations ideally, and to perform ideal measurements of single-qubit Pauli operators.
The only faulty resource is the preparation of magic states from stabilizer states, modeled by the mixed state $\rho_T(\epsilon)$.
This error is precisely the natural error defined in Eq.~\eqref{eq:natural error of the gate}, originating from the faulty implementation of the $\mT$ gate.
Under these assumptions, one can compute the effective operation corresponding to the intended $\Lambda(e^{-2i\gamma})$ rotation.
In fact, the resulting operation is not a unitary, but rather a quantum channel. Explicitly, the channel including the two errors, $\epsilon_1$ and $\epsilon_2$, in the preparataion of the $T$-states is given by
\begin{equation}
    \ket{\psi}\bra{\psi}\mapsto
    \begin{pmatrix}
        \cos^2 \theta & 
        -i a\, e^{-i\phi}\, \cos\theta \sin\theta \\[0.8em]
         i a^*\, e^{i\phi}\, \cos\theta \sin\theta & 
        \sin^2 \theta
    \end{pmatrix},
\end{equation}
where $\ket{\psi}\equiv \cos\theta \, \ket{0}+e^{i \phi}\sin\theta \, \ket{1}$,
\begin{equation}
    a=\frac{\,1 - i\sqrt{3} 
   + i\big(2i + \sqrt{3}\big)\,\epsilon_{2} 
   + \epsilon_{1}\big(-2 + i\sqrt{3} + 4\epsilon_{2}\big)}
     {-2 + \epsilon_{1} + \epsilon_{2} - 2\epsilon_{1}\epsilon_{2}},
\end{equation}
and we used the density matrix representation for the resulting state. We denote this channel by $C(\epsilon_1,\epsilon_2)$.

To quantify the performance of this channel, we can use Jozsa’s fidelity for quantum states~\cite{Jozsa01121994}
\footnote{
The Jozsa’s fidelity between two quantum states $\rho$ and $\sigma$, expressed as density matrices, is:
\[
F(\rho, \sigma) 
= \left( \operatorname{Tr} \sqrt{ \, \sqrt{\rho}\, \sigma \, \sqrt{\rho} } \right)^{2}.
\]
}
between the achieved state and the desired state.
The resulting fidelity is
\begin{equation}
\begin{split}
            F & =1-\frac{3}{4}\frac{\epsilon_1+\epsilon_2}{2-\epsilon_1-\epsilon_2+2\epsilon_1\epsilon_2}\sin^2\left(2\theta\right) \\
            & = 1-\frac{3\sin^2(2\theta)}{8}(\epsilon_1+\epsilon_2)+O(\epsilon^2).
\end{split}
\end{equation}
Hence, the error in the channel scales linearly with the leading-order error of the noisy prepared $T$-states, and the analysis presented in the previous sections is therefore sufficient.

A stronger and operationally meaningful performance metric is provided by the diamond-norm distance between the noisy channel and the ideal one. 
We first evaluate the trace norm of the channel difference $\Delta C \equiv C(\epsilon_1,\epsilon_2)-C(0,0)$.
Its trace norm is
\begin{equation}
\begin{split}
    \left\| \Delta \right. & \left. C \right\| _ 1
    = \Tr \sqrt{\left( \Delta C \right) ^ \dagger \left( \Delta C \right)}
    \\
    & = \left| a(\epsilon_1,\epsilon_2) - a(0,0) \right| \cdot | \sin (2\theta) |
    \\
    & = \sqrt{3 \frac{\left( \epsilon_1 + \epsilon_2 \right) ^2 - 2 \epsilon_1 \epsilon_2 (\epsilon_1 + \epsilon_2) + 4 \epsilon_1^2 \epsilon_2 ^2 }{\left(2-\epsilon_1 - \epsilon_2 + 2 \epsilon_1 \epsilon_2\right)^2}} \cdot | \sin (2\theta) |
    \\
    & = \sqrt{\frac{3}{4}}(\epsilon_1+\epsilon_2) | \sin (2\theta) | + O(\epsilon^2).
\end{split}
\end{equation}
The diamond norm is obtained by maximizing the output trace distance over all input states,
\begin{equation}
\begin{split}
    \left\| C \right. & \left. (\epsilon_1,\epsilon_2) - C(0,0) \right\| _\diamond 
    = \max_{\theta,\phi} \left\| \Delta C \right\| _1
    \\
    & = \sqrt{\frac{3}{4}}(\epsilon_1+\epsilon_2) + O(\epsilon^2) ,
\end{split}
\end{equation}
where the maximum is attained for $|\sin(2\theta)|=1$.
Thus, the diamond distance, which upper bounds the distinguishability between the two channels, also scales linearly with the dominant preparation errors.
The optimal single-shot probability for distinguishing the noisy channel from the ideal one is therefore
\begin{equation}
    p_{\mathrm{max}} = \frac{1}{2} + \frac{1}{4} \left\| C  (\epsilon_1,\epsilon_2) - C(0,0) \right\| _\diamond ,
\end{equation}
in direct accordance with the operational interpretation of the diamond norm.

A more suitable fidelity measure is the one proposed in~\cite{RAGINSKY200111}, which is based on Jozsa’s fidelity for quantum states. Nevertheless, the calculation presented here is representative and sufficient to demonstrate that the gate error depends linearly on the error in the $T$-states.

More generally, when losses and non-Hermitian dynamics are neglected, any $n$-qubit gate $G$ implemented in a physical realization with small imperfections can be modeled as $G e^{i \epsilon_G H_G},$ where $\epsilon_G \ll 1$ quantifies the error strength and $H_G$ is a traceless Hermitian operator, conventionally normalized such that $\Tr(H_G^2)=1/2$, specifying the error direction. Both $\epsilon_G$ and $H_G$ are determined by the particular physical realization and its underlying error mechanisms.  
Although one can consider this more general case, the calculation we presented already captures the essential point: the gate error is linear in the error of the $T$-states.

\section{Conclusion}
\label{sec:Conclusion}

We have presented a general and platform-aware method for reducing the overhead of magic state distillation (MSD) through the use of composite pulse sequences that robustly implement the non-Clifford $\mT$ gate.
While conventional composite designs often focus on Clifford gates such as $X$ or Hadamard, our approach directly targets the synthesis of the $\mT$ gate-improving the fidelity of the resulting magic states prior to any distillation protocol.

We derived both analytic and numerical constructions for symmetric three- and five-segment sequences that suppress first- and second-order global Rabi frequency errors in $X$-$Y$ control systems. Similar techniques were extended to $X$-$Z$ Hamiltonians, where we demonstrated robustness to global detuning-like imperfections. In integrated photonics, we adapted our framework to directional couplers, designing geometry-optimized segments that yield high-fidelity $\mT$ gates resilient to global correlated fabrication imperfections.

To quantify the improvement, we introduced a natural, operationally meaningful fidelity measure tailored to the $\mT$ gate, providing a direct link between physical gate performance and distillation cost. Across all platforms considered, and within the systematic-error-dominated regimes adopted in this work, our schemes reduce the required number of distillation iterations by up to three levels, resulting in corresponding exponential savings in qubit overhead. This pre-distillation approach complements existing fault-tolerant architectures, and offers a scalable route to higher-fidelity magic states at reduced cost, and paves the way toward more scalable and resource-efficient universal quantum computation.
We do not claim optimality or superiority over other robust-control strategies, as no systematic benchmarking against modern optimal-control approaches is performed.

A natural direction for extending our work is the integration of pre-distillation into fault-tolerant architectures.
Benchmarking our composite Clifford- and non-Clifford- gate designs within surface code frameworks would enable a direct assessment of reductions in logical qubit counts and space-time overhead for large-scale algorithms, while exploring their adaptation to LDPC-type quantum codes could highlight advantages for emerging high-threshold, low-overhead schemes.
Beyond single-qubit resources, composite sequences may be generalized to prepare two-qubit non-Clifford resources, such as entangled magic states required for Toffoli or controlled-$S$ gates, which are typically much more costly to distill.
This line of development naturally leads to the concept of “pre-distilled resource factories” implemented at the hardware level, where composite gates act as the entry point for generating batches of high-fidelity non-Clifford states that can be directly consumed by error-corrected computation, thereby substantially reducing overhead at scale.

\section*{Acknowledgements}
This work has been supported by the Israel Science Foundation (ISF), the Directorate for Defense Research and Development (DDR\&D) under Grant No. 3427/21, and the Tel Aviv University Center for Quantum Science and Technology.
M.G. acknowledges additional support from ISF Grant No. 1113/23 and from the US–Israel Binational Science Foundation (BSF) under Grant No. 2020072.
Y.O. acknowledges further support from the ISF Excellence Center, the BSF, and the Israel Ministry of Science.

\small
\bibliographystyle{unsrt.bst}
\bibliography{bibliography}

\appendix

\section{Magic State Distillation}
\label{app:Background and Preliminaries}

Quantum computing has emerged as a powerful paradigm that challenges classical notions of efficiency in computation. Yet, despite significant theoretical breakthroughs and experimental progress, realizing scalable quantum computers remains a daunting task. At the heart of this challenge lies the fragility of quantum information and the limitations imposed by fundamental theorems in quantum error correction. In particular, achieving both fault-tolerance and universality requires circumventing deep structural constraints on quantum codes. This section explains how one can overcome these obstacles through the technique of MSD, which supplements Clifford-based computation with carefully prepared non-stabilizer resources.

\subsection{Quantum Computing and\\Theoretical Limitations}

Quantum computing holds the promise of solving certain problems exponentially faster than classical algorithms. Landmark examples include Shor's algorithm for integer factorization~\cite{Shor1997} and Grover's algorithm for unstructured search~\cite{Grover1996}. Quantum systems harness the phenomena of superposition, entanglement, and interference, enabling entirely new paradigms of information processing. A single qubit can be described by the pure state
\begin{equation}
    \ket{\psi} = \alpha\ket{0} + \beta\ket{1}, \quad \text{with } \alpha, \beta \in \mathbb{C}, \quad |\alpha|^2 + |\beta|^2 = 1.
\end{equation}

The Clifford group, which plays a central role in quantum error correction and classical simulation of quantum circuits, is defined as the normalizer of the Pauli group within the unitary group. It is generated by the following gates:
\begin{itemize}
    \item The Hadamard gate: \( H = \frac{1}{\sqrt{2}} \begin{pmatrix} 1 & 1 \\ 1 & -1 \end{pmatrix} \),
    \item The Phase gate: \( S = \begin{pmatrix} 1 & 0 \\ 0 & i \end{pmatrix} \),
    \item The CNOT gate: \( \text{CNOT} = \ket{0}\bra{0} \otimes I + \ket{1}\bra{1} \otimes X \).
\end{itemize}

Clifford gates map Pauli operators to Pauli operators under conjugation. The Gottesman-Knill theorem~\cite{GottesmanHeisenberg,AaronsonGottesman} shows that quantum circuits composed solely of Clifford gates, Pauli measurements, and stabilizer state preparations can be efficiently simulated on a classical computer. Hence, such circuits offer no computational advantage.

Moreover, despite their theoretical power, physical quantum computers face a critical challenge: decoherence and noise.
Quantum systems interact unavoidably with their environment, leading to loss of coherence and errors. Unlike classical bits, qubits cannot be copied due to the no-cloning theorem, making traditional error correction inapplicable.

Quantum error correction offers a path forward. Stabilizer codes~\cite{GottesmanThesis}, a powerful family of quantum error-correcting codes (QECCs), encode logical qubits into entangled states of multiple physical qubits. For instance, the five-qubit code encodes one logical qubit and can correct any arbitrary single-qubit error. These codes rely on stabilizer generators-commuting Pauli operators that define the code space and protect the logical information.

A key requirement for fault-tolerant quantum computation is to prevent error propagation or to control the errors' spread. Transversal gates, which apply the same unitary to each qubit in a code block, naturally suppress correlated errors. However, the Eastin-Knill theorem~\cite{EastinKnill} proves that no QECC can implement a universal set of gates transversally. While Clifford gates often admit transversal implementations, at least one non-Clifford gate must be realized via a different, fault-tolerant mechanism.

\subsection{Magic State Distillation and Universal Computation}

Bravyi and Kitaev~\cite{BravyiKitaev} proposed \emph{magic state distillation} (MSD) as a fault-tolerant method to overcome the Eastin-Knill constraint. The key idea is to prepare special non-stabilizer states, known as \emph{magic states}, and use them to promote the Clifford group to universality via state injection. We begin by explaining how the injection of magic states enables universal computation.

\subsubsection{Universality via Ancilla-Assisted State Injection}

To achieve universality, we often need to implement non-Clifford gates such as phase rotations of the form
\begin{equation}
\Lambda(e^{i\theta}) \equiv \begin{pmatrix}
1 & 0 \\
0 & e^{i\theta}
\end{pmatrix}.
\end{equation}
That is because the subgroup of \( \text{U}(2) \) generated by the gate $\Lambda(e^{i\theta})$ and the Clifford Group for single qubits is dense in \( \text{U}(2) \). Therefore, this set provides a universal basis for quantum computation
\footnote{
    The Universal Gate Set Theorem states that any unitary operation on \( n \) qubits can be approximated to arbitrary accuracy using a finite sequence of gates drawn from:
\begin{itemize}
    \item All single-qubit unitary operations (i.e., elements of \( \text{U}(2) \)), and
    \item The two-qubit controlled-NOT gate (CNOT).
\end{itemize}
Therefore, the set consisting of all single-qubit gates and the CNOT gate is universal for quantum computation.
This result implies that any quantum algorithm can be implemented using only single-qubit operations and entangling CNOT gates, making this set of gates a \emph{universal basis} for quantum computation~\cite{NielsenChuang,Barenco1995}.
}
.

A standard technique is to use an ancilla-based circuit that probabilistically applies the desired gate by postselection.
Consider the ancilla state
\begin{equation}
    \ket{A_\theta} = \frac{1}{\sqrt{2}}(\ket{0} + e^{i\theta} \ket{1}),
\end{equation}
and assume the initial state \( \ket{\psi} \otimes \ket{A_\theta} \). A stabilizer measurement of \( \sigma_z \otimes \sigma_z \) is performed, followed by a CNOT gate with the first qubit as control.

\begin{itemize}
    \item If the measurement outcome is \( +1 \), the resulting state is
    \(
    \Lambda(e^{i\theta}) \ket{\psi} \otimes \ket{0}
    \).
    
    \item If the outcome is \( -1 \), the state becomes
    \(
    e^{i\theta} \Lambda(e^{-i\theta}) \ket{\psi} \otimes \ket{1}
    \).
\end{itemize}

Discarding the ancilla leaves the data qubit in a state that has undergone a known unitary-either \( \Lambda(e^{i\theta}) \) or \( \Lambda(e^{-i\theta}) \). Repeating this process allows one to simulate the desired non-Clifford gate with high probability.

\subsubsection{Magic States and Fault-Tolerance}

The states that can be injected and allow fault tolerant quantum computation are the magic states.
A canonical magic state is the \( T \)-state:
\begin{equation}
\begin{split}
    \ket{T} & = \cos \beta \ket{0} + e^{i\pi/4} \sin \beta \ket{1} \\
    & = \sqrt{\frac{3+\sqrt{3}}{6}}\ket{0} + e^{i\pi/4} \sqrt{\frac{3-\sqrt{3}}{6}}\ket{1},
\end{split}
\end{equation}
where \( \cos^2(2\beta) = \frac{1}{3} \). It is an eigenstate of the \( T \)-gate,
\begin{equation}
    T = e^{i\pi/4} S H = \frac{e^{i\pi/4}}{\sqrt{2}} \begin{pmatrix} 1 & 1 \\ i & -i \end{pmatrix}.
\end{equation}
We denote its two orthogonal eigenstates as
\begin{subequations}
    \begin{align}
        \ket{T_0} &= \ket{T}, \\
        \ket{T_1} &= -e^{-i\pi/4} \sin \beta \ket{0} + \cos \beta \ket{1}.
    \end{align}
\end{subequations}

Given access to \( \ket{T} \), one can prepare \( \ket{T} \otimes \ket{T} \), measure \( \sigma_z \otimes \sigma_z \), and apply a CNOT, and then a Hadamard gate on the first qubit. For a \( +1 \) measurement outcome, the postselected state is
\[
e^{i\gamma} \ket{A_{-2\gamma}}, \quad \text{with } \gamma = \frac{\pi}{12},
\]
which enables implementation of the non-Clifford unitary $\Lambda(e^{-2i\gamma})$.

\begin{figure*}[t!]

    \begin{minipage}
    {0.9\linewidth}
    \raggedright (a) \hspace{280pt} (b)
    \end{minipage}

  \centering
  \includegraphics[page=1,width=0.58\textwidth]{Figures_for_Magic_PreDistillation_final.pdf}
  \includegraphics[page=2,trim=11.5cm 0pt 0pt 0pt,clip,width=0.32\textwidth]{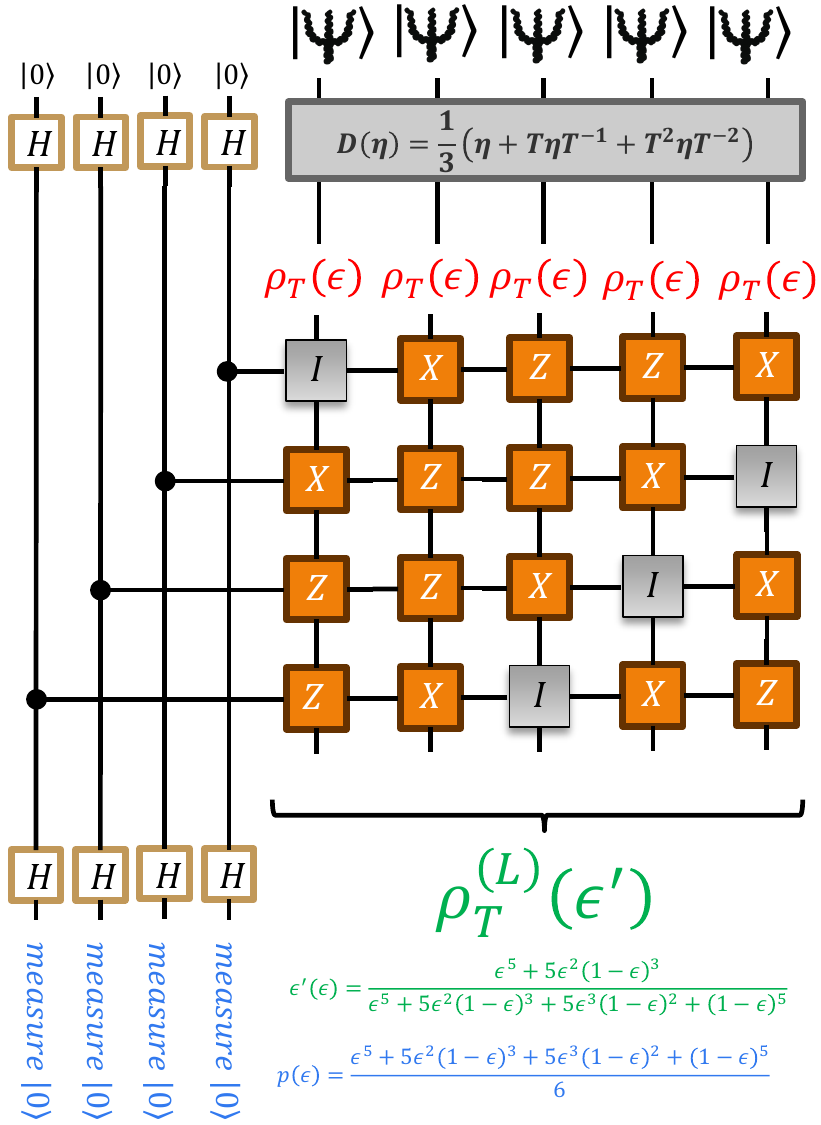}

    \begin{minipage}
    {0.9\linewidth}
    \raggedright (c) \hspace{214pt} (d)
    \end{minipage}
  
  \includegraphics[width=0.45\linewidth]{Distillation_Function.pdf}
  \includegraphics[width=0.45\linewidth]{Number_of_iterations.pdf}

    \caption{
        Five-qubit $T$-state distillation protocol.
        (a) Conceptual schematic of a single round of the $T$-magic-state distillation protocol. One round consumes five noisy copies of $\rho_T(\epsilon) \equiv (1-\epsilon)\ket{T_0}\bra{T_0} + \epsilon \ket{T_1}\bra{T_1}$ (denoted by $\rho_T$ with heavily wavy symbols to emphasize noise). With probability $p(\epsilon) = \tfrac{1}{6} + O(\epsilon)$, the protocol outputs the logical state $\rho^{(L)}_T(\epsilon')$, whose error is quadratically suppressed, $\epsilon'(\epsilon) = 5\epsilon^2 + O(\epsilon^3)$. We represent this improved state using symbols with reduced waviness, reflecting its lower noise.
        (b) Circuit-level details of the distillation protocol (see Appendix A for a full derivation). The input state is the coherent superposition $\ket{\psi} = \sqrt{1-\epsilon}\ket{T_0} + e^{i\phi}\sqrt{\epsilon}\ket{T_1}$, which, after dephasing via random application of $I$, $T$, or $T^2$ (each with probability $1/3$), reduces to the mixed state $\rho_T(\epsilon)$.
        (c) Distillation curve $\epsilon'(\epsilon)$: the output error rate as a function of the input error rate, showing quadratic suppression after one round.
        (d) Required number of recursive distillation rounds to achieve an error below the $10^{-15}$ fault-tolerance threshold, illustrating the exponential scaling of qubit overhead with the initial error $\epsilon$.
    }
  \label{fig:DistillationAppendix}
\end{figure*}

What makes the state \( \ket{T} \) \emph{magic} is not only its role in enabling universality, but also the fact that it can be distilled using the five-qubit code.
Starting from five copies of the noisy mixed state
\footnote{\label{footnote-dephasing}
Why do we consider mixed states of the form \( \rho_\epsilon = (1 - \epsilon)\ket{T_0}\bra{T_0} + \epsilon \ket{T_1}\bra{T_1} \) rather than pure states of the form \( \sqrt{1 - \epsilon}\ket{T_0} + e^{i\phi}\ket{T_1} \) for arbitrary \( \phi \)? While distillation is still possible in the latter case, the threshold \( \epsilon_c \) becomes significantly lower. To mitigate this, one can apply a dephasing transformation to each copy of the pure state, defined as
\begin{equation}
D(\eta) = \frac{1}{3} \left( \eta + T \eta T^\dagger + T^\dagger \eta T \right),
\label{eq:dephasing}
\end{equation}
thereby obtaining the mixed state \( \rho_\epsilon \) described above. The transformation \( D \) can be realized by randomly applying one of the operators \( I \), \( T \), or \( T^{-1} \), each with probability \( \frac{1}{3} \). This preprocessing makes the distillation process significantly more efficient.}
\begin{equation}
    \rho_\epsilon = (1 - \epsilon)\ket{T_0}\bra{T_0} + \epsilon \ket{T_1}\bra{T_1},
\end{equation}
a syndrome measurement projects (with probability $p(\epsilon)=\frac{\epsilon^5 + 5\epsilon^2(1 - \epsilon)^3 + 5\epsilon^3(1 - \epsilon)^2 + (1 - \epsilon)^5}{6} = \frac{1}{6} + O(\epsilon^2)$) onto the same form $\rho_{1 - \epsilon'(\epsilon)}$ with reduced noise:
\begin{equation}
\begin{split}
    \epsilon'(\epsilon) & = \frac{\epsilon^5 + 5\epsilon^2(1 - \epsilon)^3}
     {\epsilon^5 + 5\epsilon^2(1 - \epsilon)^3 + 5\epsilon^3(1 - \epsilon)^2 + (1 - \epsilon)^5} \\
     & = 5\epsilon^2 + O(\epsilon^3).
\end{split}
\end{equation}
This function is illustrated in Figure~\ref{fig:DistillationAppendix}a. The distillation works for \( \epsilon < \epsilon_c \approx 0.173 \), where
\begin{equation}
    \epsilon_c = \frac{1}{2}\left(1 - \sqrt{\frac{3}{7}}\right).
\end{equation}
So the five-qubit code, like every distillation protocol, exhibits a distillation threshold beyond which distillation fails to converge. This marks the maximum tolerable noise level for the input states under this scheme. Above this threshold, the output fidelity deteriorates with successive iterations, rendering the protocol ineffective for magic state purification.

By recursively applying the distillation procedure, through concatenation or nesting of the five-qubit code, the error can be suppressed to arbitrarily low levels, enabling the preparation of magic states with arbitrarily high fidelity.
Each iteration corresponds to a level of code nesting; that is, performing \( n \) iterations implies that each logical qubit requires \( 5^n \) physical qubits due to recursive encoding.

We denote by \( \epsilon_{c_i} \) the critical value of \( \epsilon \) at which the protocol transitions from requiring \( i{-}1 \) iterations to \( i \) iterations.
In Figure~\ref{fig:DistillationAppendix}b, we present the required number of distillation iterations, using the five-qubit code, to reduce the physical error rate \( \epsilon \) below the fault-tolerance threshold of \( 10^{-15} \), plotted on a logarithmic scale.
The following are the critical error values \( \epsilon_{c_i} \) for the first ten iteration thresholds:
\begin{equation}
\begin{split}
    \epsilon_{c_1} & =10^{-15} \quad \quad \quad , \ \epsilon_{c_2}=1.414 \times 10^{-8},\\
    \epsilon_{c_3}&=5.318 \times 10^{-5},\ \epsilon_{c_4}=3.251 \times 10^{-3}, \\
    \epsilon_{c_5}&=2.490 \times 10^{-2},\ \epsilon_{c_6}=6.676 \times 10^{-2}, \\
    \epsilon_{c_7}&=1.072 \times 10^{-1},\ \epsilon_{c_8}=1.353 \times 10^{-1}, \\
    \epsilon_{c_9}&=1.520 \times 10^{-1},\ \epsilon_{c_{10}}=1.615 \times 10^{-1} .
\end{split}
\label{eq:thresholds}
\end{equation}

These thresholds illustrate the sharp trade-off between input noise and resource overhead: as \( \epsilon \) increases, the number of required iterations (and thus the number of physical qubits per logical qubit) increases rapidly.
For instance, an initial error just below \( 10^{-2} \) requires four levels of distillation, resulting in \( 5^4 = 625 \) physical qubits per logical qubit, while an error around \( 10^{-1} \) would require seven or more iterations, consuming over \( 78,000 \) qubits.
This exponential growth in resource overhead underscores the importance of reducing the initial magic state error as much as possible prior to distillation, as even small improvements can drastically reduce the number of required physical qubits.

Figure~\ref{fig:Distillation} provides an overview to illustrate these ideas. In panel (a), five noisy input states of the form $\rho_T(\epsilon) = (1-\epsilon)\ket{T_0}\bra{T_0} + \epsilon \ket{T_1}\bra{T_1}$ are consumed in a single round of distillation. With probability $p(\epsilon) = \tfrac{1}{6} + O(\epsilon)$, the protocol outputs a logical state $\rho^{(L)}_T(\epsilon')$ whose error is quadratically suppressed, $\epsilon'(\epsilon) = 5\epsilon^2 + O(\epsilon^3)$. Thus, the protocol achieves both error reduction and probabilistic success, at the cost of increased resource overhead.
The technical details of the distillation circuit are depicted in panel (b). The input state can be written in the coherent form $\ket{\psi} = \sqrt{1-\epsilon}\ket{T_0} + e^{i\phi}\sqrt{\epsilon}\ket{T_1}$. After a random dephasing step, implemented by applying $I$, $T$, or $T^2$ with equal probability, it reduces to the mixed form $\rho_T(\epsilon)$. This mixed state serves as the standard input to the stabilizer-based distillation procedure. Panels (c) and (d) characterize the performance of the protocol. Panel (c) shows the distillation curve $\epsilon'(\epsilon)$, making explicit how the output error depends on the input. Panel (d) highlights the recursive structure of MSD: by concatenating multiple rounds, one can reduce the error below the fault-tolerance threshold (here taken to be $10^{-15}$). However, the required number of rounds grows rapidly with the initial error, reflecting the exponential qubit overhead inherent to this process.

Another notable single-qubit magic state is the Hadamard eigenstate:
\begin{equation}
    \ket{H} = \cos\left(\frac{\pi}{8}\right)\ket{0} + \sin\left(\frac{\pi}{8}\right)\ket{1}.
\end{equation}
Similar to \( \ket{T} \), this state can be distilled and used to implement non-Clifford operations. In particular, noting that
\begin{equation}
    \ket{H} = e^{i\pi/8} S^3 H \ket{A_{-\pi/4}},
\end{equation}
reveals that \( \ket{H} \) enables the injection of the non-Clifford unitary
\begin{equation}
\Lambda(e^{-i\pi/4}) = 
\begin{pmatrix}
1 & 0 \\
0 & e^{-i\pi/4}
\end{pmatrix},
\end{equation}
which lies outside the Clifford group and is commonly referred to as the \( T \) gate in other contexts.

\begin{figure}[h!]
    \centering
    \includegraphics[trim=3.1cm 3cm 3.1cm 3cm,clip,width=0.6\linewidth]{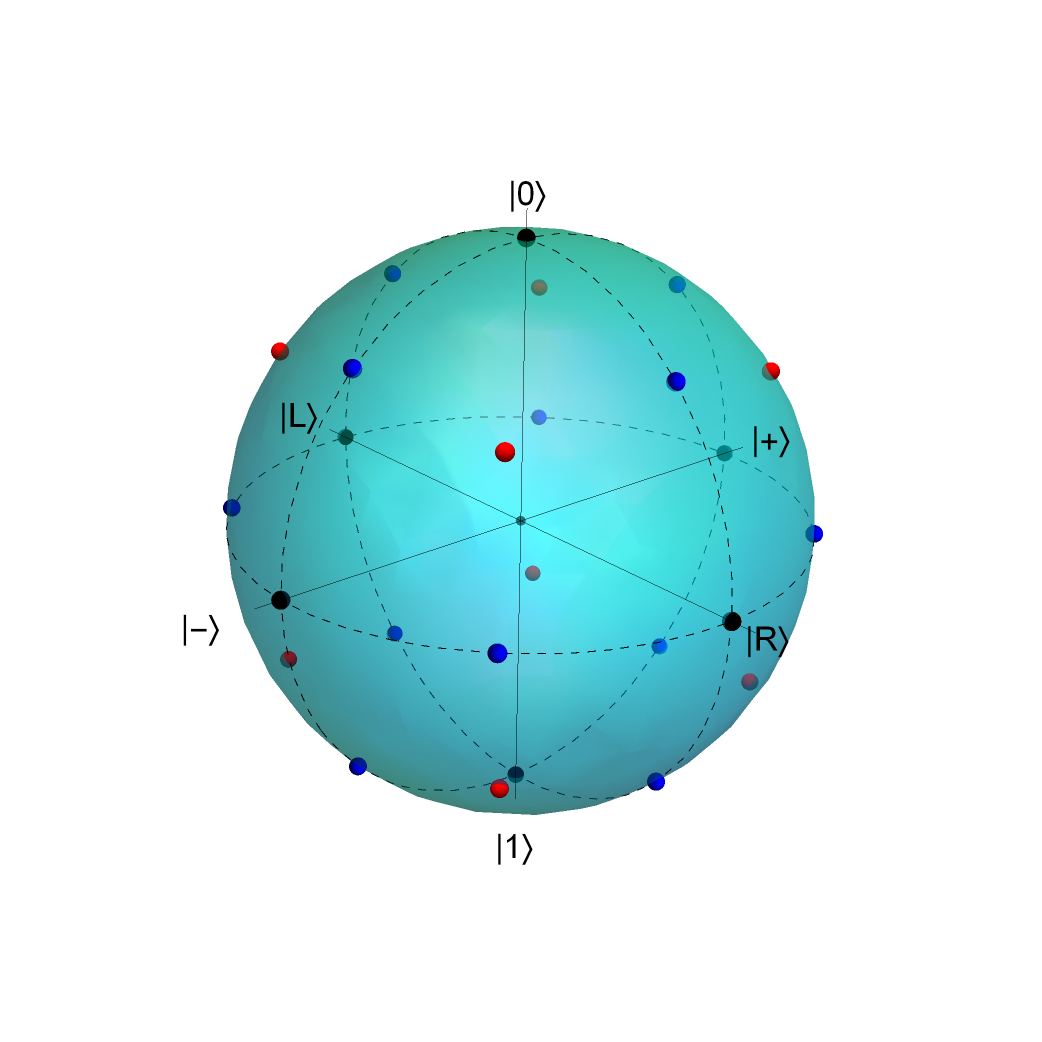}
    \caption{
    Representation of stabilizer and magic states for a single qubit plotted on the Bloch sphere. 
    Black points denote stabilizer states, which form the vertices of the stabilizer polytope. 
    Red points correspond to the non-stabilizer $T$-states, and blue points represent the $H$-states, both of which are essential resources for universal quantum computation. 
    These magic states lie outside the stabilizer polytope and enable non-Clifford operations via state injection.
}
    \label{fig:Single Qubit Magic States}
\end{figure}

The set of magic states for single qubits consists of \( \ket{T} \), \( \ket{H} \), and their Clifford equivalents. These are visualized on the Bloch sphere in Figure~\ref{fig:Single Qubit Magic States}.

\subsection{Toward Fault-Tolerant Universality}

To summarize, combining the following ingredients enables fault-tolerant, universal quantum computation:
\begin{itemize}
    \item Stabilizer codes to protect quantum information,
    \item Clifford gates implemented transversally,
    \item MSD to generate high-fidelity non-stabilizer states,
    \item State injection to realize non-Clifford gates.
\end{itemize}

While the resource overhead of MSD remains substantial, ongoing research seeks more efficient protocols. MSD continues to play a central role in the quest for scalable, fault-tolerant quantum computing.

\section{Physical Realizations of Qubits and Operations for Quantum Computing}
\label{app:Realizations}

In this appendix, we provide a brief overview of several leading physical platforms for realizing qubits and implementing single-qubit operations. These include superconducting circuits, trapped ions, neutral atoms, and integrated photonic devices. Each platform offers distinct control mechanisms -from microwave or laser-driven rotations in matter-based systems to coherent interference in photonic waveguides- and each suffers from characteristic error sources such as amplitude miscalibration, detuning, crosstalk, or fabrication imperfections. Understanding these platforms and their dominant error models establishes the foundation for the composite schemes introduced in the main text, which are designed to mitigate such systematic errors and enable more robust implementations of non-Clifford gates like $\mT$ and $\mH$.

\subsection{Superconducting Qubits}

In superconducting qubits (e.g., transmons), quantum states are manipulated via resonant microwave drives.
A typical driven Hamiltonian in the rotating frame is
\begin{equation}
H(t) = \frac{\Omega(t)}{2} \left( \cos\phi \, X + \sin\phi \, Y \right),
\end{equation}
where \( \Omega(t) \) is the Rabi frequency envelope (determined by pulse amplitude), and \( \phi \) is the microwave phase.
The resulting unitary is a rotation:
\begin{equation}
U(\theta, \phi) = \exp\left(-i \frac{\theta}{2} (\cos\phi\, X + \sin\phi\, Y)\right),
\end{equation}
with total angle \( \theta = \int \Omega(t) dt \). To implement $\mT$, one chooses $\theta = 2\beta$ and $\phi = 3\pi/4$ to realize a rotation in the equatorial plane with the desired complex phase. The \(\mH\) gate, involving a rotation by \(\pi/4\), is typically synthesized through a sequence of calibrated pulses, e.g., using a combination of $R_x(\theta)$ and $R_z(\phi)$ rotations.

Typical errors in the physical realization of quantum gates include amplitude errors (\(\theta \to \theta + \delta\theta\)), which are caused by inaccuracies in pulse duration or power and result in over- or under-rotation; phase errors (\(\phi \to \phi + \delta\phi\)), which arise from miscalibration of the phase reference frame or IQ imbalance in the control electronics; and leakage errors, which correspond to unintended transitions to noncomputational energy levels outside the qubit subspace.
Composite pulse schemes like BB1 or SK1 suppress systematic amplitude errors. For example, the BB1 sequence corrects a rotation \(R_\phi(\theta)\) using a sequence of four additional rotations:
\small
\begin{equation}
\begin{split}
    U_{\text{BB1}} = & R_\phi(\theta/2) R_{\phi+\arccos(-\theta/4\pi)}(\pi)
    \\ & R_{\phi+3\arccos(-\theta/4\pi)}(2\pi) R_{\phi+\arccos(-\theta/4\pi)}(\pi) R_\phi(\theta/2).
\end{split}
\end{equation}
\normalsize
The amplitude error is largely global across composite gates (same \(\delta\theta/\theta\)), but phase errors may be gate-specific depending on the control electronics.

\subsection{Trapped Ions}

In trapped ion systems, qubits are encoded in the internal energy levels of ions (e.g., $^{40}$Ca$^+$ or $^{171}$Yb$^+$).
Laser pulses drive Rabi oscillations, enabling high-fidelity single-qubit gates.
The effective Hamiltonian for a Raman transition is:
\begin{equation}
H = \frac{\Omega}{2} \left( e^{i\phi} \ket{0}\bra{1} + e^{-i\phi} \ket{1}\bra{0} \right),
\end{equation}
leading to evolution:
\begin{equation}
U(\theta, \phi) = 
\begin{pmatrix}
\cos(\theta/2) & -i e^{i\phi} \sin(\theta/2) \\
-i e^{-i\phi} \sin(\theta/2) & \cos(\theta/2)
\end{pmatrix}.
\end{equation}
Here, the laser phase $\phi$ and pulse area $\theta$ control the rotation axis and angle.
By tuning the laser parameters- detuning, amplitude, and phase- rotations like those required for \(\mT\) and \(\mH\) can be implemented.
A Raman beam pair can create an effective Hamiltonian that induces the desired unitary operation with a phase corresponding to $e^{i\pi/4}$, thus implementing \(\mT\).
The \(\mH\) gate is often realized via composite pulse sequences or through direct single-pulse methods.

Typical errors in the physical realization of quantum gates in trapped-ion systems include laser intensity fluctuations, where variations in laser power modify the effective Rabi frequency \(\Omega\), resulting in rotation angle errors (\(\theta \to \theta + \delta\theta\)); laser detuning (off-resonant driving), where frequency mismatch between the laser and the qubit transition introduces over- or under-rotation and undesired phase accumulation, often modeled as spurious \(Z\) rotations; and AC Stark shifts, where off-resonant couplings to auxiliary levels induce systematic shifts in the qubit energy levels, effectively contributing additional unwanted phases to the gate evolution.
Robust sequences like CORPSE or SCROFULOUS cancel detuning and amplitude errors. For instance, CORPSE decomposes a \(\pi\) pulse into three rotations with over- and under-rotation to cancel detuning effects.
Errors like detuning are typically global to all gate components, but intensity noise may vary slightly across segments, especially in pulse-shaping scenarios.

\subsection{Neutral Atoms}

Neutral atoms trapped in optical tweezers or lattices encode qubits in their hyperfine ground states.
Microwave or Raman transitions are used to implement single-qubit gates.
The Hamiltonian has a similar form to trapped ions:
\begin{equation}
H = \frac{\Omega(t)}{2} \left( \cos\phi \, X + \sin\phi \, Y \right),
\end{equation}
where the control parameters are again laser phase and pulse area. Arbitrary gates are synthesized using:
\begin{equation}
U = R_z(\gamma) R_y(\theta) R_z(\delta),
\end{equation}
where $R_y(\theta)$ and $R_z(\phi)$ are physical operations: $R_y$ by microwave pulses, and $R_z$ often via virtual frame updates.
Arbitrary unitaries like \(\mT\) and \(\mH\) are realized using laser pulses with precisely controlled phase and amplitude.
The $\mT$ gate can be realized by choosing angles to match a rotation axis with complex coefficients (e.g., a rotation in the $XY$-plane with global phase $e^{i\pi/4}$).

Typical errors in the physical realization of quantum gates with neutral atom qubits include intensity fluctuations, where variations in laser or microwave power alter the effective pulse area and lead to rotation angle errors (\(\theta \to \theta + \delta\theta\)); frequency drift, where temporal fluctuations in the driving field frequency result in detuning from the qubit transition, effectively introducing unwanted \(Z\) rotations and dephasing; and crosstalk, where imperfect spatial addressing causes unintentional excitation of neighboring atoms, leading to correlated errors and decoherence.
To mitigate systematic amplitude and detuning errors, composite pulse sequences such as BB1, CORPSE, and Walsh-modulated schemes are commonly employed. These techniques are designed under the assumption that the dominant errors are coherent and approximately global across the entire pulse train, thereby enabling robust error cancellation through constructive interference of control segments.

\subsection{Integrated Photonics and Directional Couplers}
\label{sec:Integrated Photonics}

In integrated photonic quantum computing, qubits are frequently encoded in the path degree of freedom of single photons using planar waveguide structures. A central component in such systems is the \emph{directional coupler}, which enables coherent interference between two waveguides via evanescent coupling.

Under coupled-mode theory, the evolution of the optical field amplitudes \( E_1(t) \) and \( E_2(t) \) in a directional coupler of fixed cross-section is governed by the unitary propagator:
\begin{equation}
    \begin{pmatrix}
        E_1(t) \\
        E_2(t)
    \end{pmatrix}
    =
    e^{-i \frac{t}{2}(\Omega X - \Delta Z)}
    \begin{pmatrix}
        E_1(0) \\
        E_2(0)
    \end{pmatrix} ,
    \label{eq:coupler}
\end{equation}
where \( \Omega \) is the coupling coefficient (primarily dependent on the inter-waveguide spacing), \( \Delta \) denotes the detuning or phase mismatch (determined by differences in propagation constants), and \( t \) is the propagation length along the coupler.

The parameters \( \Omega \) and \( \Delta \) are real-valued functions determined by the physical geometry. For silicon-on-insulator rib waveguides, they depend on the widths of the individual waveguides (\( w_1, w_2 \)), their heights (\( H_1, H_2 \)), the gap between the waveguides (\( g \)), the etch depths (\( h_1, h_2 \)), and the coupling length (\( t \)).
When the heights, etch depths, and gap are fixed and symmetric, the dependence simplifies to:
\begin{subequations}
\begin{align}
    \Omega &= \Omega(w_1, w_2), \\
    \Delta &= \Delta(w_1, w_2).
\end{align}
\end{subequations}

This correlated and nontrivial dependence makes traditional composite pulse schemes, which typically assume independent control over rotation axes, nontrivial to apply. Nevertheless, photonic implementations employ alternative strategies to enhance robustness. Post-fabrication tuning using integrated heaters or phase shifters can dynamically adjust phase and coupling to correct static errors. Symmetric design, such as balanced Mach-Zehnder interferometers (MZIs), reduces sensitivity to fabrication deviations. Robust architectures incorporate circuit redundancy and heralding techniques to mitigate the impact of parameter variations and losses. Finally, programmable photonics employs configurable interferometers with adjustable elements to enable adaptive calibration and real-time error correction.

Composite directional coupler schemes offer an alternative approach by improving error resilience through careful static design, without relying on active control. Although these designs increase circuit footprint and complexity, they are well-suited for passive platforms where dynamic reconfiguration is limited or undesired.

Crucially, due to the generator structure in Eq.~\eqref{eq:coupler}, only \( X \) and \( Z \) appear in the effective Hamiltonian. Implementing gates involving \( Y \), such as the \( \mT \) gate, thus requires at least two segments with varying parameters. In Section~\ref{sec:Pre-Distillation in Integrated Photonics}, we introduce composite designs for the \( \mT \) gate that is robust against fully correlated width-dependent errors, following the construction developed in~\cite{KaplanErew}.

\section{Plots for the T-Magic Errors in the Composite Schemes}
\label{app:Plots for the T-Magic Errors in the Composite Schemes}

To provide a deeper understanding of the role composite pulse sequences play in improving non-Clifford gate performance, we include here additional figures that focus specifically on the behavior of the T-magic error. These plots offer a more direct and operationally meaningful view of gate quality, particularly in relation to magic state distillation thresholds.

\begin{figure}[H]
    \centering
    \includegraphics[width=\linewidth]{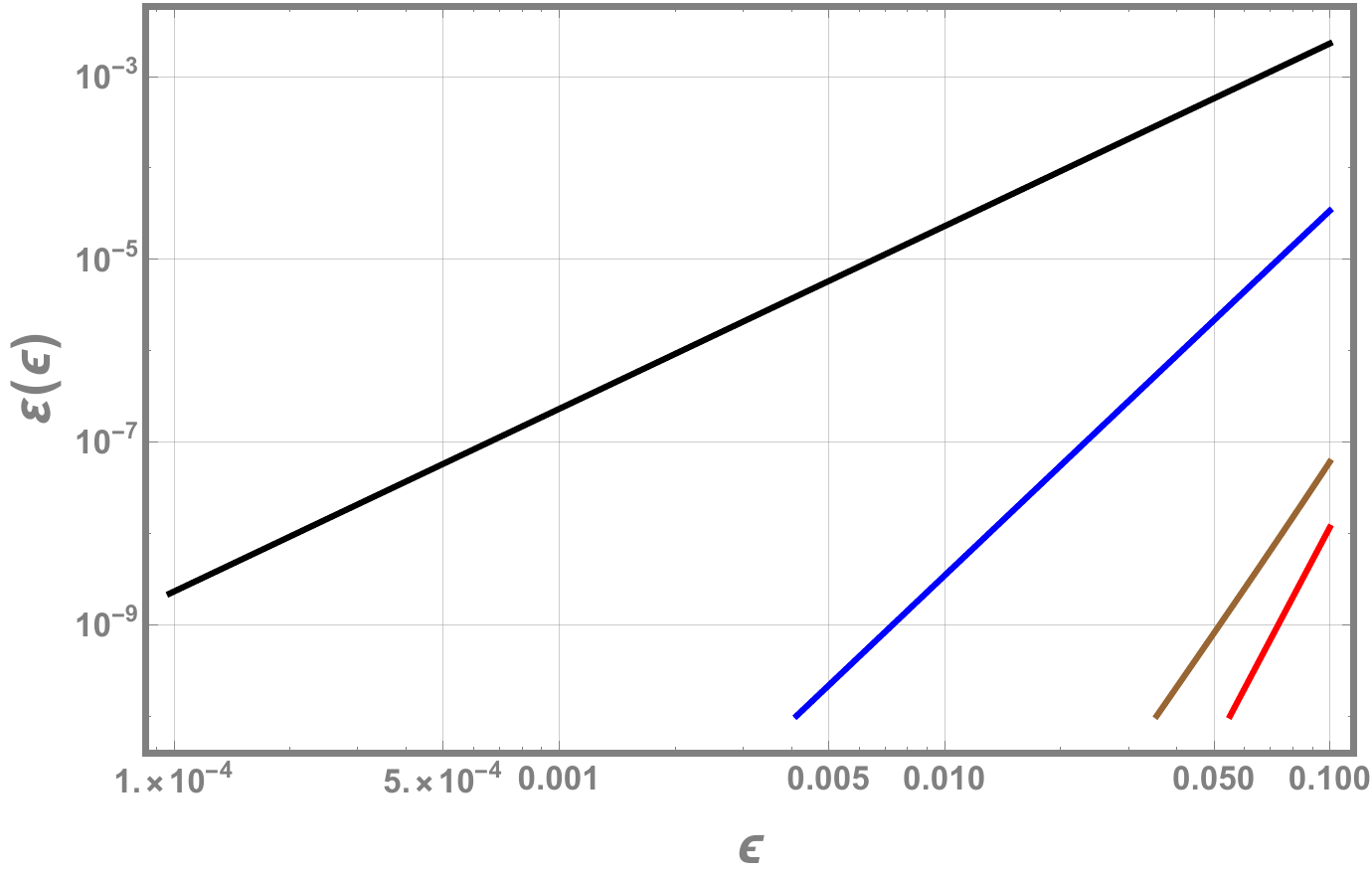}
    \caption{
        T-magic error for the $\mT$ gate as a function of the global physical error $\epsilon$ in $X$-$Y$ control systems with global Rabi frequency errors.
        Each curve corresponds to a composite sequence with a different number of pulses: single-pulse (black), three-pulse (blue), five-pulse (red), and seven-pulse (brown).
        All values are plotted on logarithmic scales.
        Composite pulse sequences significantly reduce the T-magic error, particularly in the low-error regime, thereby improving gate fidelity before any distillation is applied.
    }
    \label{fig: X-Y systems - log e135 vs log e}
\end{figure}

\begin{figure}[H]
    \centering
    \includegraphics[width=\linewidth]{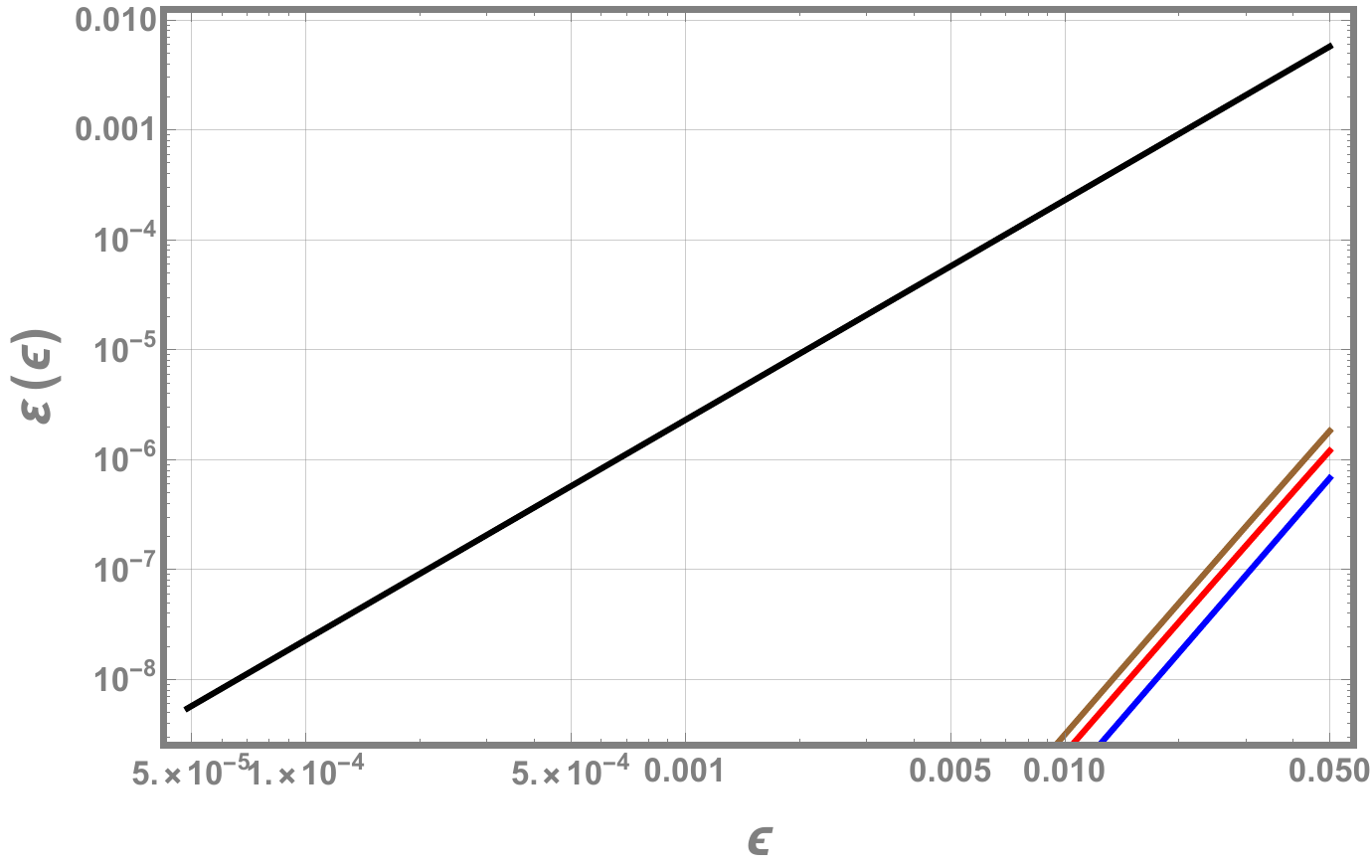}
    \caption{
        T-magic error as a function of the global physical error $\epsilon$ under correlated $Z$ errors in $X$-$Z$ Hamiltonians.
        The black curve represents the two-segment baseline, while the red, blue, and brown curves correspond to three distinct three-pulse composite designs labeled (a), (b), and (c).
        Composite sequences achieve notable suppression of the T-magic error across a broad range of error values, outperforming the baseline design.
    }
    \label{fig: Example for XZ - log e135 vs log e}
\end{figure}

\begin{figure}[H]
    \centering
    \includegraphics[width=\linewidth]{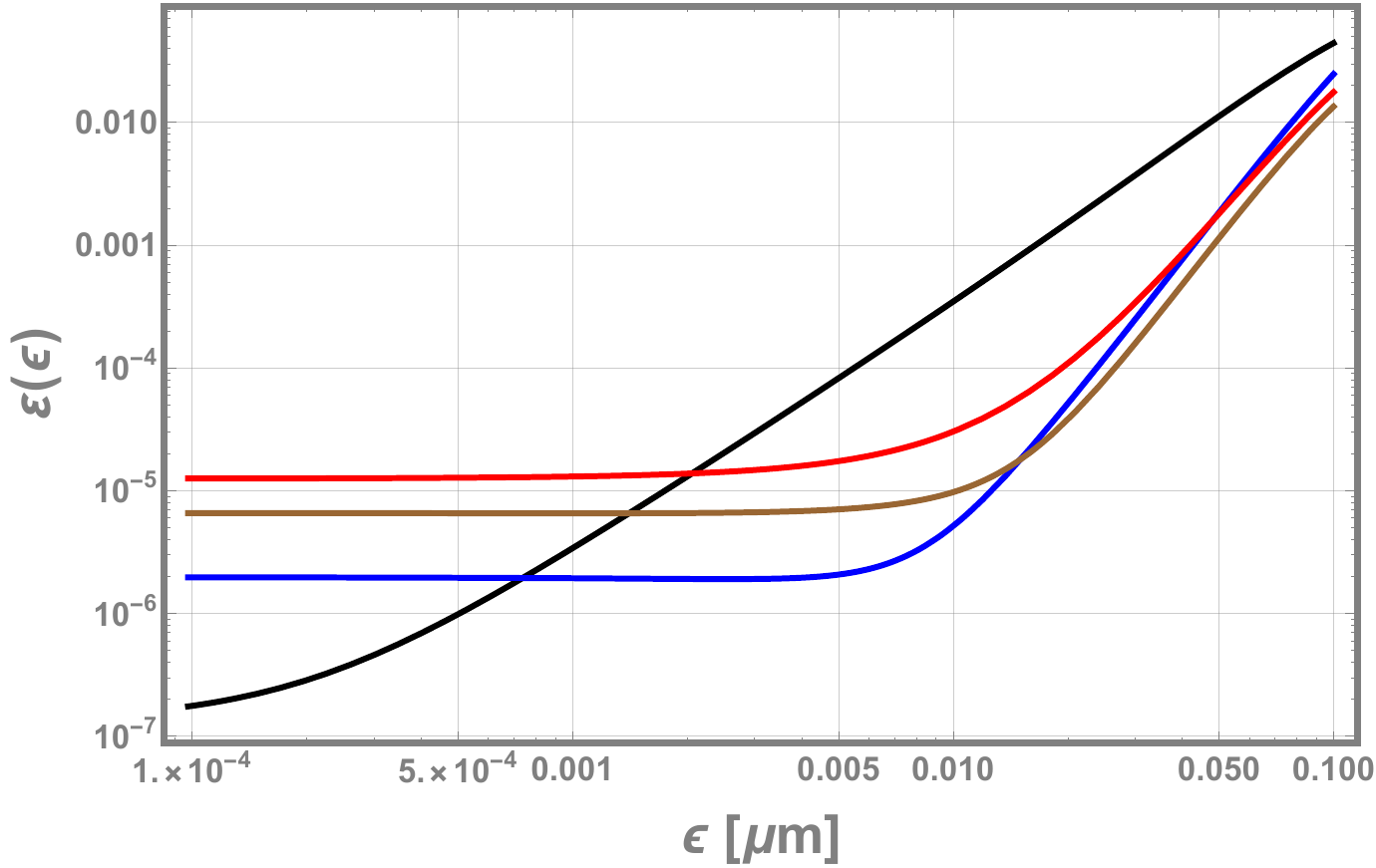}
    \caption{
        T-magic error plotted versus the global physical error $\epsilon$ for directional coupler implementations in integrated photonics, where imperfections in width introduce global systematic errors.
        The black curve shows the performance of the two-segment design, and the blue, red, and brown curves represent four-segment composite schemes (a), (b), and (c), respectively.
        Although the composite schemes may slightly underperform the baseline at very small $\epsilon$, they demonstrate superior robustness across the physically relevant error range.
    }
    \label{fig: COUPLERS - log e135 vs log e}
\end{figure}

Figures~\ref{fig: X-Y systems - log e135 vs log e},~\ref{fig: Example for XZ - log e135 vs log e}, and~\ref{fig: COUPLERS - log e135 vs log e} demonstrate the effectiveness of composite pulse schemes in enhancing the fidelity of the \( \mT \) gate under global systematic errors. Each plot shows the T-magic error as a function of the global physical error \( \epsilon \), both on logarithmic scales. The figures correspond to the designs developed in Sections~\ref{sec:X-Y based systems},~\ref{sec:Pre-Distillation in $X$-$Z$ Systems ; example}, and~\ref{sec:Pre-Distillation in Integrated Photonics}, respectively, with color schemes matched to the design-specific curves introduced in each section. In all cases, the composite pulse sequences exhibit significantly improved error suppression compared to the single-pulse implementation, especially in the low-error regime.

\begin{figure}[H]
    \centering
    \includegraphics[width=0.95\linewidth]{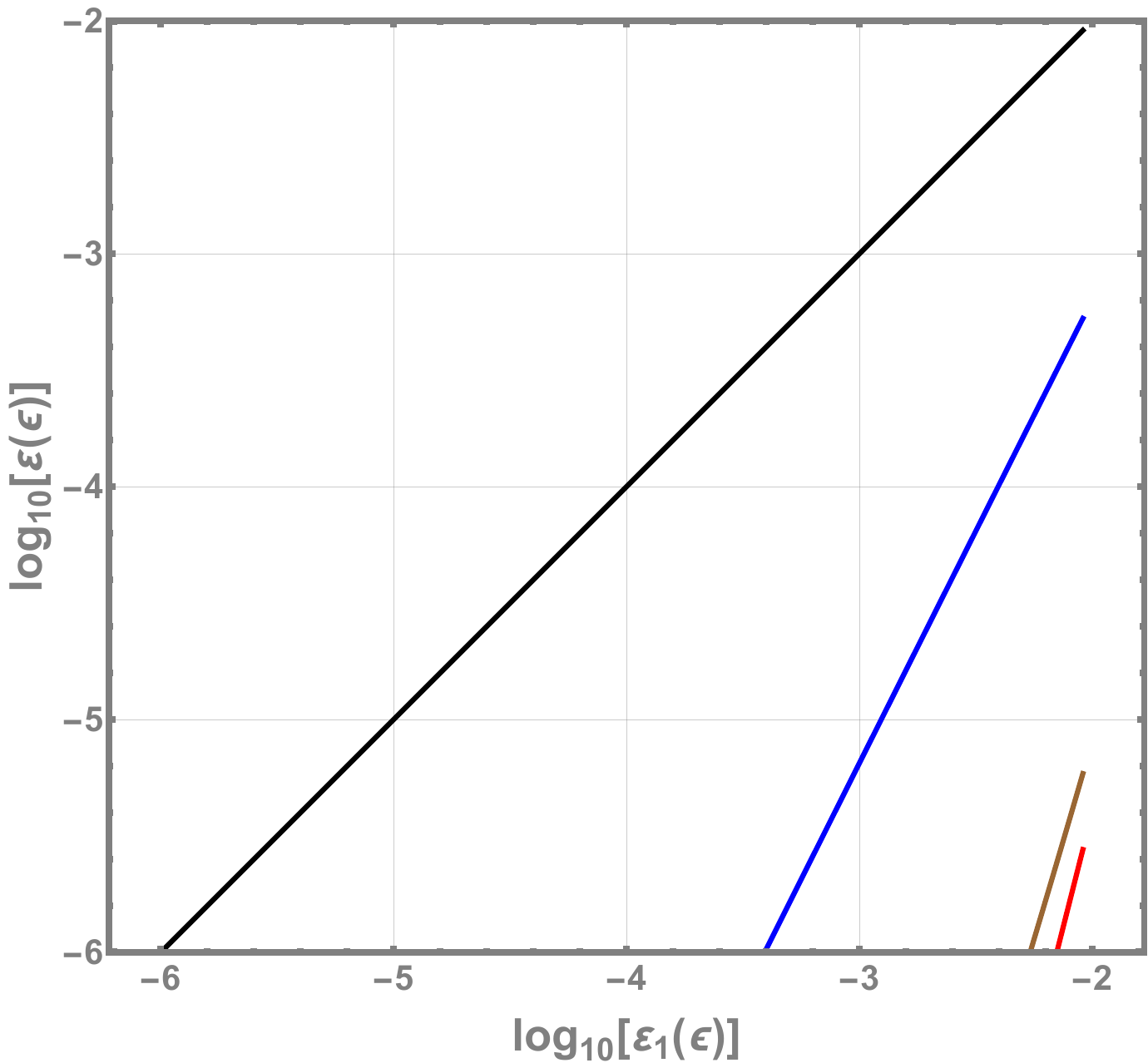}
    \caption{
        Log-log plot comparing the T-magic error of composite $\mT$ gate implementations to that of the single-pulse gate in $X$-$Y$ control systems with global Rabi frequency errors.
        The $x$-axis shows the base-10 logarithm of the error for the single-pulse implementation, and the $y$-axis shows the logarithm of the T-magic error for each design.
        The steeper slopes of the multi-pulse curves indicate improved error scaling and robustness, essential for reducing distillation overhead in practical settings.
    }
    \label{fig: X-Y systems - log e135 vs log e1}
\end{figure}

\begin{figure}[H]
    \centering
    \includegraphics[width=0.95\linewidth]{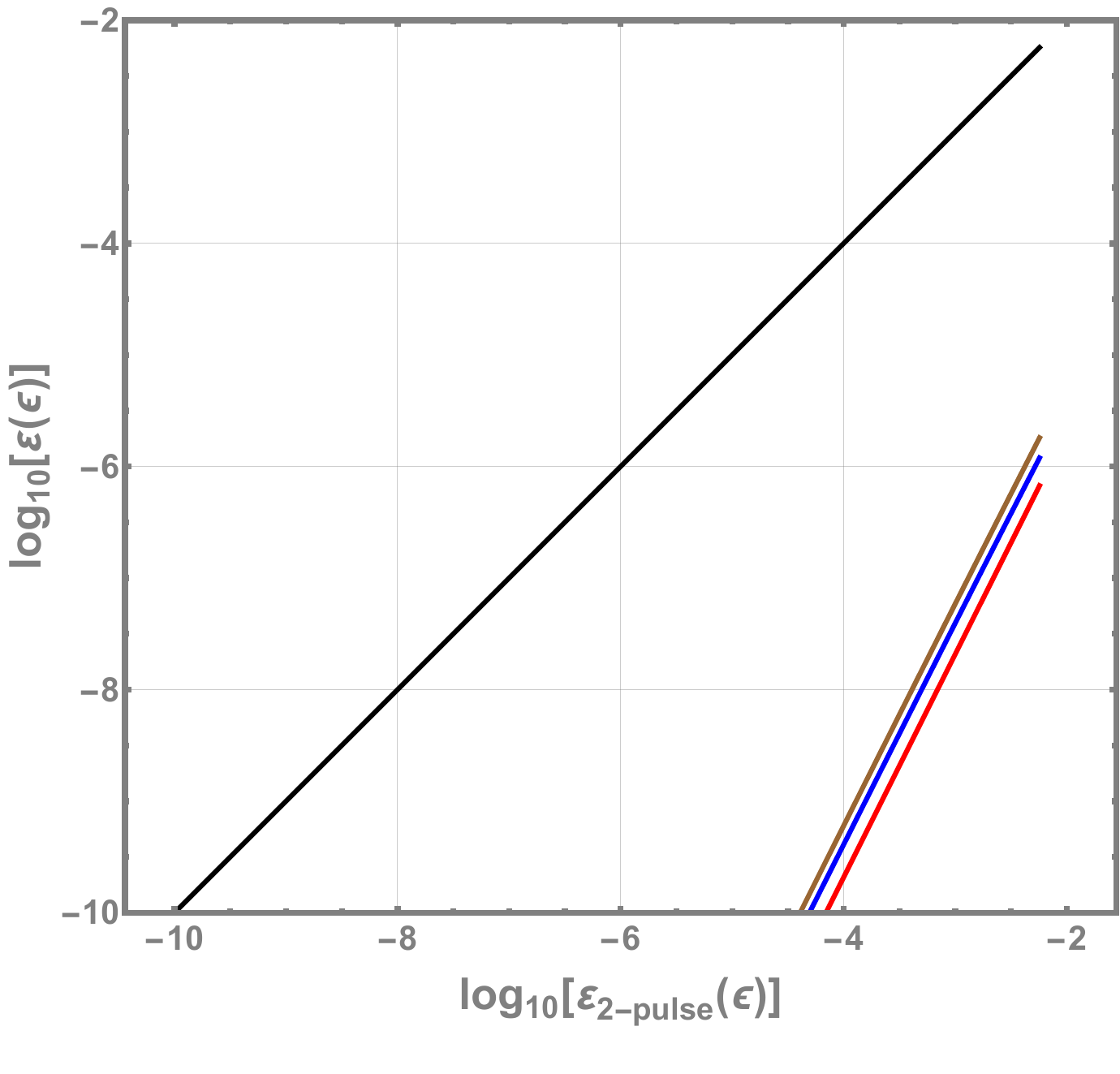}
    \caption{
        Log-log comparison of the T-magic error for three-pulse composite sequences versus the baseline two-segment gate in $X$-$Z$ systems with global $Z$ errors.
        Each point reflects the composite scheme’s T-magic error plotted against that of the corresponding single-pulse gate.
        The improved slope indicates enhanced scaling behavior of the composite designs, enabling a lower distillation burden in quantum error correction protocols.
    }
    \label{fig: Example for XZ - log e135 vs log e1}
\end{figure}

\begin{figure}[H]
    \centering
    \includegraphics[width=0.7\linewidth]{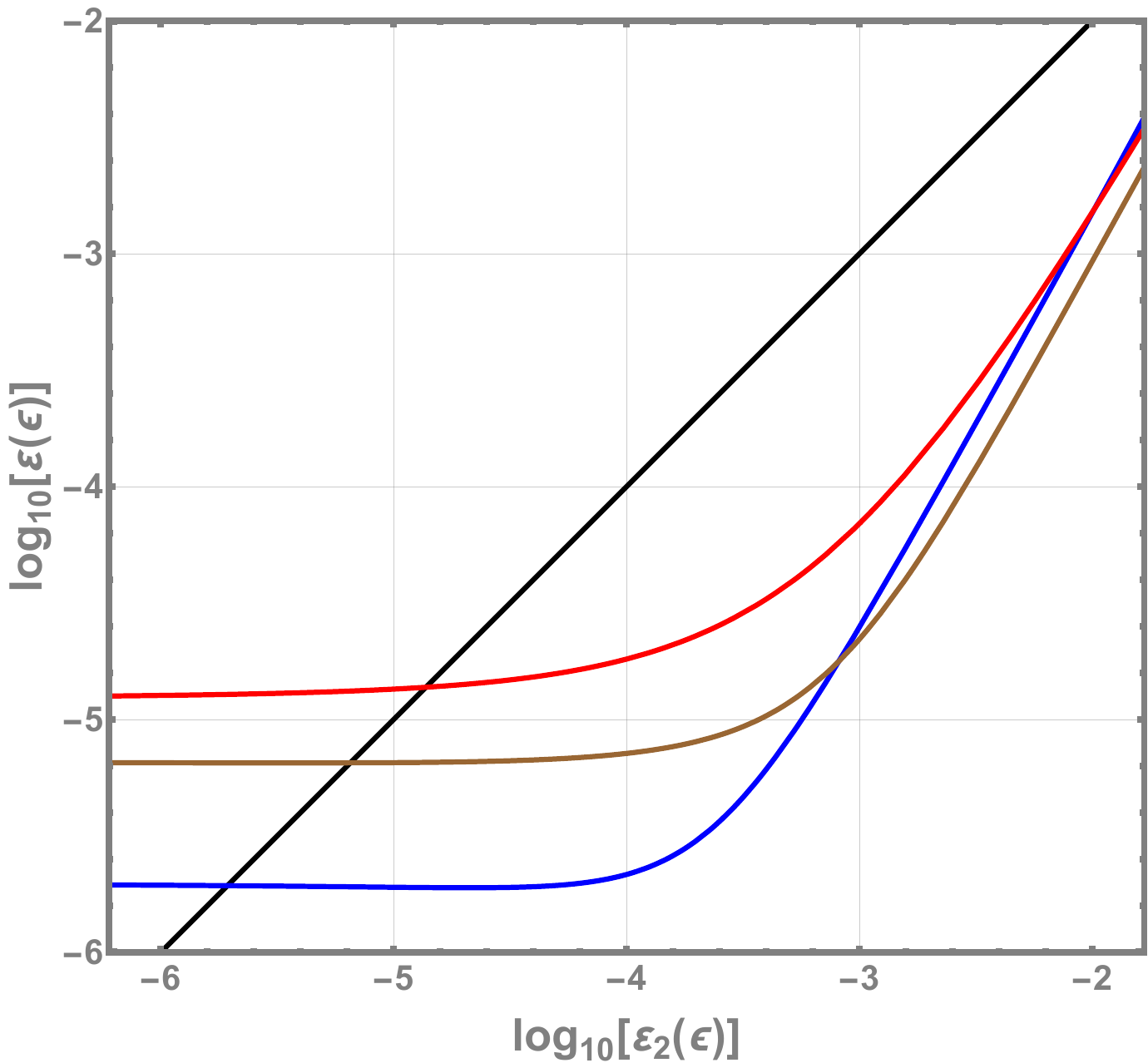}
    \caption{
        Logarithmic comparison of T-magic error scaling for $\mT$ gate implementations in integrated photonic circuits with directional couplers.
        The plot compares composite pulse schemes (a, b, c) to the baseline two-segment design in terms of T-magic error suppression relative to the single-pulse gate error.
        Despite similar performance in the low-error limit, the composite schemes (especially scheme (c)) show better scaling and reduced distillation demands across the full error range as discussed in the main text.
    }
    \label{fig: COUPLERS - log e135 vs log e1}
\end{figure}

Figures~\ref{fig: X-Y systems - log e135 vs log e1},~\ref{fig: Example for XZ - log e135 vs log e1}, and~\ref{fig: COUPLERS - log e135 vs log e1} provide a complementary perspective by plotting the base-10 logarithm of the \( \mT \)-gate T-magic error (see Eq.~(\ref{eq:natural error of the gate})) against the base-10 logarithm of the error associated with the corresponding single-pulse implementation. The steeper slopes observed for the composite pulse curves indicate superior error scaling, underscoring the effectiveness of composite control in improving gate robustness- a critical requirement for fault-tolerant quantum computation.

These six figures complement the results presented in the main text, where we plotted the Frobenius distance fidelity, the Frobenius overlap fidelity, and the corresponding reduction in the number of distillation iterations required to reach a target gate error of \(10^{-15}\). That distillation overhead is directly determined by the T-magic error, which is the quantity plotted in the figures shown here.

\section{Plots for the Frobenius Fidelity and the T-Magic Fidelity}
\label{app:Plots for the Frobenius Fidelity and the T-Magic Fidelity}

In this appendix, we illustrate the behavior of the newly introduced fidelity measure, the \emph{T-magic fidelity}, by comparing it to the conventional Frobenius fidelity across a broad range of error values. This comparison is intended to provide the reader with intuition for how this operationally motivated measure behaves and how it differs from standard fidelity metrics. Notably, the T-magic fidelity does not necessarily exhibit monotonic behavior with respect to traditional gate fidelities, highlighting its distinct functional character and the specific insights it offers. Recall that the T-magic fidelity is designed to quantify how effectively a given implementation of the $\mT$ gate prepares the target magic state when acting on the initial state $\ket{0}$ for MSD.

\begin{figure}[H]
    \centering
    \includegraphics[width=0.7\linewidth]{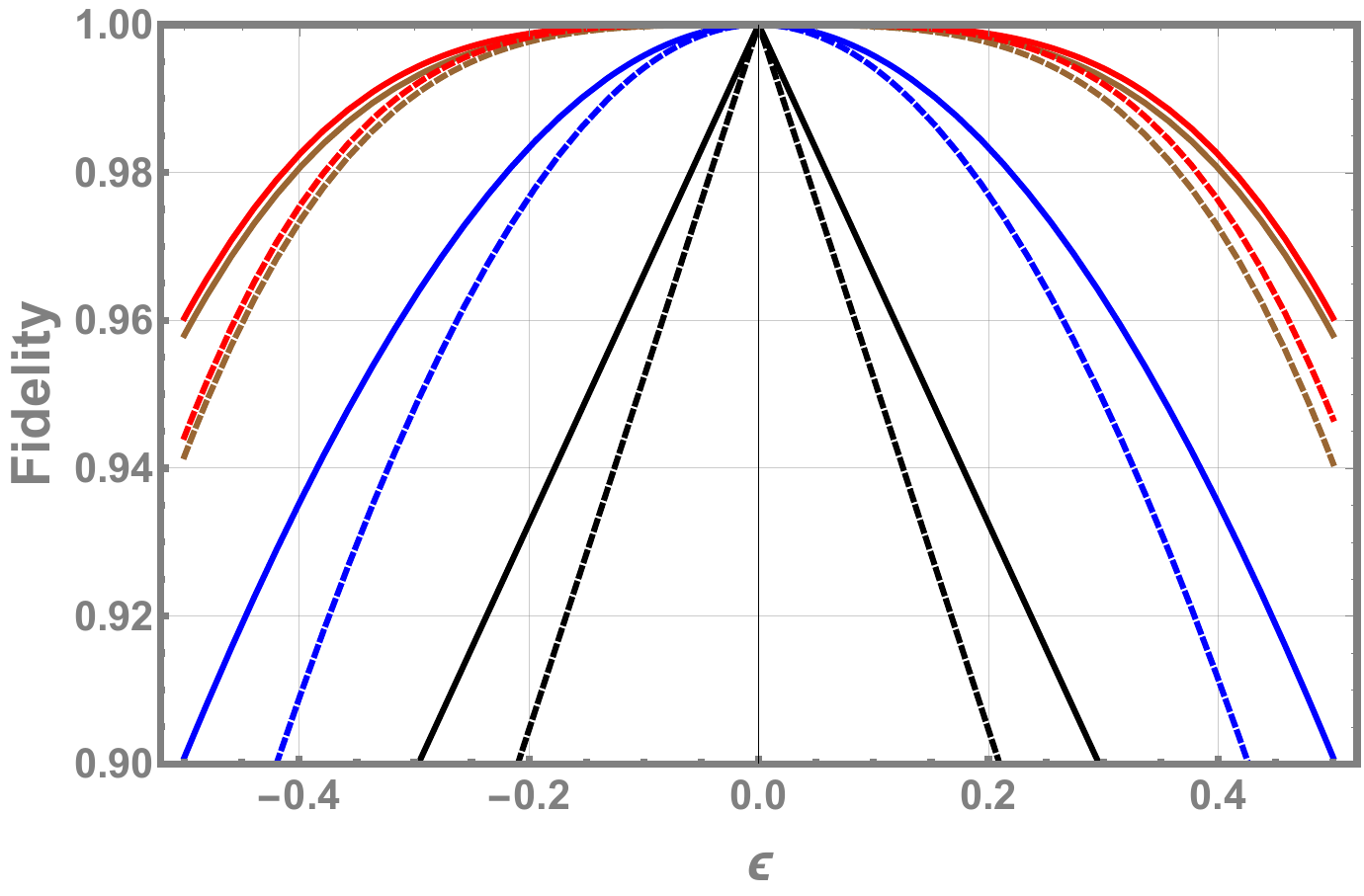}
    \caption{
        Comparison between the Frobenius fidelity (solid curves) and the T-magic fidelity (dashed curves) for $\mT$ gate implementations in $X$-$Y$ control systems with global Rabi frequency errors.
        Each curve corresponds to a composite sequence with a different number of pulses: single-pulse (black), three-pulse (blue), five-pulse (red), and seven-pulse (brown).
    }
    \label{fig: magic fidelity - XY}
\end{figure}

\begin{figure}[H]
    \centering
    \includegraphics[width=0.7\linewidth]{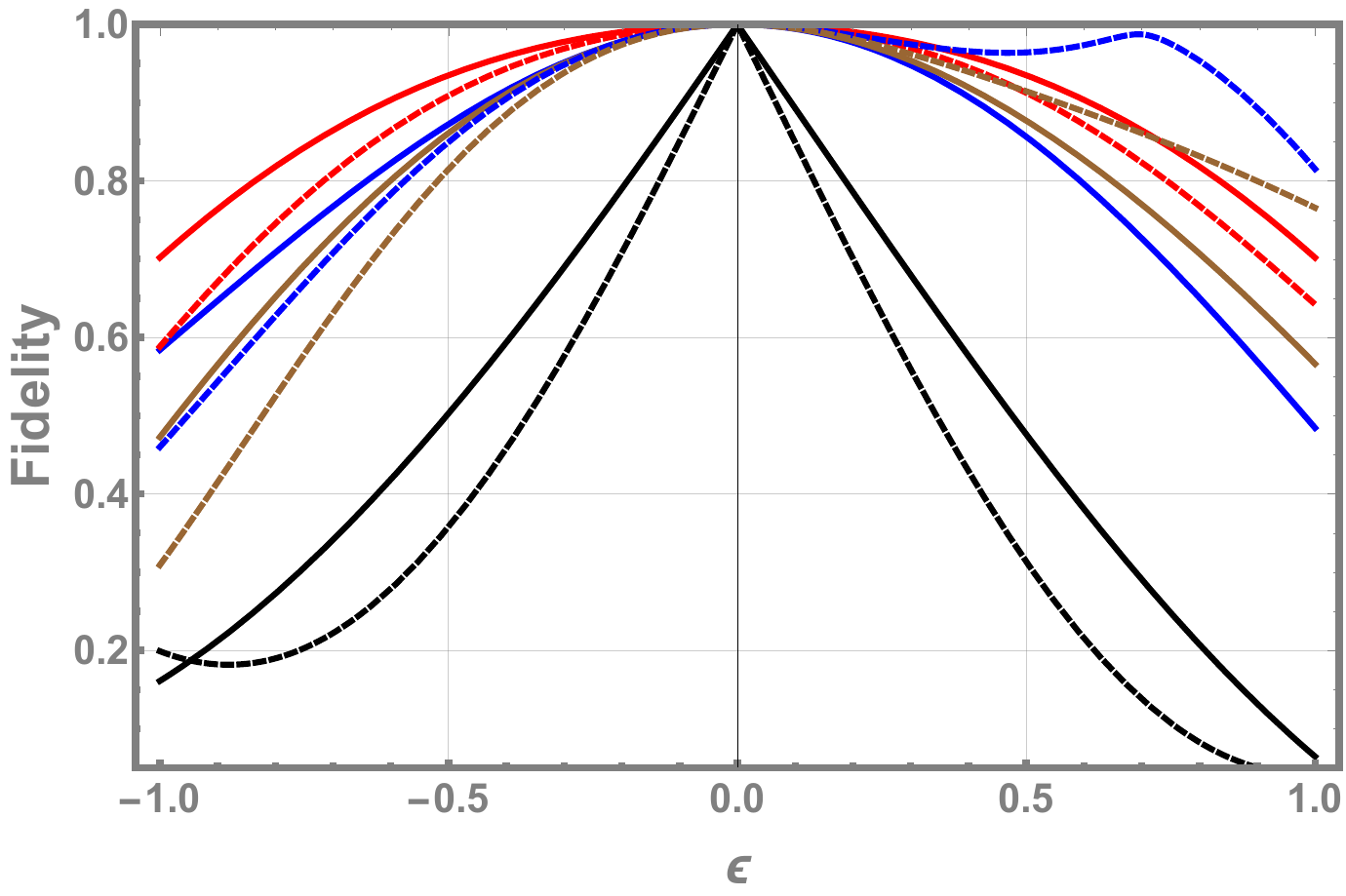}
    \caption{
        Comparison of Frobenius fidelity (solid lines) and T-magic fidelity (dashed lines) for $\mT$ gate implementations under correlated systematic $Z$ errors in $X$-$Z$ Hamiltonians.
        The black curve represents the two-segment baseline, while the red, blue, and brown curves correspond to three distinct three-pulse composite designs labeled (a), (b), and (c).
    }
    \label{fig: magic-fidelity - Example}
\end{figure}

\begin{figure}[H]
    \centering
    \includegraphics[width=0.7\linewidth]{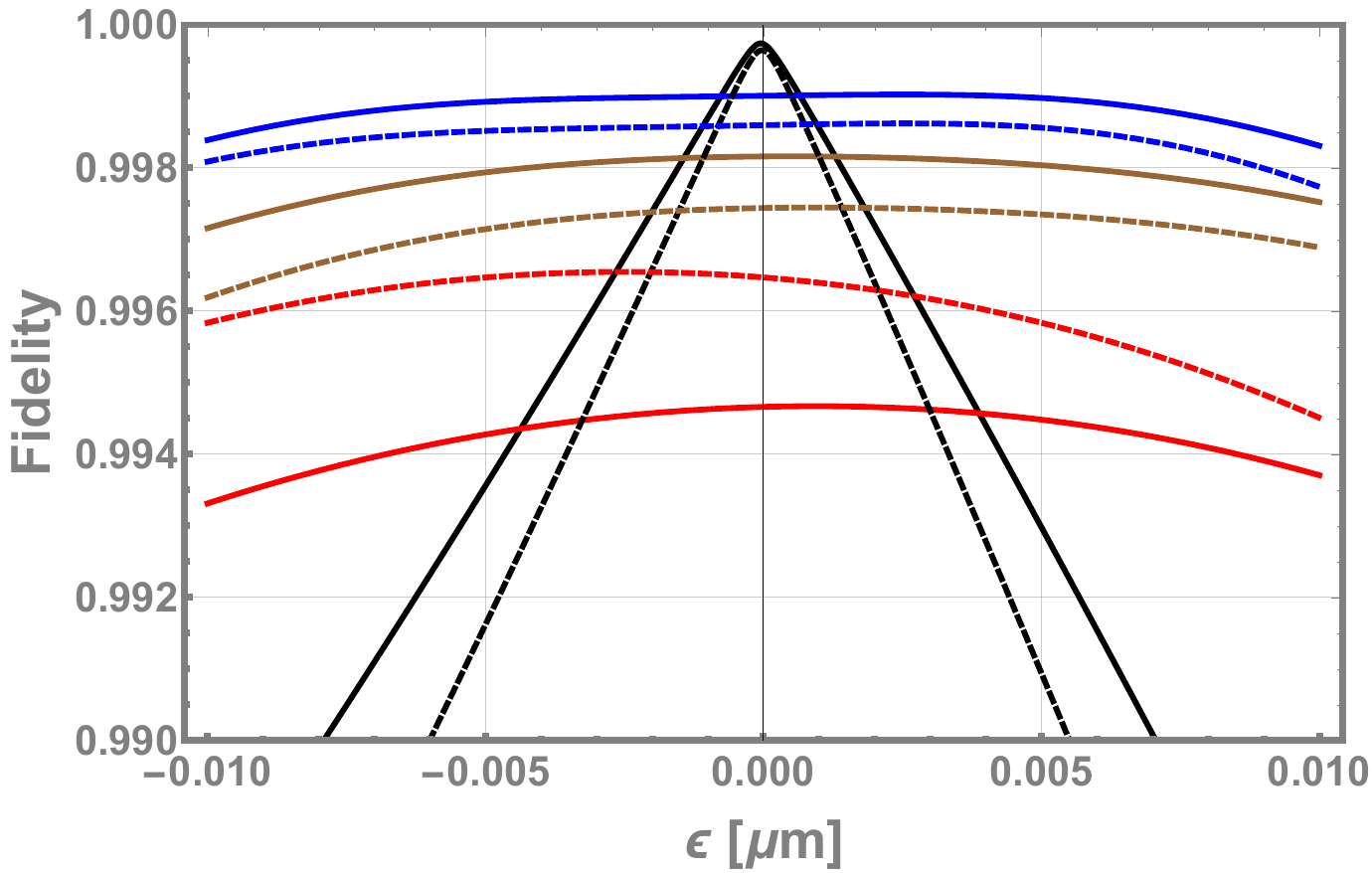}
    \caption{
        Frobenius fidelity (solid) versus T-magic fidelity (dashed) for $\mT$ gate implementations using directional couplers in integrated photonics, where global systematic width errors dominate.
        The black curve shows the performance of the two-segment design, and the blue, red, and brown curves represent four-segment composite schemes (a), (b), and (c), respectively.
    }
    \label{fig: magic fidelity - COUPLERS}
\end{figure}

Figures~\ref{fig: magic fidelity - XY},~\ref{fig: magic-fidelity - Example}, and~\ref{fig: magic fidelity - COUPLERS} show this comparison under global systematic errors for the composite designs discussed in Sections~\ref{sec:X-Y based systems},~\ref{sec:Pre-Distillation in $X$-$Z$ Systems ; example}, and~\ref{sec:Pre-Distillation in Integrated Photonics}, respectively. In each case, solid lines represent the Frobenius fidelity, while dashed lines correspond to the T-magic fidelity, with color coding consistent with the composite pulse schemes introduced in the respective sections. These plots clearly reveal the non-trivial relationship between the two measures and underscore the unique role of the T-magic fidelity in evaluating the gate’s suitability for MSD.

\end{document}